\documentclass{gGAF2e}

\usepackage{multiequations}
\newcommand*{\be}{\begin{equation}}
\newcommand*{\ee}{\end{equation}}
\newcommand*{\bme}{\begin{multiequations}}
\newcommand*{\eme}{\end{multiequations}}
\newcommand*{\bse}{\begin{subequations}}
\newcommand*{\ese}{\end{subequations}}

\renewcommand{\vec}[1]{\mbox{\boldmath $#1$}}

\newcommand{\dbtilde}[1]{\tilde{\raisebox{0pt}[0.85\height]{$\tilde{#1}$}}}

\begin{document}

\jvol{00} \jnum{00} \jyear{2012} %\jmonth{February}

%\markboth{A. Giesecke et al.}{Kinematic dynamo action of a precession driven flow}
\markboth{A. Giesecke et al.}{GEOPHYSICAL AND ASTROPHYSICAL FLUID DYNAMICS}
\title{Kinematic dynamo action of a precession driven flow based on the results of water experiments and hydrodynamic simulations.}
\author{Andr{\'e} Giesecke$^\ast$\thanks{$^\ast$Corresponding author. Email: a.giesecke@hzdr.de\vspace{6pt}}, Tobias Vogt, Thomas Gundrum, and Frank Stefani\\\vspace{6pt}
  Helmholtz-Zentrum Dresden-Rossendorf, Bautzner Landstrasse 400, D-01328 Dresden, Germany\\
\vspace{6pt}\received{v1.2 released ??? 2017}}

\maketitle

\begin{abstract}
The project DRESDYN (DREsden Sodium facility for DYNamo and
thermohydraulic studies) conducted at Helmholtz-Zentrum
Dresden-Rossendorf (HZDR) provides a new platform for a variety of
liquid sodium experiments devoted to problems of geo- and
astrophysical magnetohydrodynamics. The most ambitious experiment
within this project is a precession driven dynamo experiment that
currently is under construction. It consists of a cylinder filled
with liquid sodium that simultaneously rotates around two axes. The
experiment is motivated by the idea of a precession-driven flow as a
complementary energy source for the geodynamo or the ancient lunar
dynamo.

In the present study we address numerical and
experimental examinations in order to identify parameter regions where
the onset of magnetic field excitation will be most probable.  Both
approaches show that in the strongly nonlinear regime the flow is
essentially composed of the directly forced primary Kelvin mode and
higher modes in terms of standing inertial waves that arise from
nonlinear self-interactions.  A peculiarity is the resonance-like
emergence of an axisymmetric mode that represents a double roll
structure in the meridional plane, which, however, only occurs in a
very limited range of the precession ratio. This axisymmetric mode
turns out to be beneficial for dynamo action, and kinematic
simulations of the magnetic field evolution induced by the
time-averaged flow exhibit magnetic field excitation at critical
magnetic Reynolds numbers around ${\rm{Rm}}^{\rm{c}}\approx 430$,
which is well within the range of the planned liquid sodium
experiment.
\end{abstract}


\section{Introduction}

Precession driven flows are found in technical applications, e.g. in
flying and rotating objects with liquid payloads (spacecrafts,
rockets, satellites), as well as in geophysical problems, like
cyclonic structures in the Earth’s atmosphere that form hurricanes or
tornados.  Precession driven flows are also relevant in the liquid
part of the Earth's core \citep{stewartson1963}, which more than 100
years ago was already a motivation for the pioneering study of Henri
Poincar{\'e}, who showed that the inviscid base flow in a precessing
spheroid is described by a constant vorticity solution, the spin-over
mode \citep{poincare1910}. The extension of the calculation to the
weakly nonlinear regime including thin boundary layers yields the
Busse solution which describes the orientation of the effective
rotation axis of the fluid \citep{busse1968}.  Early experiments
showed that precession is an efficient mechanism to drive a flow in a
rotating container and provides powerful flows on the laboratory scale
without making use of propellers or pumps
\citep{leorat2003,leorat2006}.  In this context it is often speculated
whether the Earth's magnetic field or the ancient magnetic field of
the moon were powered by a flow driven by precession instead of or in
addition to convection \citep{malkus1968,vanyo1991,noir2013}. Dynamos
generated by mechanical driving such as precession or tidal forcing
have become popular in recent years, and experiments focussing on
precession driven flows were conducted in various labs
\citep{mcewan1970,gans1970,vanyo1971,manasseh1992,kobine1995,kobine1996,noir2001a,noir2003,goto2007,lagrange2011,mouhali2012,lin2014,herault2015,horimoto2017}.

\begin{figure}[h!]
\begin{center}
\resizebox*{10cm}{!}%
{\includegraphics{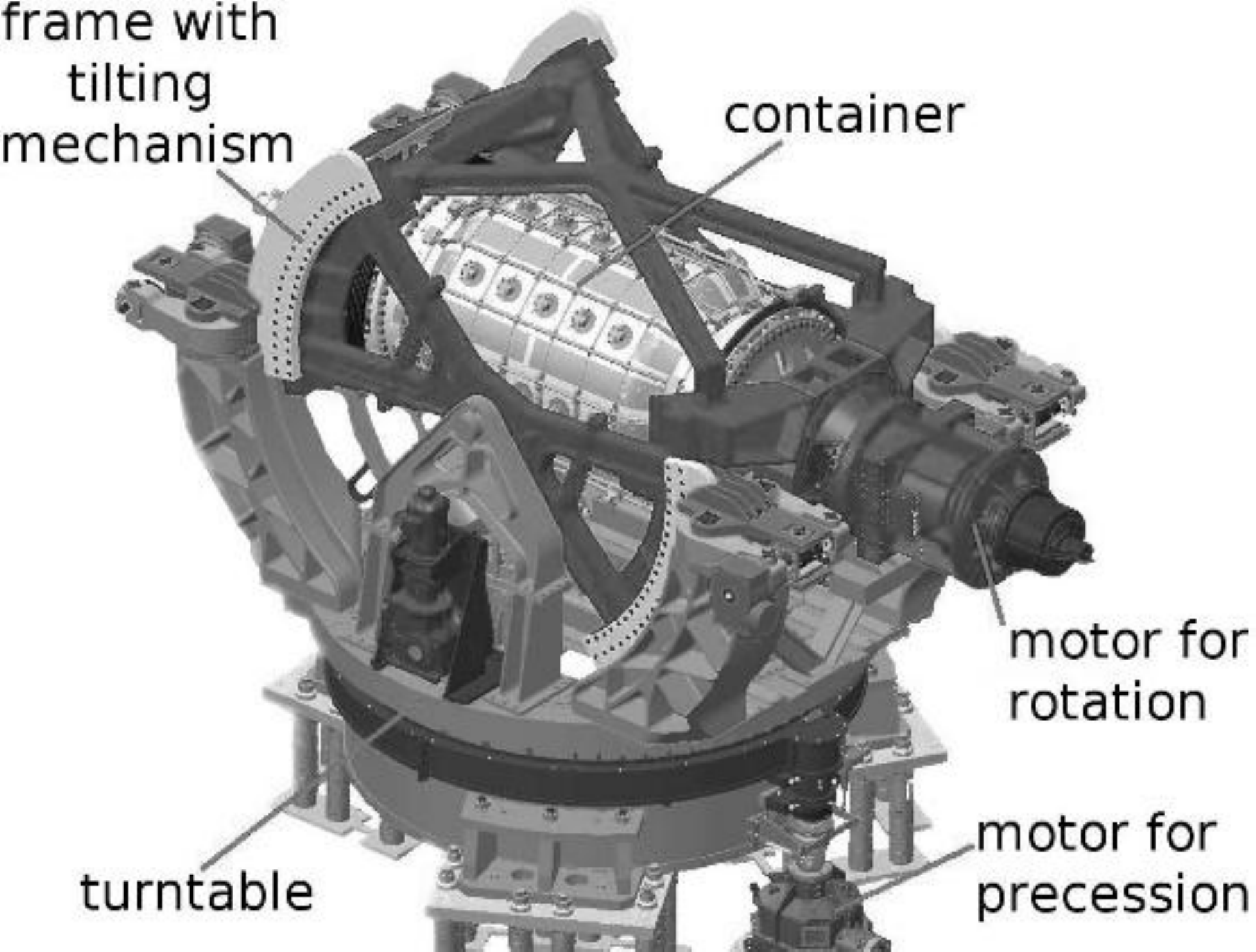}}
\caption{\label{fig::sketch_sodiumdevice} %
Sketch of the planned precession experiment. The container (central
cylinder) with a height and diameter of $2\mbox{ m}$ will rotate with
up to $10\mbox{ Hz}$. The device is mounted on a turntable that in
turn can rotate with up to $1\mbox{Hz}$. The angle between rotation
axis and precession axis, i.e., the nutation angle, can be varied from
$\alpha=45^{\circ}$ to $\alpha=90^{\circ}$ (Figure courtesy SBS
B{\"u}hnentechnik GmbH).}
\end{center}
\end{figure}

At Helmholtz-Zentrum Dresden-Rossendorf (HZDR) a dynamo experiment is
presently under construction in which a precessing flow of liquid
sodium is used in order to excite dynamo action \citep{stefani2012}.
The experiment consists of a cylinder with radius $R = 1 \mbox{ m}$
and height $H = 2 \mbox{ m}$ that will rotate around its symmetry axis
with a frequency of up to $f_{\rm{c}} = 10 \mbox{ Hz}$ and precess
around a second axis with up to $f_{\rm{p}} = 1 \mbox{ Hz}$
(figure~\ref{fig::sketch_sodiumdevice}).  A cylindrical geometry has
been chosen because the pressure forces that arise in the corners at
the end caps provide a more efficient transfer of kinetic energy into
the fluid compared to the spherical case where this only happens by
viscous coupling at the walls.  The device represents a considerably
enlarged version of the liquid metal experiment conducted by
\cite{gans1971} who achieved an amplification of an applied magnetic
field by a factor of 3. However, in these experiments, the cylindrical
container was far to small ($R = 0.125 \mbox{ m}$) to provide a
sufficiently large magnetic Reynolds number required for the onset of
dynamo action.

First water test-experiments at HZDR focussing on the mechanical
stability of the device and on the properties of the flow are
scheduled for late 2019.  In the very dynamo runs the cylinder will be
filled with liquid sodium which is the best known liquid
conductor. The typical critical magnetic Reynolds number that will be
achievable in the dynamo experiment will be around
${\rm{Rm}}={\varOmega}_{\rm{c}}R^2/\eta\approx 700$ where
$\varOmega_{\rm{c}}=2\pi f_{\rm{c}}$ is the angular frequency of the
rotation of the cylinder and $\eta\approx 0.1\mbox{ m}^2/\mbox{s}$ is
the magnetic diffusivity of liquid sodium at a temperature of
$T\approx 120^{\circ}\mbox{ C}$.
  
The present study is an extension of the results presented in
\cite{giesecke2018}. Here we detail numerical and experimental
investigations of the hydrodynamics of a precession driven
flow. Subsequently, the resulting flow data are applied to kinematic
dynamo models which are conducted in preparation of the large scale
experiment in order to estimate the parameter regimes that may be
suited for the onset of dynamo action.  Based on the growth rates for
the magnetic field energy obtained for different velocity fields, we
explicitly show that a dynamo occurs at experimentally attainable
rotation and precession frequencies, although only in a very narrow
regime.  In addition to our previous results, we examine the impact of
a more realistic numerical set-up that considers an outer wall with
finite thickness with a finite electrical conductivity that differs
from the conductivity of the fluid in the interior.

The hydrodynamic water experiments are conducted with a down-scaled
device with the same aspect ratio $\varGamma=H/R=2$ as in the planned
large scale facility.  The experiment has been in operation at HZDR
for several years to provide detailed information on the flow behavior
in dependence of the rotation rate parameterized by the Reynolds
number $Re = \varOmega_{\rm{c}}R^2/\nu$ and the precession ratio
${\rm{Po}} = \varOmega_{\rm{p}}/\varOmega_{\rm{c}}$ with $\nu$ the
viscosity and $\varOmega_{\rm{p}}=2\pi f_{\rm{p}}$ the angular frequency
of the precession.  Note that water has similar density and viscosity
as liquid sodium so that the resulting outcome can well be applied for
scaling to the flow behavior in the large scale liquid sodium device.
\begin{figure}[h!]
\begin{center}
\includegraphics[width=0.7\textwidth]{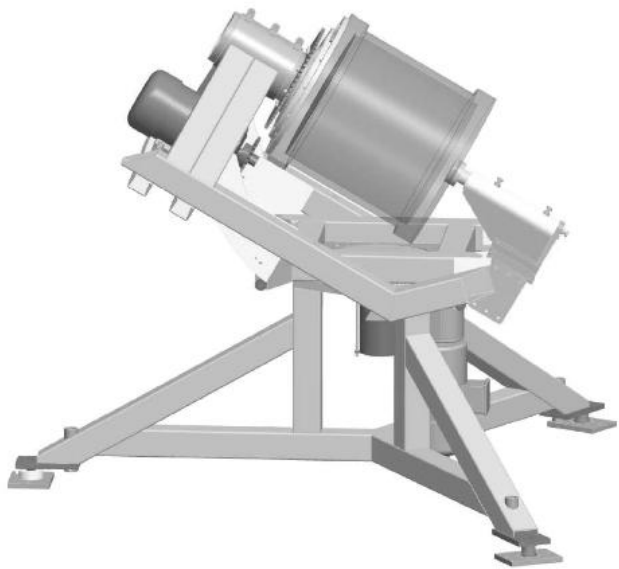}
\caption{\label{fig::sketch_water}
Sketch of the water experiment. Compared to the planned
large scale experiment the water device is smaller
by a factor of 6.}
\end{center}
\end{figure}

It is well-known that a fluid flow forced by precession is instable
\citep{kerswell1993} and allows complex three-dimensional velocity
structures that ultimately may end in a fully turbulent state.  Such
instabilities in precession driven flows have been found in
experiments and simulations
\citep{noir2003,hollerbach1995,noir2001a,lorenzani2001,lorenzani2003,noir2001b,goto2014,horimoto2018}.
Moreover, simulations in various geometries (spherical, ellipsoidal,
cylindrical and cubic) have shown that the precession driven
generation of magnetic fields is possible with a critical magnetic
Reynolds number ${\rm{Rm}}^{\rm{c}}\sim O(10^3)$
\citep{tilgner2005,hullermann2011,wu2009,lin2016,goepfert2016,nore2011},
which has strongly encouraged the construction of our precession
dynamo experiment.

The preliminary experimental investigations are accompanied by
hydrodynamic simulations with the code {\tt{SEMTEX}} that applies
quadrilateral spectral elements for the direct numerical simulation of
the incompressible Navier Stokes equation
\citep{blackburn2004}\footnote{The code including documentation and
examples is available at
{\tt{http://users.monash.edu.au/$\sim$bburn/semtex.html}}.}.
The code has been thoroughly validated against analytical solutions
and shows an excellent agreement with experimental data (see below).
The two complementary approaches provide extensive information about
the flow in the precessing cylinder and thus allow an estimation of
the system's capability to obtain dynamo action in the planned large
scale sodium device.  For this purpose, the flow fields obtained in
the hydrodynamic simulations serve as input for kinematic dynamo
simulations that allow an assessment in which parameter ranges the
occurrence of dynamo action is most likely.

The present paper is organized as follows: In the next section we
briefly describe the theory and give a qualitative overview about the
various characteristics of a precession driven flow in different
regimes defined by the Poincar{\'e} number ${\rm{Po}}$, i.e., the
precession ratio (the ratio of precession frequency and the rotation
frequency) ${\rm{Po}}=\varOmega_{\rm{p}}/\varOmega_{\rm{c}}$.

The third section presents the results of the hydrodynamic experiment
and simulations, which are applied in kinematic dynamo models that are
detailed in the fourth section.  In the final section we conclude our
results and present an outlook for the large scale experiment.

%%%%%%%%%%%%%%%%%%%%%%%%%%%%%%%%%%%%%%%%%%%%%%%%%%%%%%%%%%%%%%%%%%%%%%%%%%%%%%%%%%%%

\section{Inertial waves in a precessing fluid}

In a precessing reference frame (we will call this in the following the
mantle frame) the fluid flow is described by the Navier-Stokes
equation including additional terms for a Coriolis force and the
Poincar{\'e} force \citep{tilgner1998}:
\begin{equation}
\frac{\upartial\vec{u}}{\upartial t} +\vec{u}{\bm \cdot}{\bm \nabla}\vec{u}\,
=\,-\,{\bm \nabla} P\,\underbrace{-\,2(\vec{\varOmega_{\rm{p}}}
+\vec{\varOmega_{\rm{c}}})\times\vec{u}}_{\mbox{Coriolis force }\vec{F}_{\!\!\rm{c}}}\,\,
\underbrace{-\,(\vec{\varOmega_{\rm{p}}}
\times\vec{\varOmega_{\rm{c}}})
\times\vec{r}}_{\mbox{Poincar{\'e} force } \vec{F}_{\!\!\rm{p}}}
\,+\,\nu\nabla^2\vec{u}\,.
\label{eq::navier}
\end{equation}
%
%In equation~(\ref{eq::navier})
Here $\vec{u}$ is the velocity field (obeying
${\bm \nabla}{\bm \cdot}\vec{u}=0$), $P$ the modified pressure (including the
centrifugal terms), $\vec{\varOmega}_{\rm{p}}$ the angular frequency
due to the precession, $\vec{\varOmega}_{\rm{c}}$ the angular
frequency of the rotation of the container, and $\nu$ the viscosity.
In the mantle frame $\vec{\varOmega}_{\rm{p}}$ is time dependent and
given by
\begin{equation}
\vec{\varOmega}_{\rm{p}}(t)\,=\,
{\varOmega}_{\rm{p}}\bigl[\sin\alpha\cos(\varOmega_{\rm{c}}t+\varphi)\vec{{\hat{r}}}\,
-\,\sin\alpha\sin(\varOmega_{\rm{c}}t+\varphi)\vec{\hat{\varphi}}+\cos\alpha\vec{{\hat{z}}}\bigr].
\end{equation}
The Poincar{\'e} force on the right hand side of~(\ref{eq::navier})
results from the temporal change of the 
orientation of the rotation axis and provides a volume force that
directly drives a fluid flow.  In cylindrical systems the dynamical
relevant part of this Poincar{\'e} force is
\begin{equation}
\vec{F}_{\!\!\rm{p}}\,=\,
-\varOmega_{\rm{c}}\varOmega_{\rm{p}}r\sin\alpha
\cos(\varOmega_{\rm{c}}t+\varphi)\vec{\hat{z}}
\label{eq::poincare}
\end{equation}
with $\vec{\hat{z}}$ the unit vector along the rotation axis of the
cylinder and $\alpha$ the nutation angle, i.e. the angle between
precession axis and rotation axis.  The spatial structure of the
Poincar{\'e} force specifies the fundamental pattern of the directly
driven flow which consists of a non-axisymmetric structure with the
azimuthal wave number $m=1$ ($\propto \cos\varphi$) and an odd axial
wave number $k$, i.e., anti-symmetric with respect to the equatorial
plane.  In order to characterize the precession driven flow we utilize
the eigenfunctions that result from the linear inviscid
approximation. These solutions are inertial waves and read
\citep{kelvin1880,greenspan}
\begin{equation}
\vec{U}_{j}(r,z,\varphi,t)\,=\,
\vec{u}_{j}(r,z)e^{{\mathrm i}(\omega_{j}t+m\varphi)}\,+\,\mbox{c.c.},\label{eq::kelvinmode}
\end{equation}
where the single index $j$ comprises three numbers $m,k,n$, which
denote the dependence on the azimuthal wave number $m$, the axial wave
number $k$ and a third number $n$ that counts the roots of the
dispersion relation.  This dispersion relation determines the
eigenfrequency of a Kelvin mode and reads
\bme
\label{eq::dispersion}
\be
\omega_{j}\lambda_{j} {\mathrm J}_{m-1}(\lambda_{j}) +
m\left({2}-\omega_{j}\right){\mathrm J}_m(\lambda_{j})\,=\,0
\hskip 7mm \mbox{ with } \hskip 7mm
%\omega_{{j}}=\pm 2 \sqrt{\left(1 +\left(\frac{\lambda_{{j}}}{\varGamma k\pi}\right)^2\right)^{-1}}
\omega_{{j}}\,=\,\pm 2 \Biggl[1 +\biggl(\frac{\lambda_{{j}}}{\varGamma k\pi}\biggr)^{\!2}\Biggr]^{\!-1/2}, \hskip 5mm
\ee
\eme
%
%with
${\mathrm J}_m$ the Bessel function of order $m$, $\varGamma$ the aspect ratio
defined by $\varGamma=H/R$ and $\lambda_{j}$ a radial wave number with
its position in the sequence of zeros of (\ref{eq::dispersion}$a$)
corresponding to the number of half-cycles in the radial direction.
An inertial mode becomes resonant when the frequency of the precession
matches the natural frequency of the corresponding linear solution of
the problem.

Experimental investigations have shown that the linear theory, which
in addition to the structure of the main flow also allows the
calculation of the associated amplitude, essentially provides a good
representation of the observations when the nutation angle $\alpha$
remains small and the system is not close to a resonance of an
inertial wave \citep{kobine1995,meunier2008}.  Linear corrections for
the structure of inertial waves by viscous boundary layers were
computed by \cite{herreman2009phd} for the simplified case of no-slip
boundary conditions at the lateral cylinder walls whereas free-slip
conditions were still assumed for the lids.  A consideration of
no-slip condition on all boundaries requires more complicated models,
which, however, are only available for the directly driven $m=1$ mode
\citep{liao2012}.  In any case, even for small precession ratios
nonlinear effects are important, which may trigger e.g. the resonant
collapse, i.e., the breakdown of laminar-like structure into a chaotic
flow observed repeatedly in precession experiments at small nutation
angles $\alpha$ \citep{mcewan1970,manasseh1992,manasseh1994}, or the
breaking of parity with respect to the equatorial plane
\citep{tilgner2005}.  Nonlinear self-interactions and viscous boundary
layers will also be important for the planned experiment at HZDR with
the aspect ratio $\varGamma=H/R=2$ being close to the theoretical value
$\varGamma=1.9898174$ for the resonance of the simplest Kelvin mode with
$m=1$, $k=1$ and $n=1$.  However, the corresponding nonlinear theory
is a complex issue providing solutions only for particular conditions
in the weakly nonlinear regime \citep{meunier2008,lagrange2011}.  The
difficulty for nonlinear models results from the suppression of
nonlinear self-interactions of inertial modes at first order
\citep{greenspan1969} which hence must happen either at higher order,
within the boundary layers \citep{busse1968}, or at internal
singularities \citep{tilgner2007b}.  Nevertheless, it is precisely
these interactions that are responsible for the major flow
contributions beyond the directly forced mode and for the considerable
modification of the azimuthal circulation \citep{kobine1995} which is
always oriented opposite to the (initial) solid body rotation, giving
rise to a ``braking'' of the flow in the bulk of the cylinder.  A
direct consequence is the detuning of free inertial modes that appear
in terms of triadic resonances
\citep{giesecke2015b,lopez2016,herault2018,albrecht2018} and the
development of a strong shear layer in the vicinity of the lateral
cylinder walls, which may become unstable provoking the abrupt
transition into a turbulent flow regime if the forcing is sufficiently
strong \citep{herault2015,marques2015}. Recent simulations have shown
that such transient phenomena can indeed be explained by nonlinear
models that involve a cascade of triadic resonances
\citep{albrecht2018}.

\begin{figure}[t!]
\begin{center}
\includegraphics[width=\textwidth]{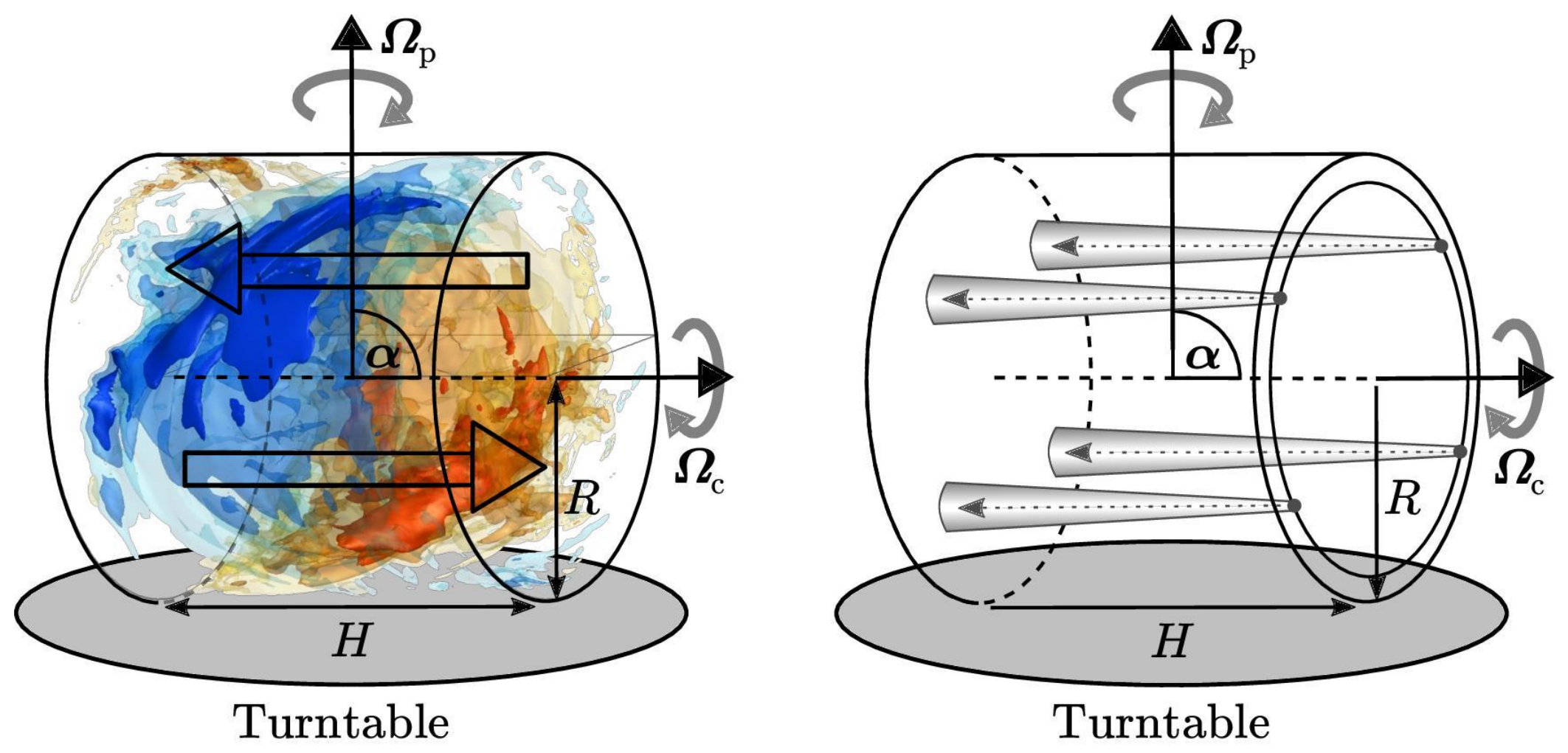}
\caption{\label{fig::sketch_udv} %
Left: Flow field obtained from simulations at ${\rm{Re}}=10^4$ and
${\rm{Po}}=0.1$. The arrows denote the overall orientation of the
axial velocity component (colour online).
Right: sketch of the water experiment with four ultrasound probes
mounted at one end cap of the cylinder. 
}
\end{center}
\end{figure}

%%%%%%%%%%%%%%%%%%%%%%%%%%%%%%%%%%%%%%%%%%%%%%%%%%%%%%%%%%%%%%%%%%%%%%%%%%%%%%%%%%%%%%%

\section{The hydrodynamic flow in the water experiment}

\subsection{Setup}

Let us briefly describe the setup of the precession water experiment
and the measurement techniques that are used to determine the flow
structure and amplitude.  The setup in the down-scaled water
experiment consists of a cylindrical vessel having an inner length of
$H = 326 \mbox{ mm}$ and an inner radius of $R = 163 \mbox{ mm}$.
Similar to the planned liquid sodium device, the down-scaled vessel
can rotate with a maximum rotation rate of $10 \mbox{ Hz}$ around its
axis and is mounted on a turntable frame that allows for an additional
rotation around a second axis with a maximum rotation rate of $1
\mbox{ Hz}$.  The nutation angle between the two axes of rotation is
kept at $90^{\circ}$ throughout this study. One end wall of the vessel
is equipped with 9 ultrasound transducers (TR0408SS, Signal Processing
SA, Lausanne) whereby 6 transducers are arranged in a radial array
ranging from $r/R = 0$ to $0.92$, equally spaced. Four transducers are
located at $r/R = 0.92$ having a $90^{\circ}$ spacing in between in
order to estimate the azimuthal symmetry of the flow. All transducers
are aligned parallel to the vessel axis and capture the instantaneous
axial velocity distribution between $z/H = 0$ and $1$. The transducers
are connected to an ultrasound Doppler velocimeter (Dop 3010, Signal
Processing SA, Lausanne) that records the velocity profiles with a
temporal resolution of about $10 \mbox{ Hz}$.

\subsection{Response of the fluid to precession}

\begin{figure}[t!]
\begin{center}
\includegraphics[width=0.9\textwidth]{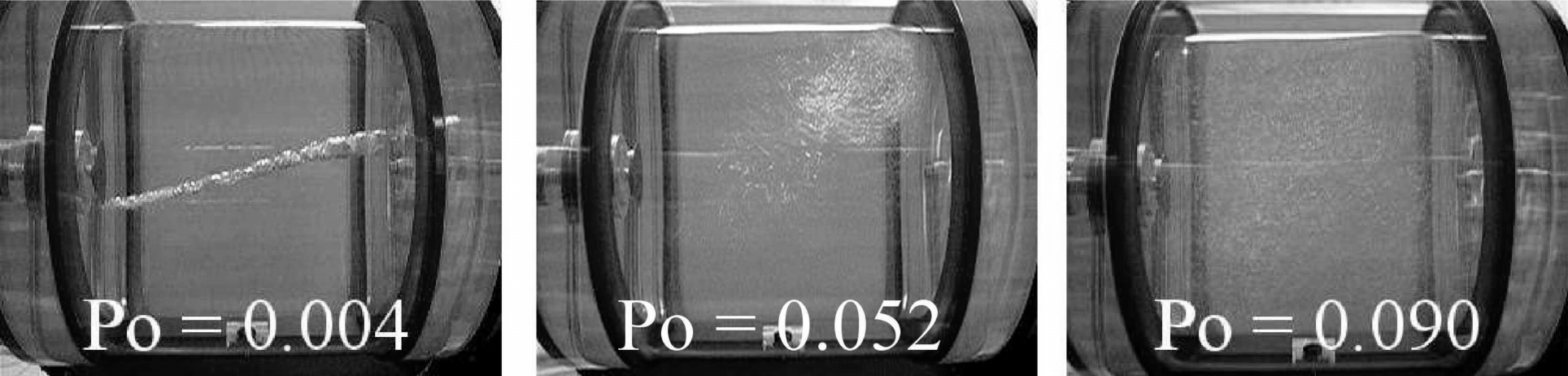}
\caption{\label{fig::water_exp_photo} 
Pictures of the flow behavior in three different regimes. The gas
bubbles give an overall impression of the flow behavior. Left: Linear
regime, weak precession ratios. Center: Nonlinear regime with large
scale flow. Note the breaking of the equatorial symmetry. Right:
Turbulent regime with no large scale structures recognizable in the
dynamics of the bubbles.}
\end{center}
\end{figure}

We now discuss the behavior of the precession driven flow in the water
experiment when the rotation rate remains fixed and the precession
frequency is successively increased.  A qualitative characterization
of the flow behavior can easily be done by injecting a little amount
of gas into the fluid and observing the distribution of the evolving
gas bubbles within the precessing fluid.  In the weekly precessing
regime, these bubbles align along the minimum pressure line which
follows an S-shaped tube that resembles the superposition of the
nearly resonant forced mode and the solid body rotation thus
indicating the effective fluid rotation axis (left panel in
figure~\ref{fig::water_exp_photo}).  In the intermediate range
(approximately for $0.03 \lesssim {\rm{Po}} \lesssim 0.06$) we observe
a weakly nonlinear behavior characterized by erratic fluctuations of
the minimum pressure line
\citep[similar to the observations of][in a precessing sphere]{noir2003}
and by a breaking of the equatorial symmetry (central panel in
figure~\ref{fig::water_exp_photo}).  In this regime measurements of
pressure fluctuations show periodic signatures that indicate the
presence of two free Kelvin modes with azimuthal wave numbers $m=5$
and $m=6$ which constitute a triadic resonance with the forced mode
\citep{herault2018}.  Triadic resonances have also been found in
numerical simulations. However, since the Reynolds number in these
computations is much lower, they are damped by viscous effects and
could only be identified at aspect ratios away from the fundamental
resonance \citep{giesecke2015b} where detuning effects like those
observed in \cite{herault2018} are less significant.  In all cases, in
simulations as well as in experiments, the flow energy beyond the
fundamental forced mode remains weak, and it seems to be rather
difficult to inject kinetic energy in addition to the forced mode via
a triadic resonance mechanism \citep{giesecke2015b}.

When attaining large precession ratios with ${\rm{Po}} \gtrsim 0.07$,
we ultimately observe an abrupt transition of the flow into a
turbulent behavior with the large scale flow structures becoming less
significant (right picture in figure~\ref{fig::water_exp_photo}). This
transition becomes only evident at sufficiently fast rotation of the
cylinder, i.e. at sufficiently large ${\rm{Re}}$.  At the rather low
values for ${\rm{Re}}$ that can be achieved in our simulations
(maximum value ${\rm{Re}}=10^4$) or for which flow measurements are
practicable with UDV (maximum value for reliable quantitative
measurements ${\rm{Re}}=10^5$) the flow maintains large scale
structures for all precession ratios and the transition into the
turbulent regime characterized in \cite{herault2015} can only be
identified by secondary criteria, like the abrupt drop of the
amplitude of the forced mode.

\newcommand{\za}{-1.5cm}
\newcommand{\siz}{5.1cm}

\begin{figure}[h!]
\begin{center}
\includegraphics[width=\siz]{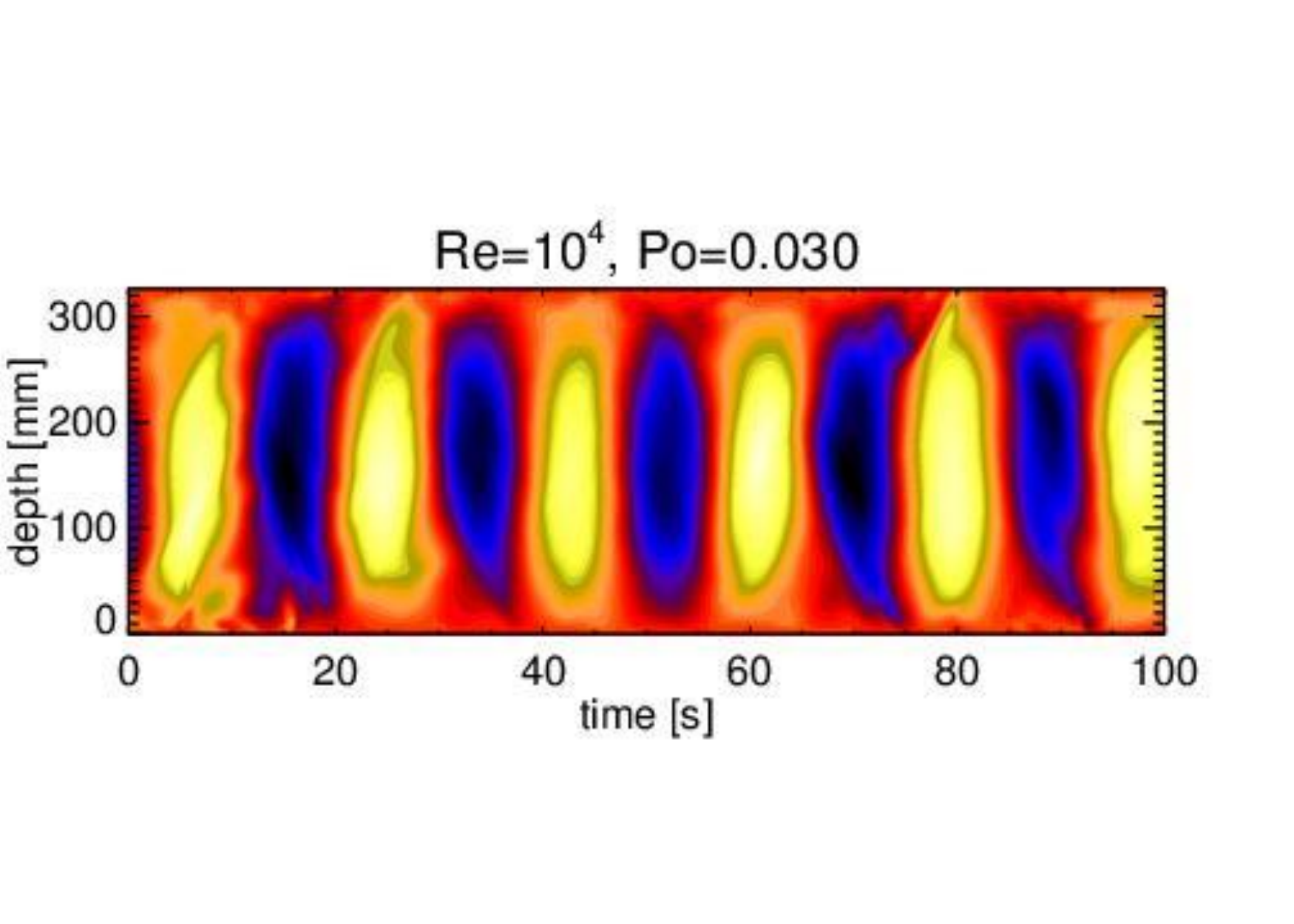}
\includegraphics[width=\siz]{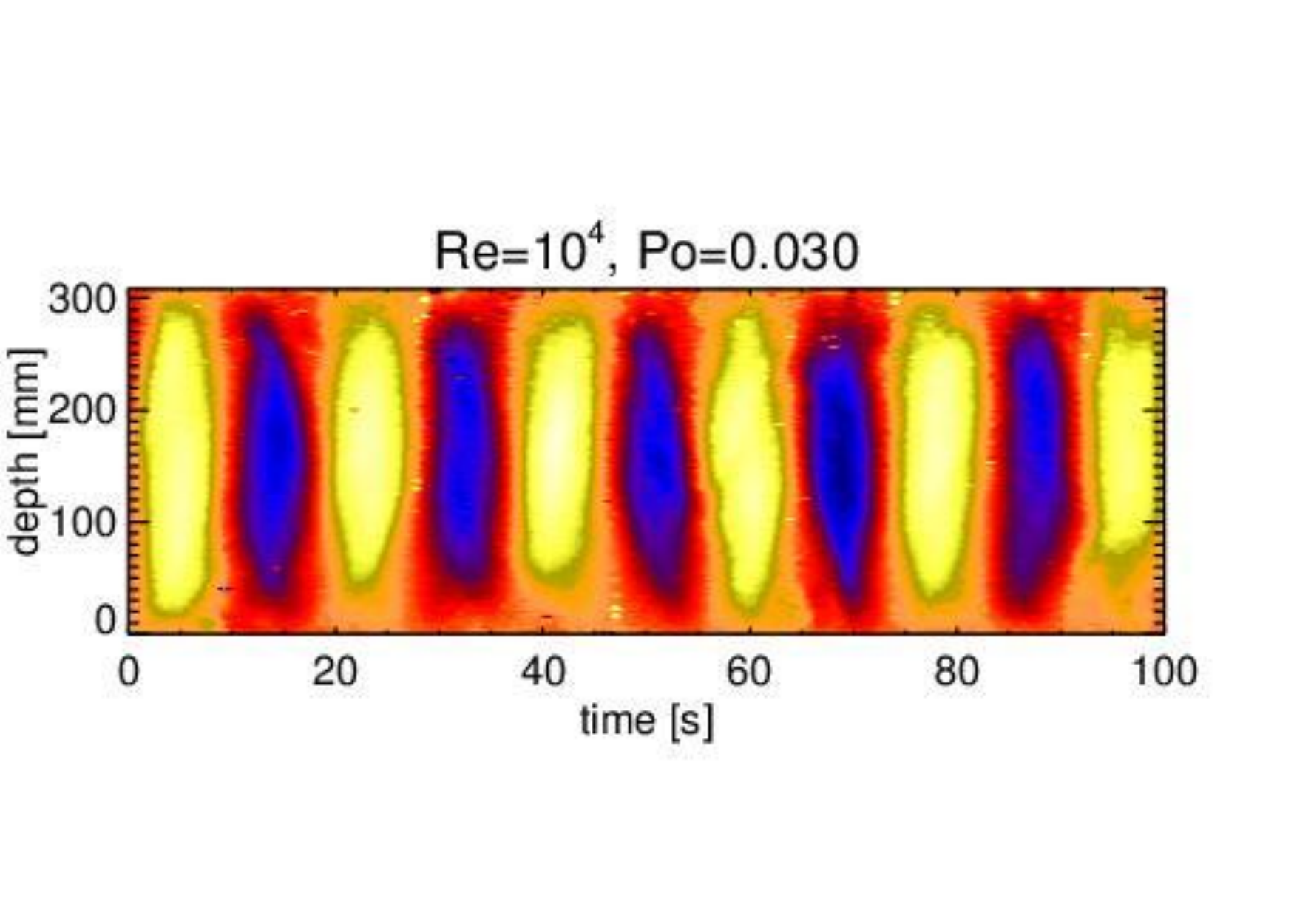}
\includegraphics[width=\siz]{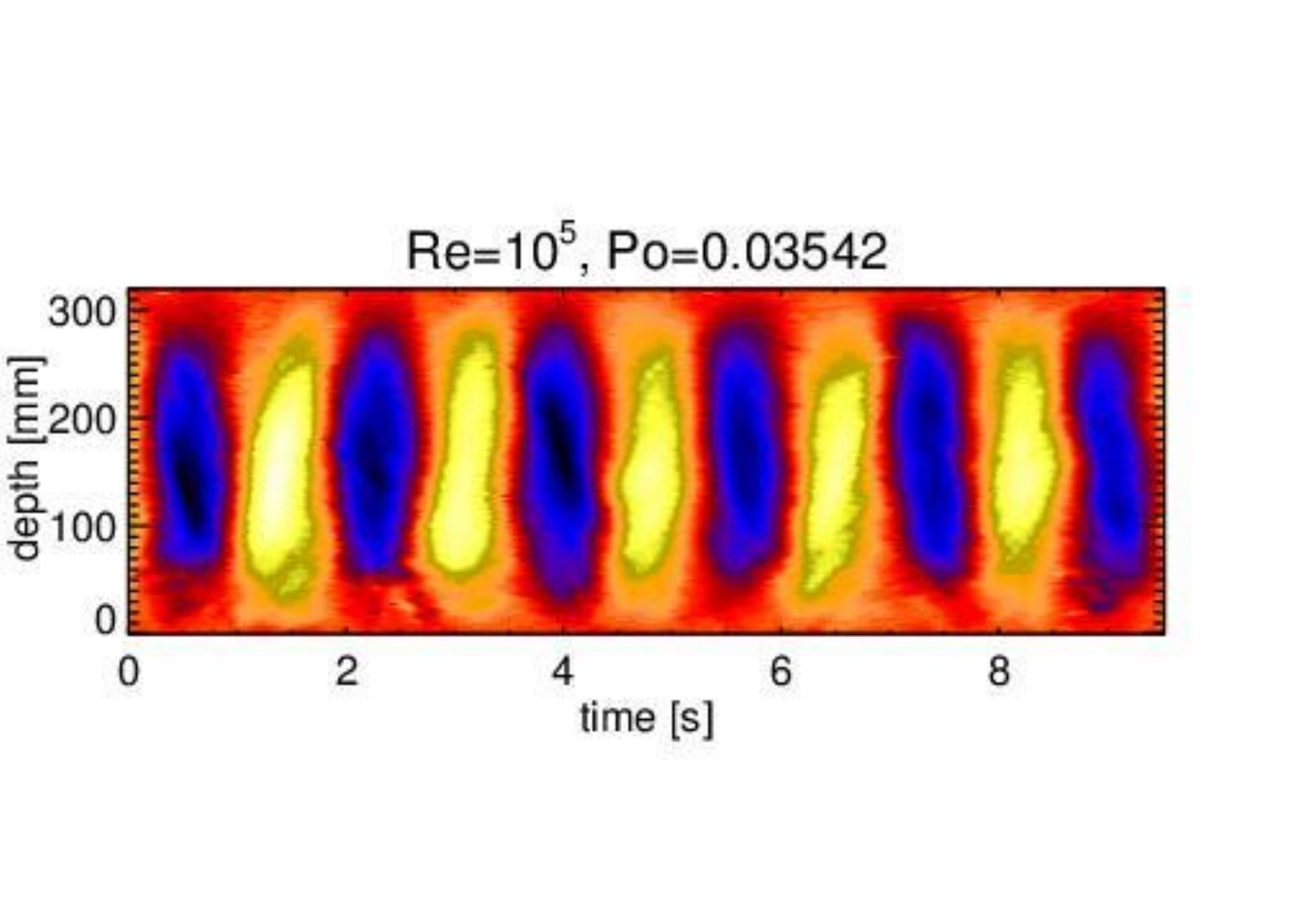}
\\[\za]
\includegraphics[width=\siz]{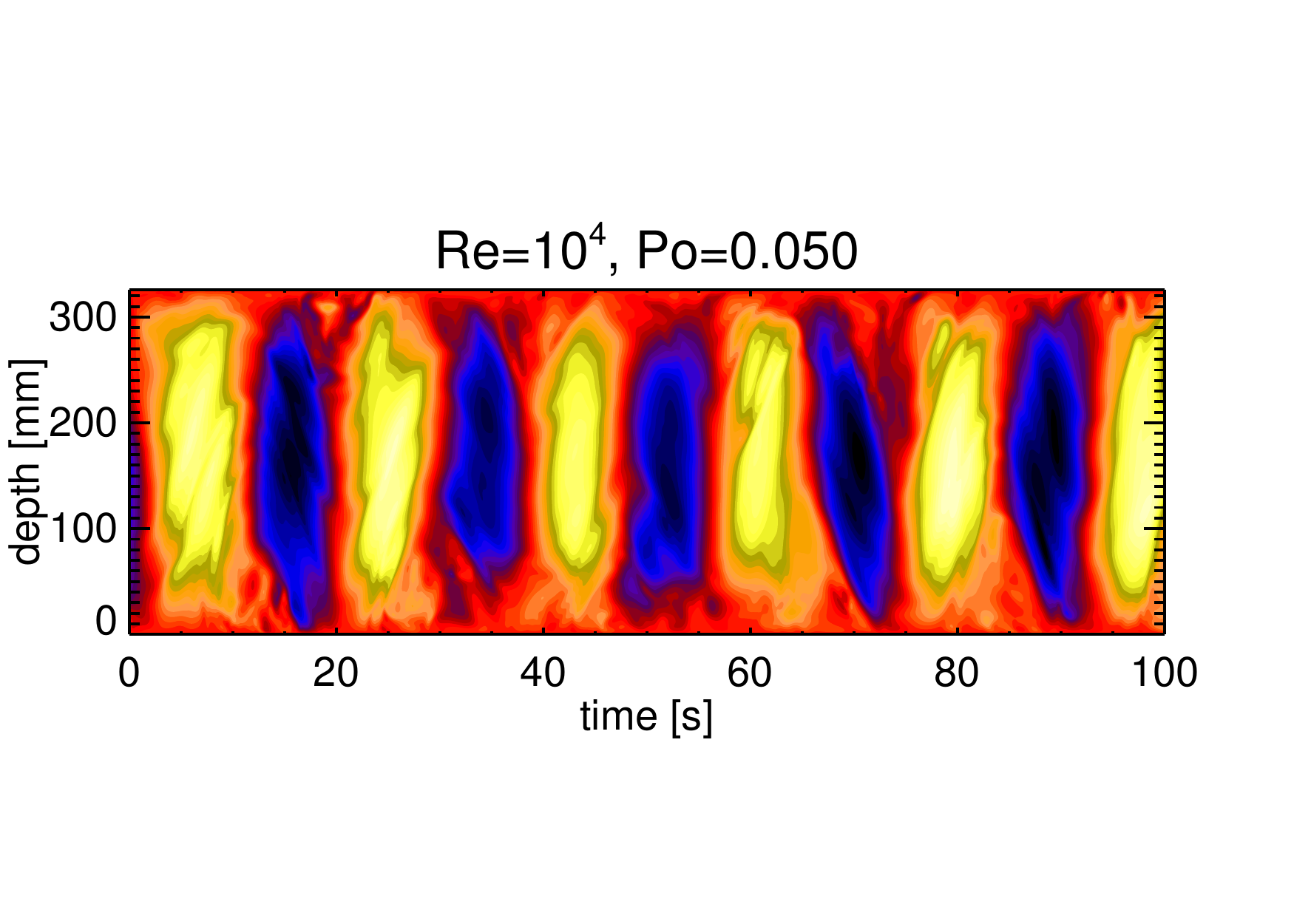}
\includegraphics[width=\siz]{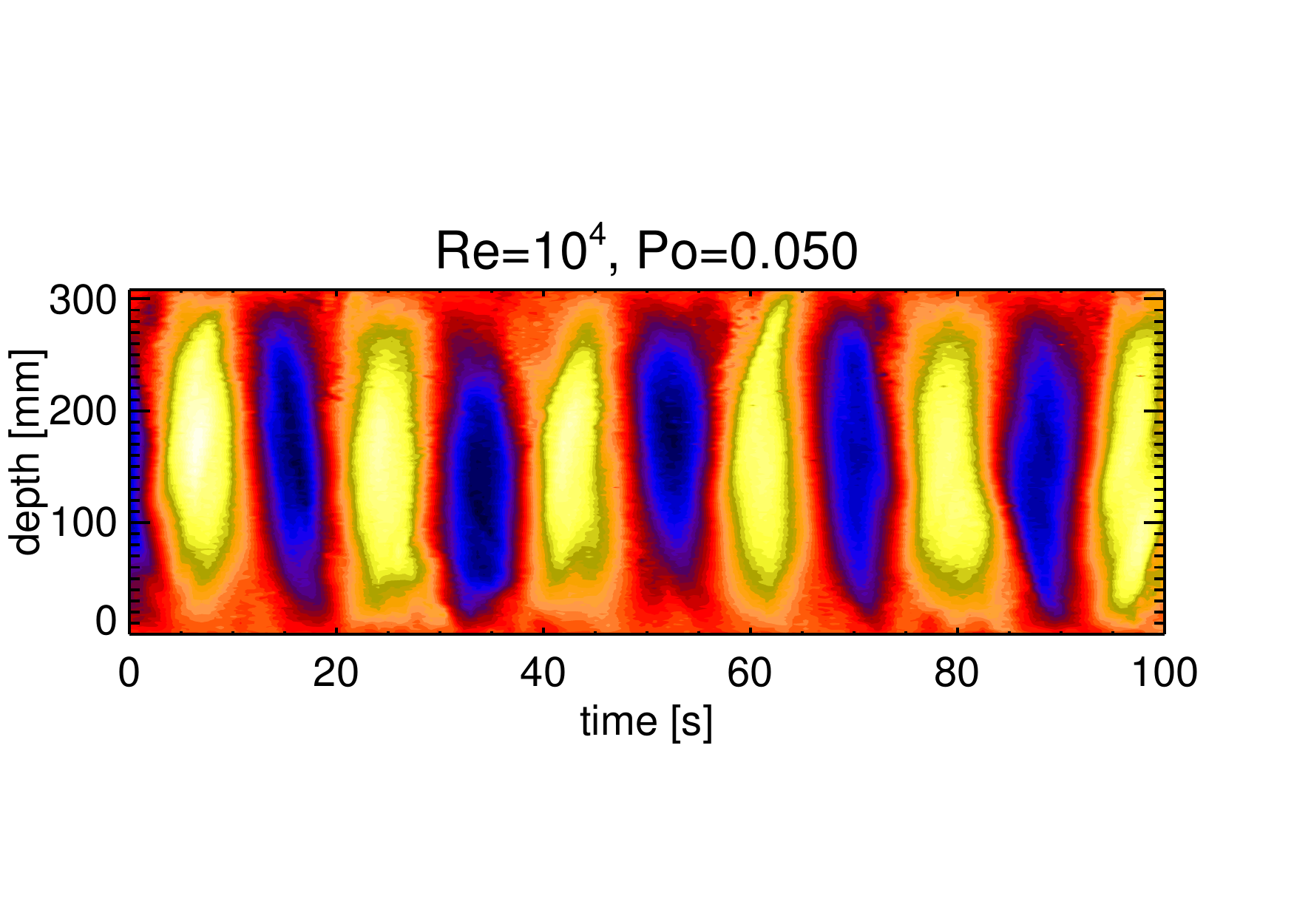}
\includegraphics[width=\siz]{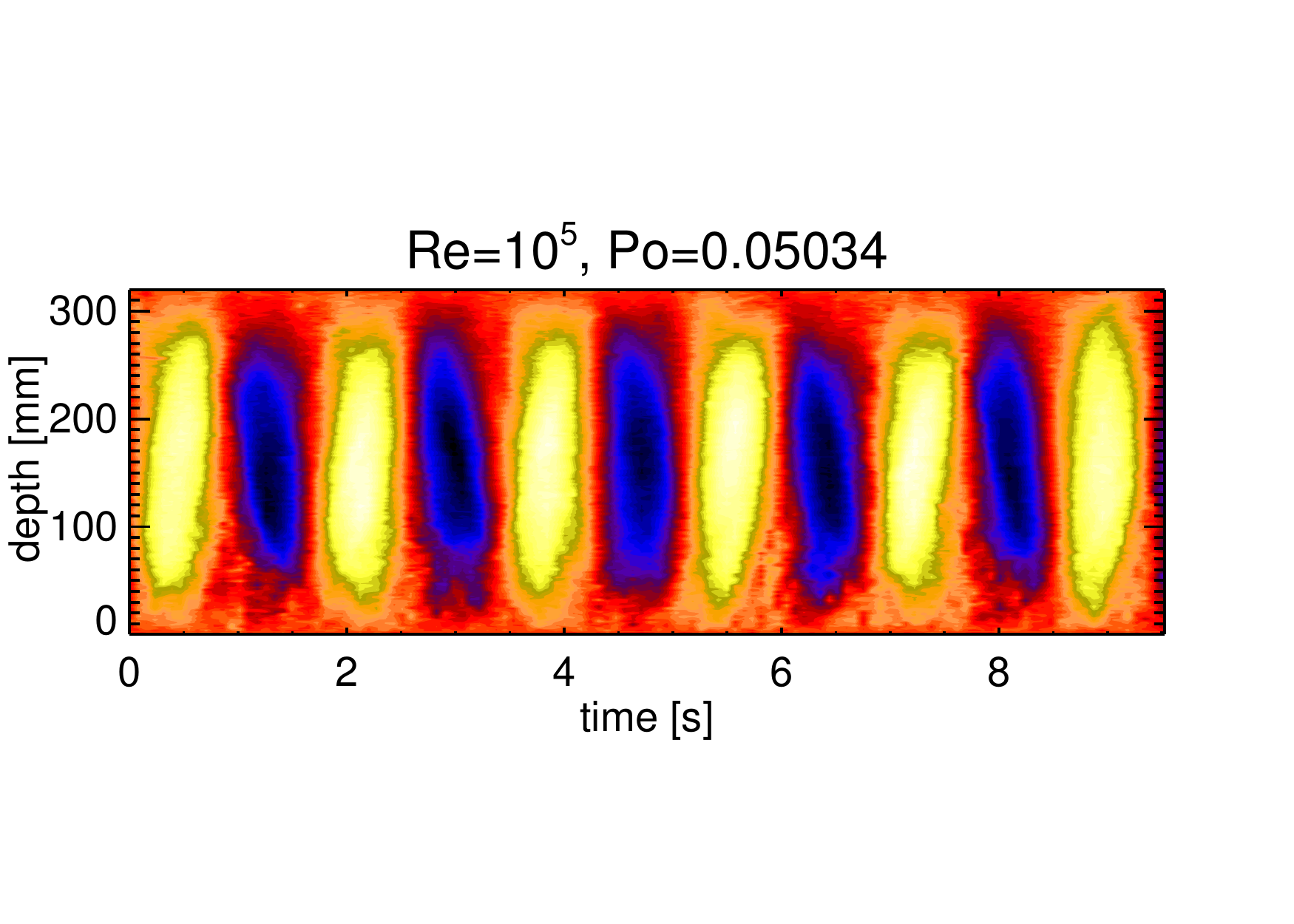}
\\[\za]
\includegraphics[width=\siz]{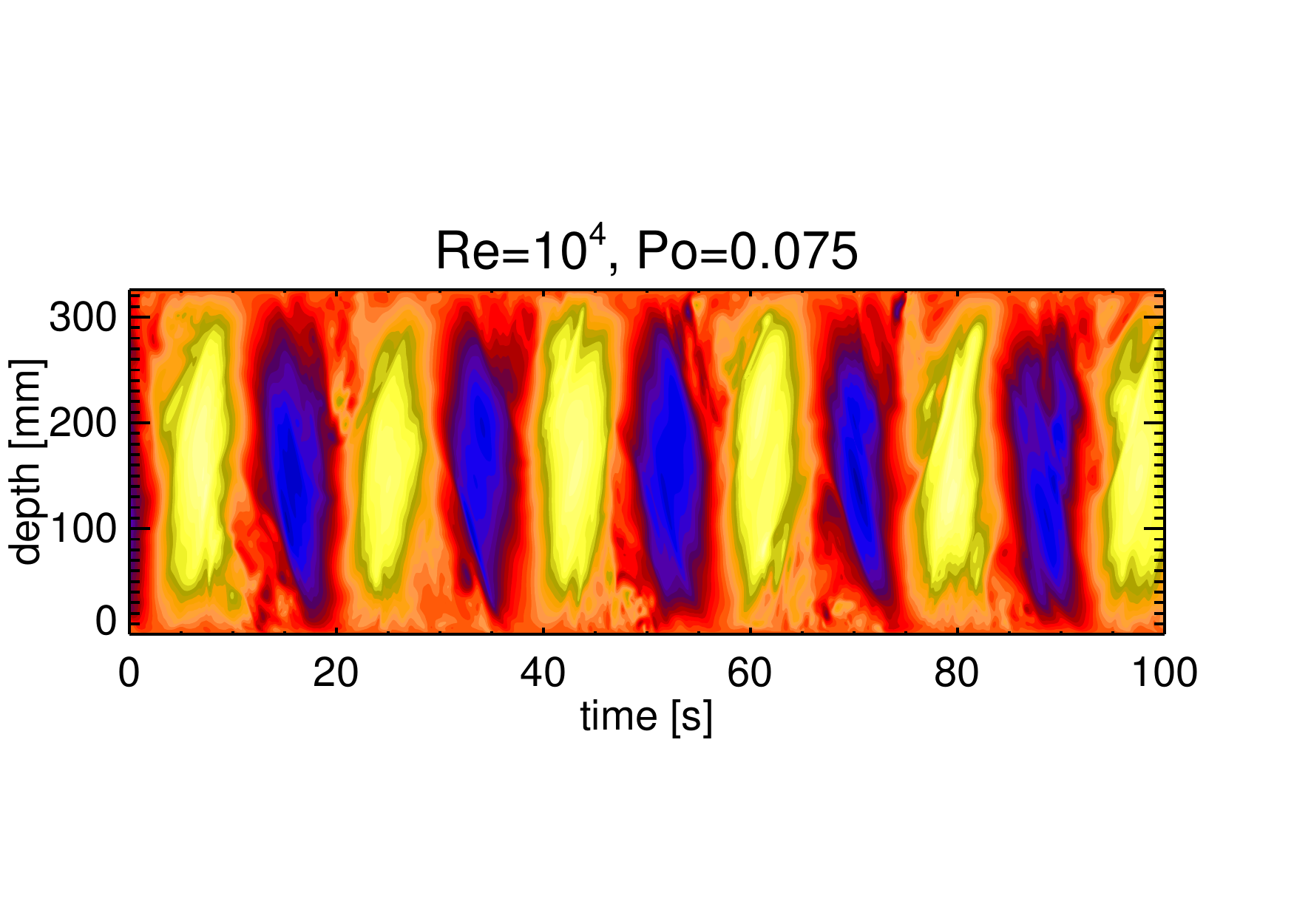}
\includegraphics[width=\siz]{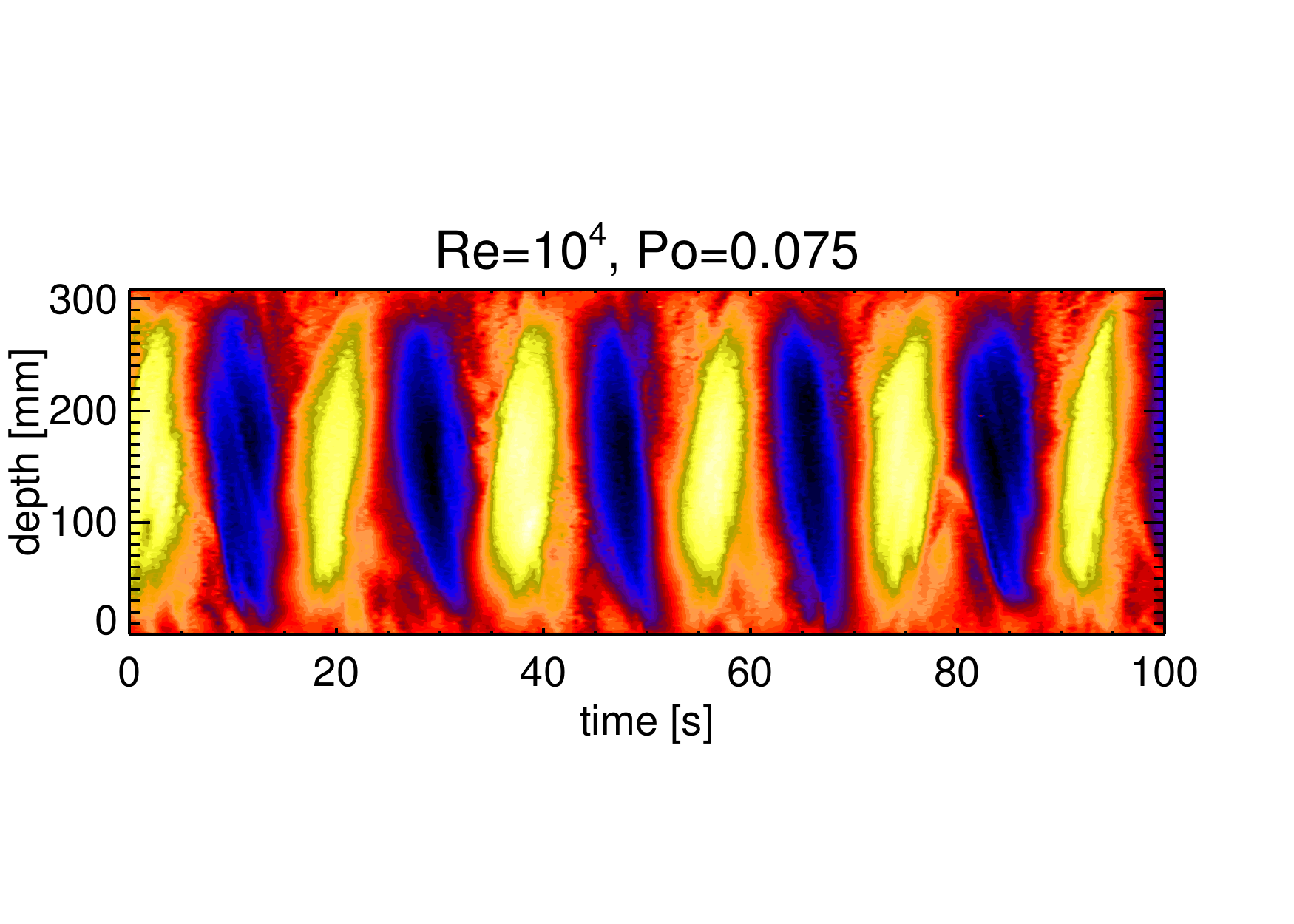}
\includegraphics[width=\siz]{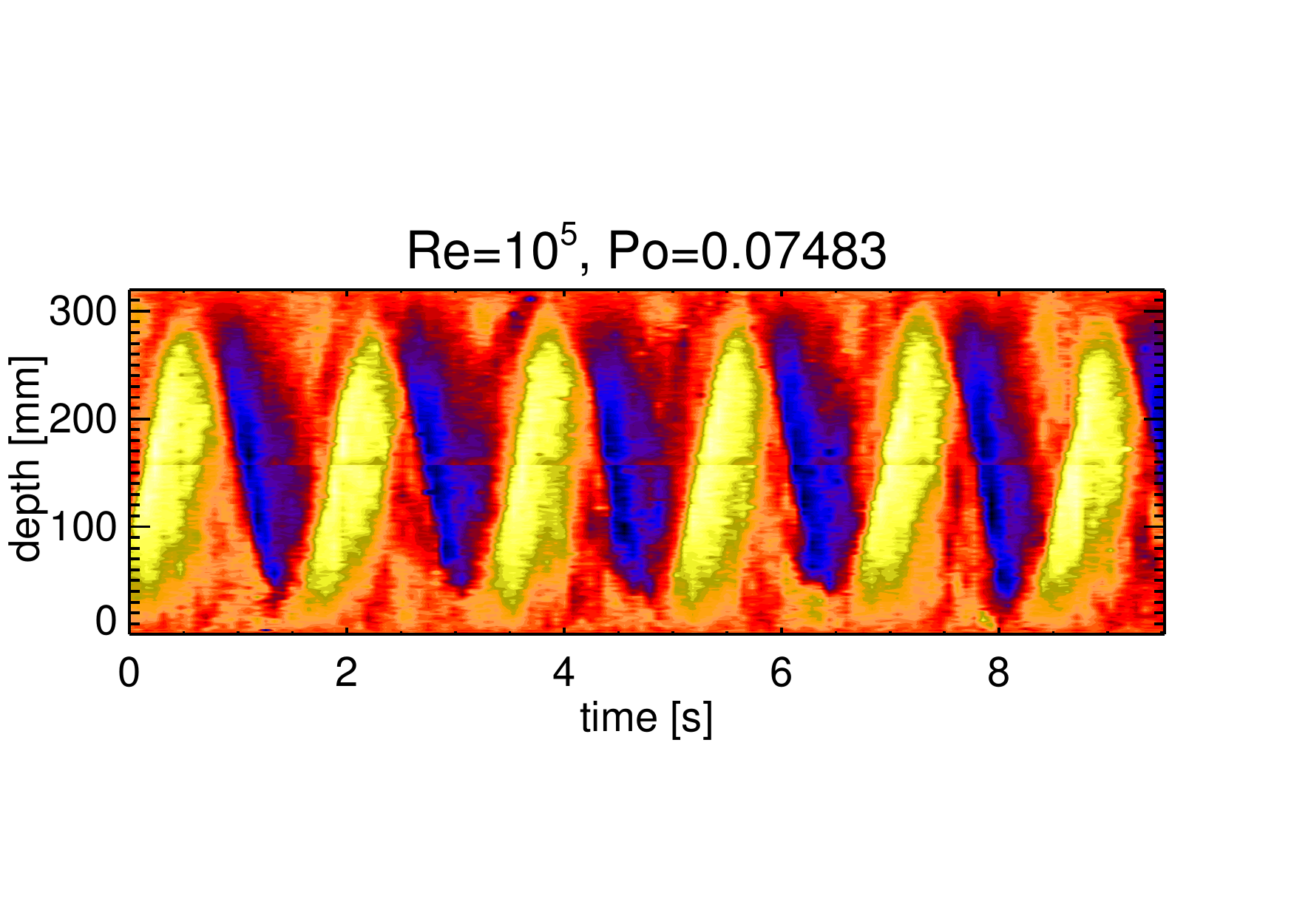}
\\[\za]
\includegraphics[width=\siz]{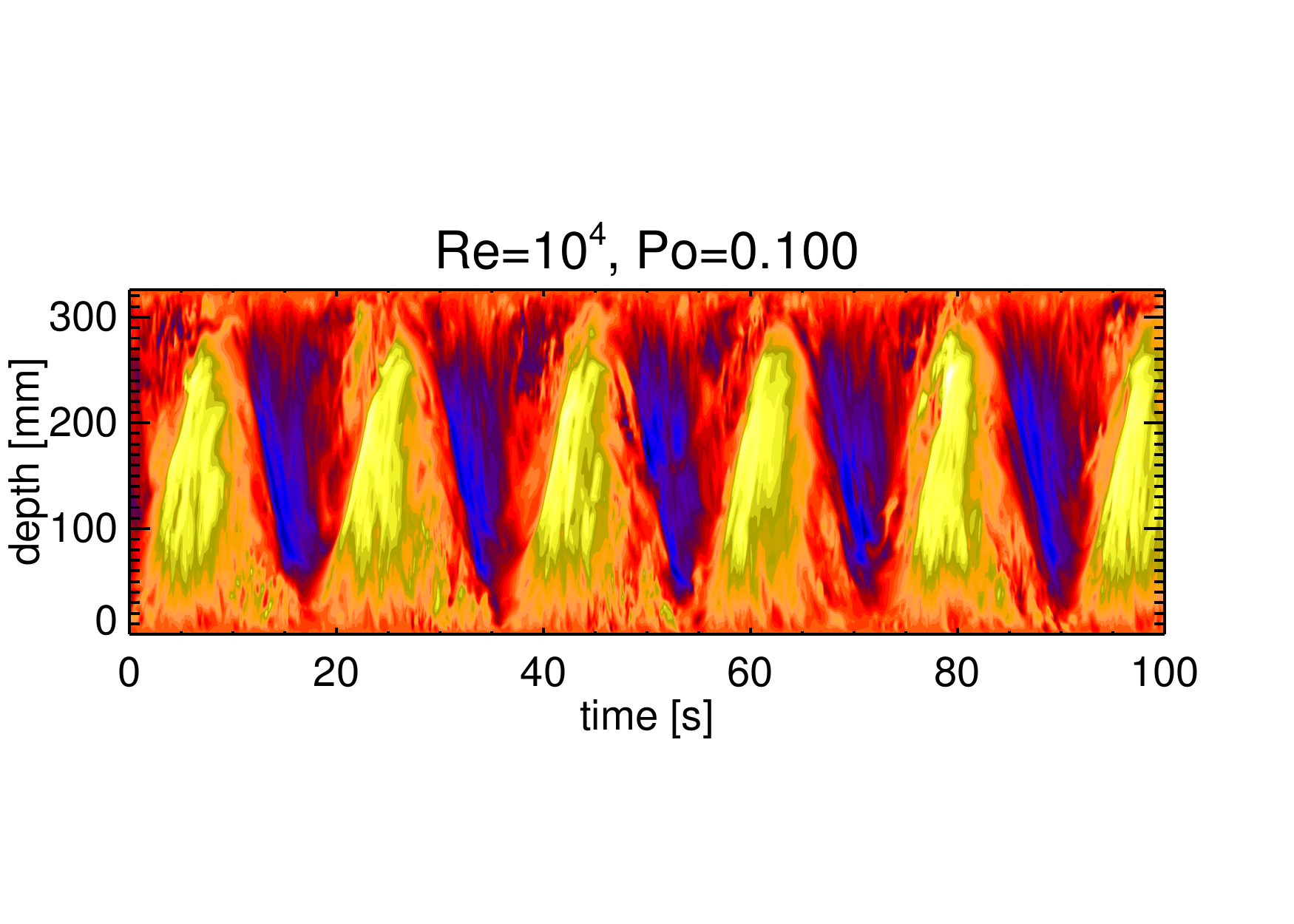}
\includegraphics[width=\siz]{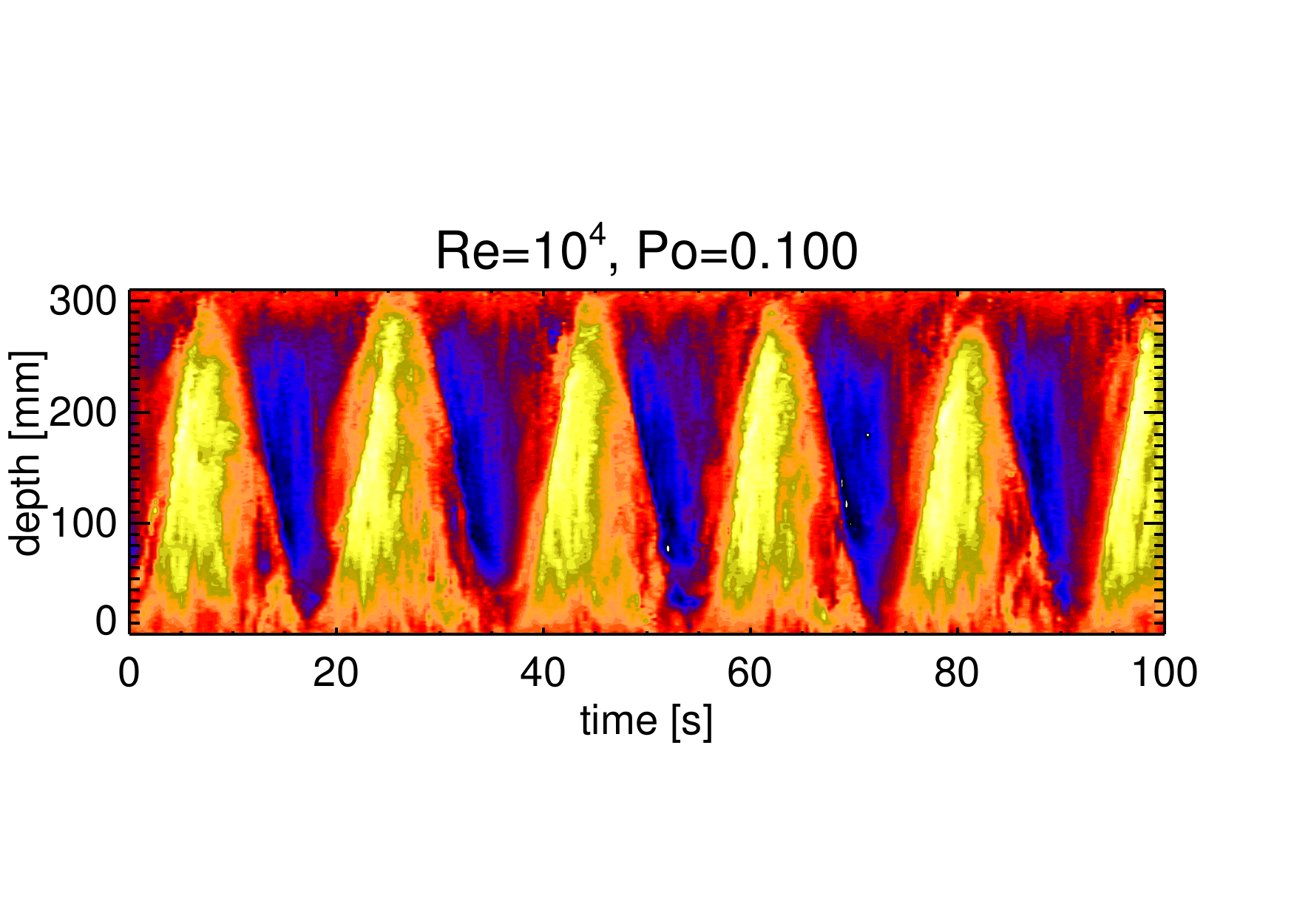}
\includegraphics[width=\siz]{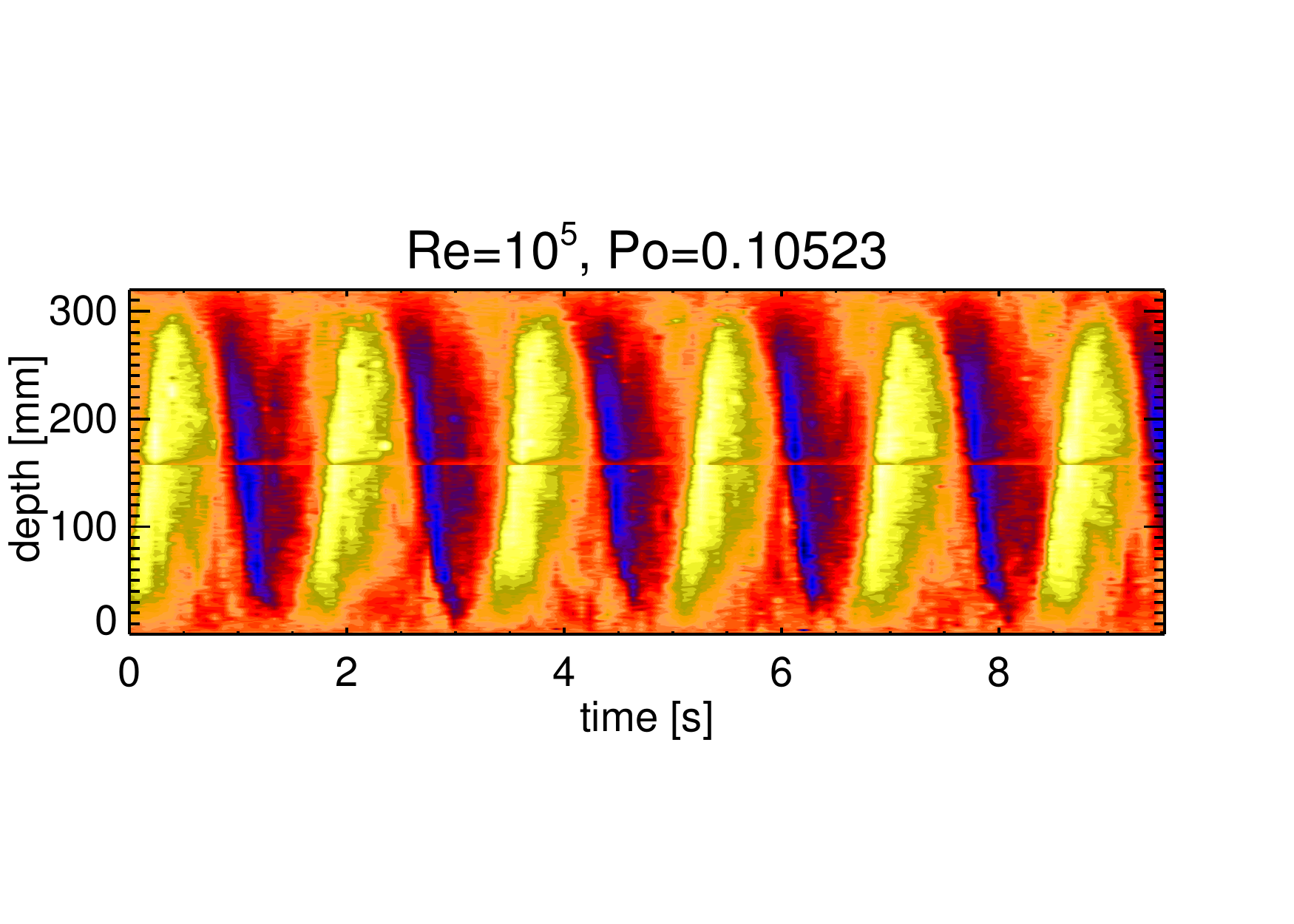}
\caption{Velocity pattern for ${\rm{Re}}=10^4$ (left and center) and
${\rm{Re}}=10^5$ (right).  From top to bottom: ${\rm{Po}}=0.03, 0.05,
0.075, 0.10$.  The left column shows results from numerical
simulations and the central and the right column show results from
experimental measurements (colour online).
\label{fig::pattern}}
\end{center}
\end{figure}

Typical flow patterns obtained in simulations and/or experiments are
shown in figure~\ref{fig::pattern}. The plots present the evolution of
the axial profile of the axial velocity $u_z(z)$ versus time $t$ at a
fixed point in a co-rotating reference frame. The left column shows
the results from simulations at ${\rm{Re}}=10^4$, and the central and
the right column show the results from experimental measurements (at
${\rm{Re}}=10^4$ and ${\rm{Re}}=10^5$, respectively).  The alternating
pattern with particular periodicity determined by the rotation
frequency of the cylinder denotes the fundamental forced mode, the
standing inertial wave with $(m,k)=(1,1)$ sampled by the rotating
probe which is attached to the container wall.

Above a certain precession ratio the fundamental pattern changes and
the dominant structures exhibit a considerable tilt with respect to
the vertical axis. This tilt is the implication of other inertial
modes -- in particular $(m,k)=(2,2)$ and $(m,k)=(0,2)$. The transition
occurs at a critical precession ratio ${\rm{Po}}^{\rm{c}}$ which in
turn depends on the Reynolds number. A further characteristic that
changes at this transition is the amplitude of the directly driven
mode, which decreases abruptly.

\subsection{Spectra and amplitudes}

Quantitative values for these amplitudes are obtained from the spectra
of individual inertial modes $(m,k)$ that are computed by a
decomposition of the axial velocity profile in $z$ direction according
to the typical dependence of an analytical inertial wave which behaves
$\propto \sin(\pi zk/H)$.  At a fixed timestep $t_n$ a $k$-mode
$\tilde{u}_z^k$ (with the axial wave number $k$ being an integer) is
calculated from the discretized data according to
\begin{equation}
\tilde{u}_z^k(r,\varphi_j,t_n)\,=\,
\displaystyle\frac{1}{N_z}\sum\limits_{l=0}^{N_z-1}
u_z(r,\varphi_j,z_l,t_n)\,
\sin\!\left(\frac{k\pi z_l}{H}\right),\label{eq::kdecomp}
\end{equation}
where the radial coordinate $r$ is fixed at $r=0.92$, the position of
the UDV sensor in the experiment, and $\varphi_j, z_l, t_n$ are the
discretized values for azimuthal coordinate, axial coordinate and time
taken from the simulations. The axial modes $\tilde{u}_z^k$ are
further decomposed by performing a 2D Fourier transformation in
azimuth and in time with the resulting amplitudes characterized by the
azimuthal wave number $m$ and the frequency $\omega$. This
transformation explicitly reads
\begin{equation}
\dbtilde{u}_z^{km\omega}(r)\,=\,
\frac{1}{N_{\varphi}N_t}\sum\limits_{n=0}^{N_{t\vphantom{\varphi}}-1}\sum\limits_{j=0}^{N_{\varphi}-1}
\tilde{u}_z^k(r,\varphi_j,t_n)\,{\mathrm e}^{-{\mathrm i}(m\varphi_j+\omega t_n)}\, .
\label{eq::decomp}
\end{equation}
The decomposition~(\ref{eq::decomp}) involves a 2D FFT, which provides
the amplitude spectrum for modes with fixed $m$ and $k$ in dependence
of the frequency $\omega$ including the sign of $\omega$ that allows
the distinction of prograde ($+$) and retrograde ($-$) components.

Figure~\ref{fig::spec} shows characteristic spectra for the dominant
modes obtained from simulations at ${\rm{Re}}=10^4$ and
${\rm{Po}}=0.01,0.03,0.05,0.875,0.10,0.125$ (from top to bottom).  In
these plots we switch from the mantle system into the turntable system
in which the observer follows the precession and looks at the spinning
cylinder.  This means that a mode that in the mantle frame rotates
with the frequency of the container, as it is the case for the
directly forced flow, has the frequency $\omega=0$ in the turntable
frame thus being a standing wave.  The reason for the change of the
reference frame becomes clear when regarding the spectra of individual
modes in figure~\ref{fig::spec}, which are always dominated by one
single peak located at $\omega=0$.  Thus, the main contribution of the
flow results from modes that are stationary in the turntable reference
frame.
\newcommand{\sizfigsix}{4.925cm}
\begin{figure}[t!]
\begin{center}
\includegraphics[width=\sizfigsix]{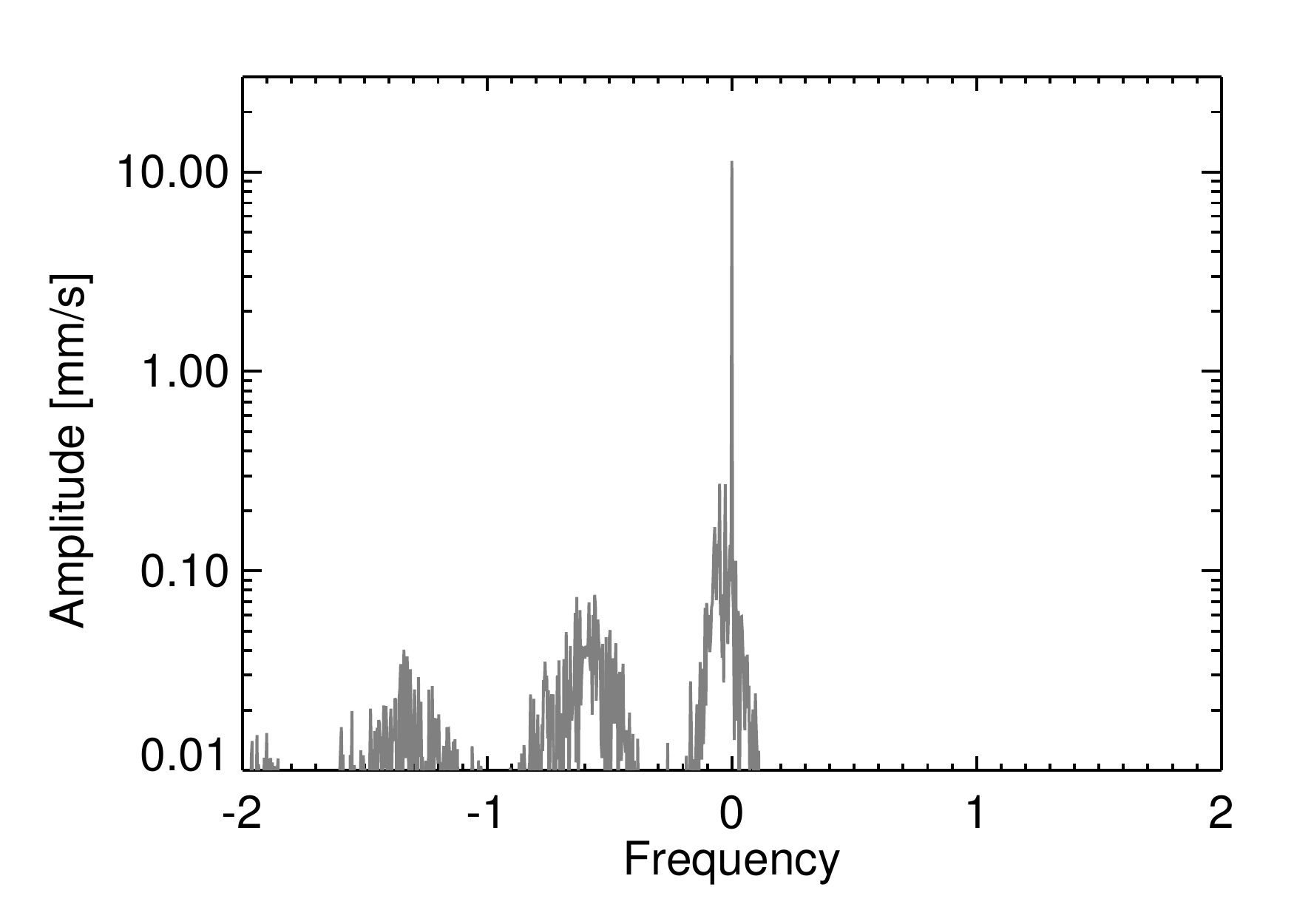}
\includegraphics[width=\sizfigsix]{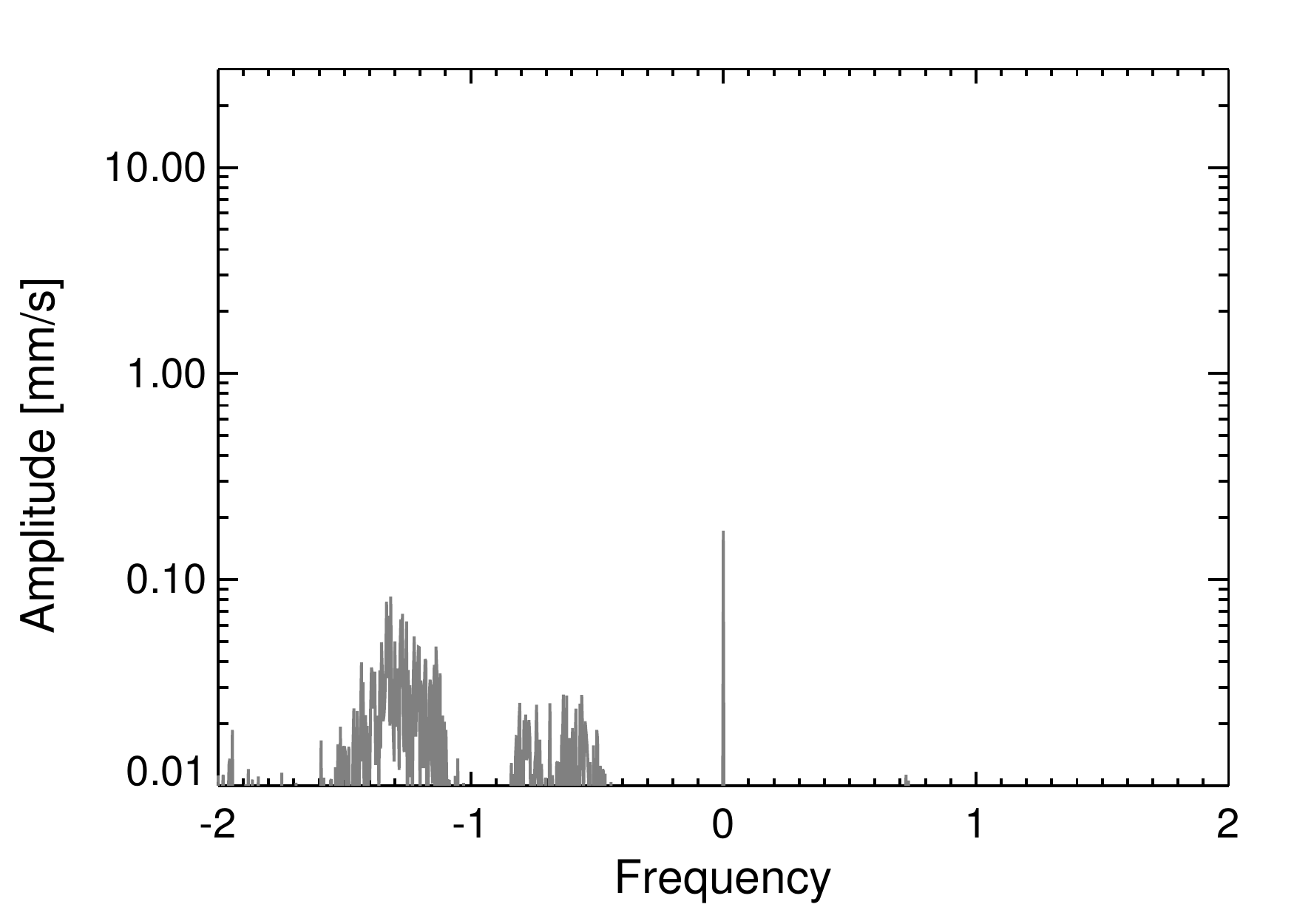}
\includegraphics[width=\sizfigsix]{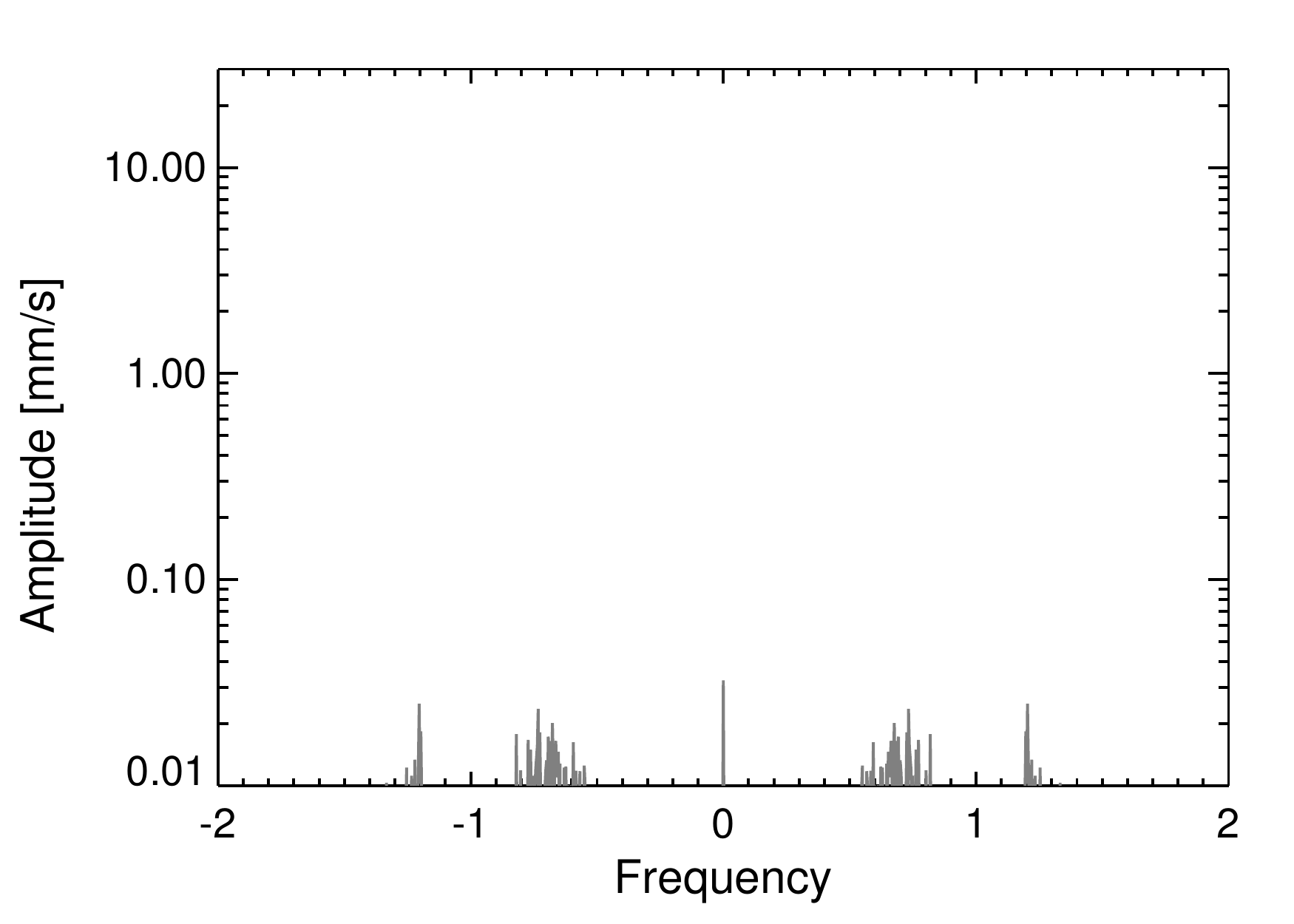}
\\
\includegraphics[width=\sizfigsix]{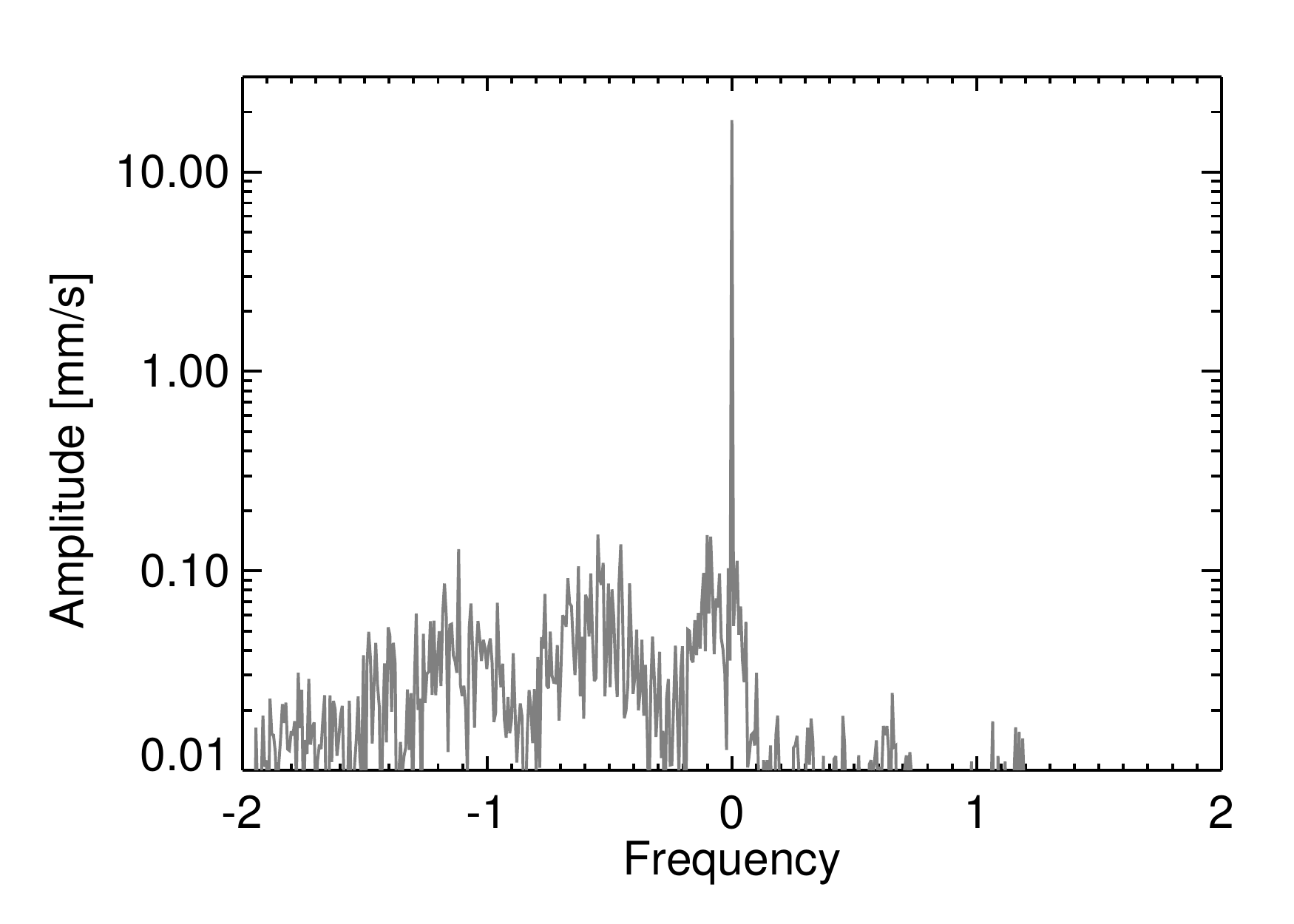}
\includegraphics[width=\sizfigsix]{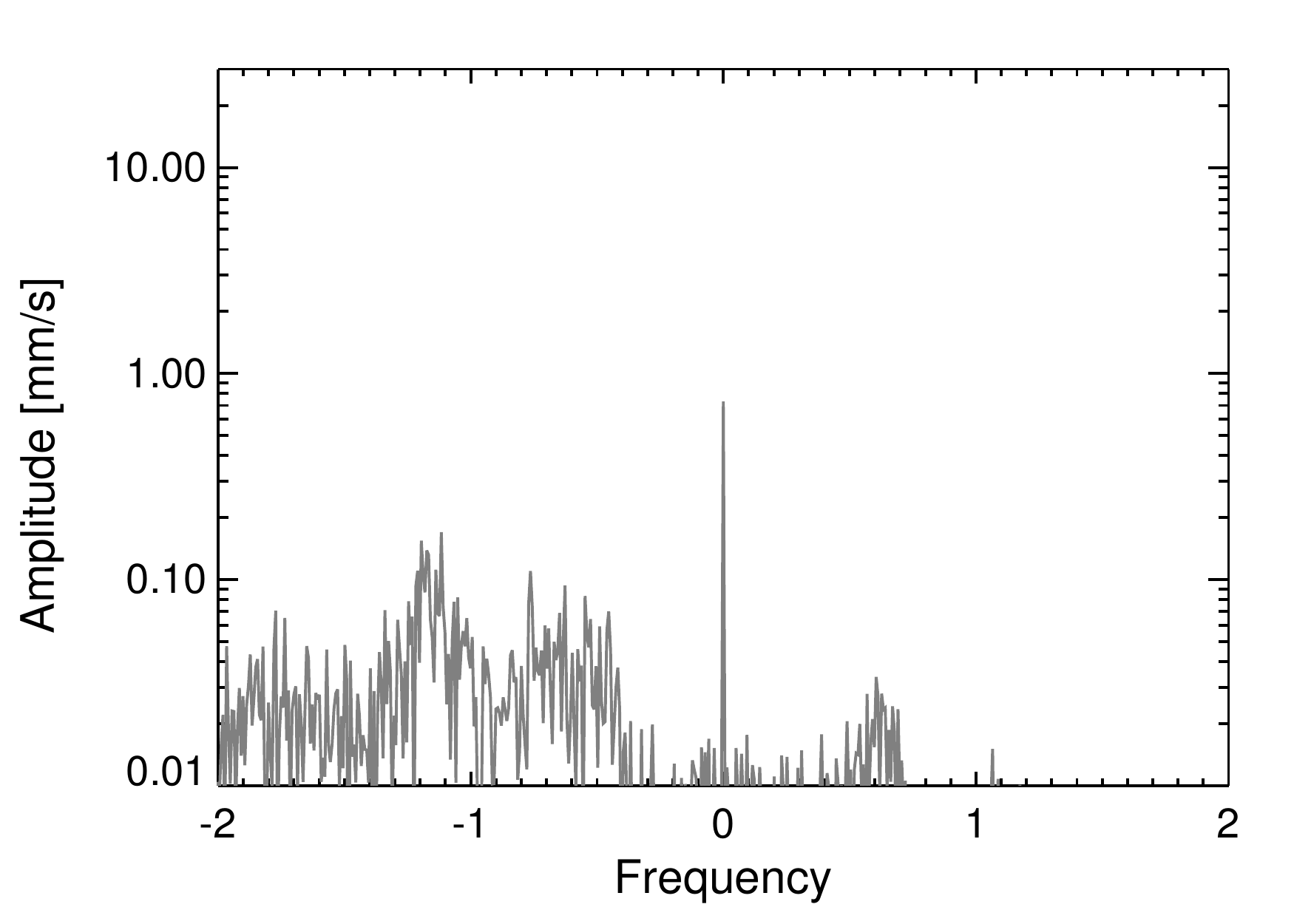}
\includegraphics[width=\sizfigsix]{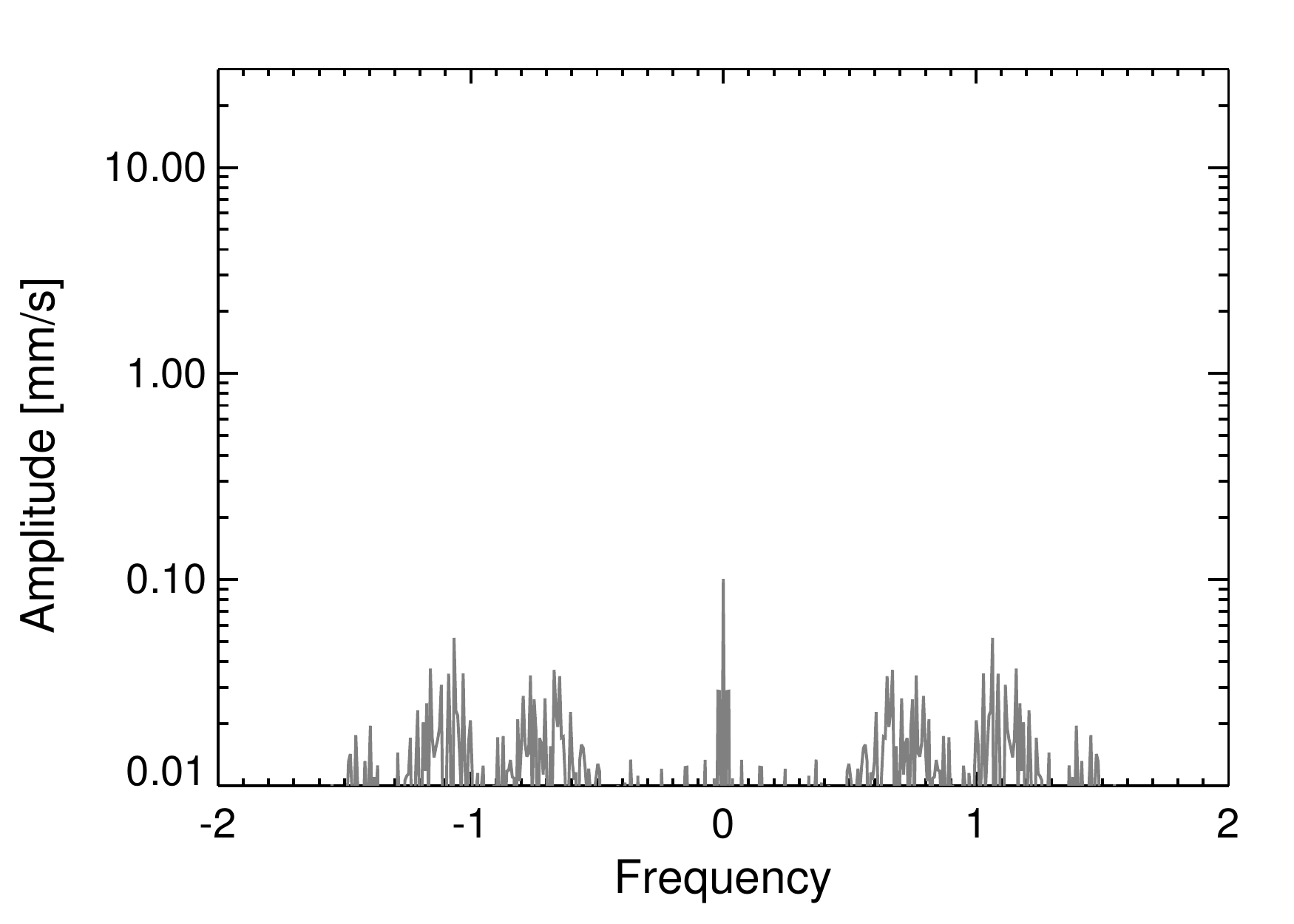}
\\
\includegraphics[width=\sizfigsix]{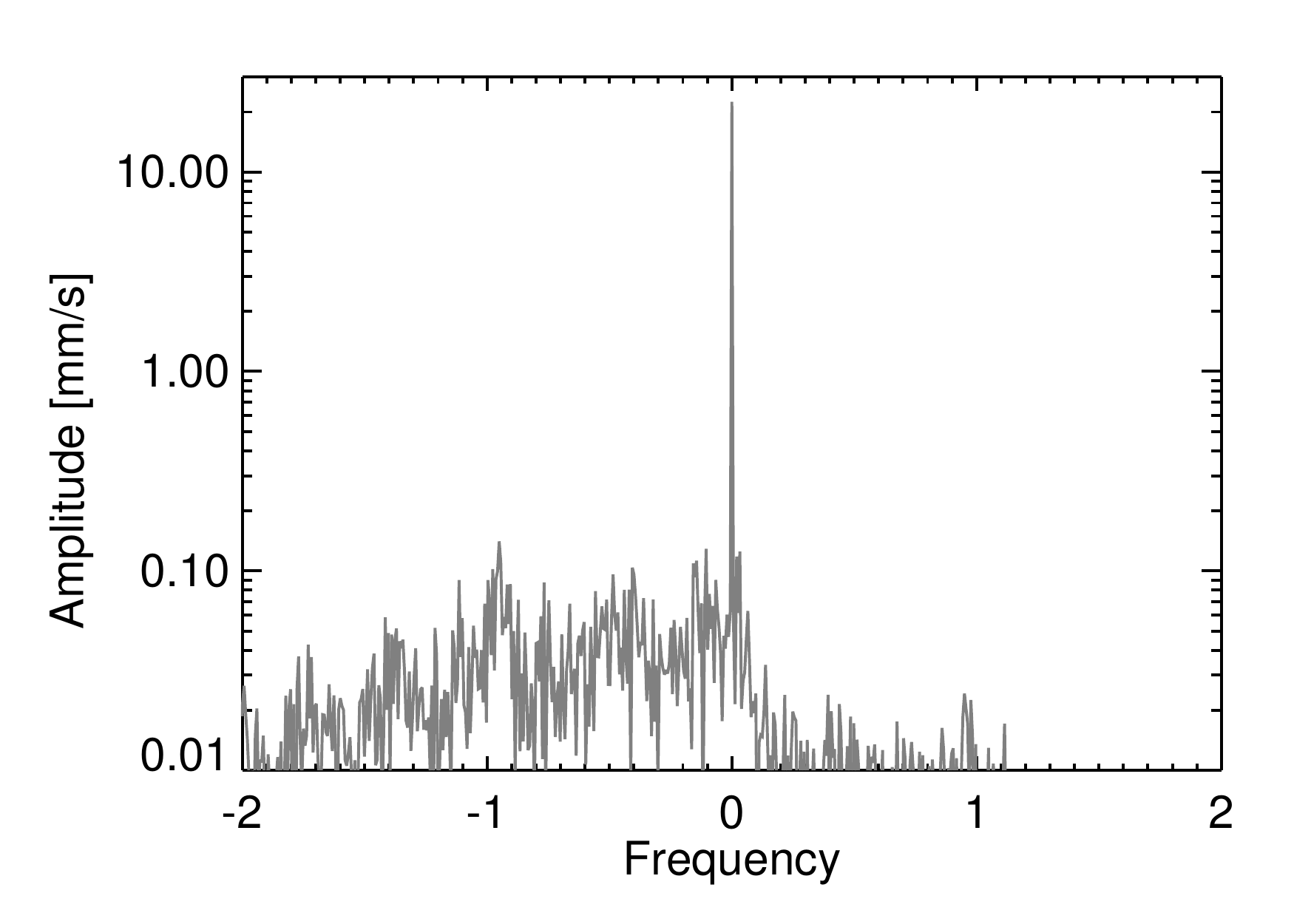}
\includegraphics[width=\sizfigsix]{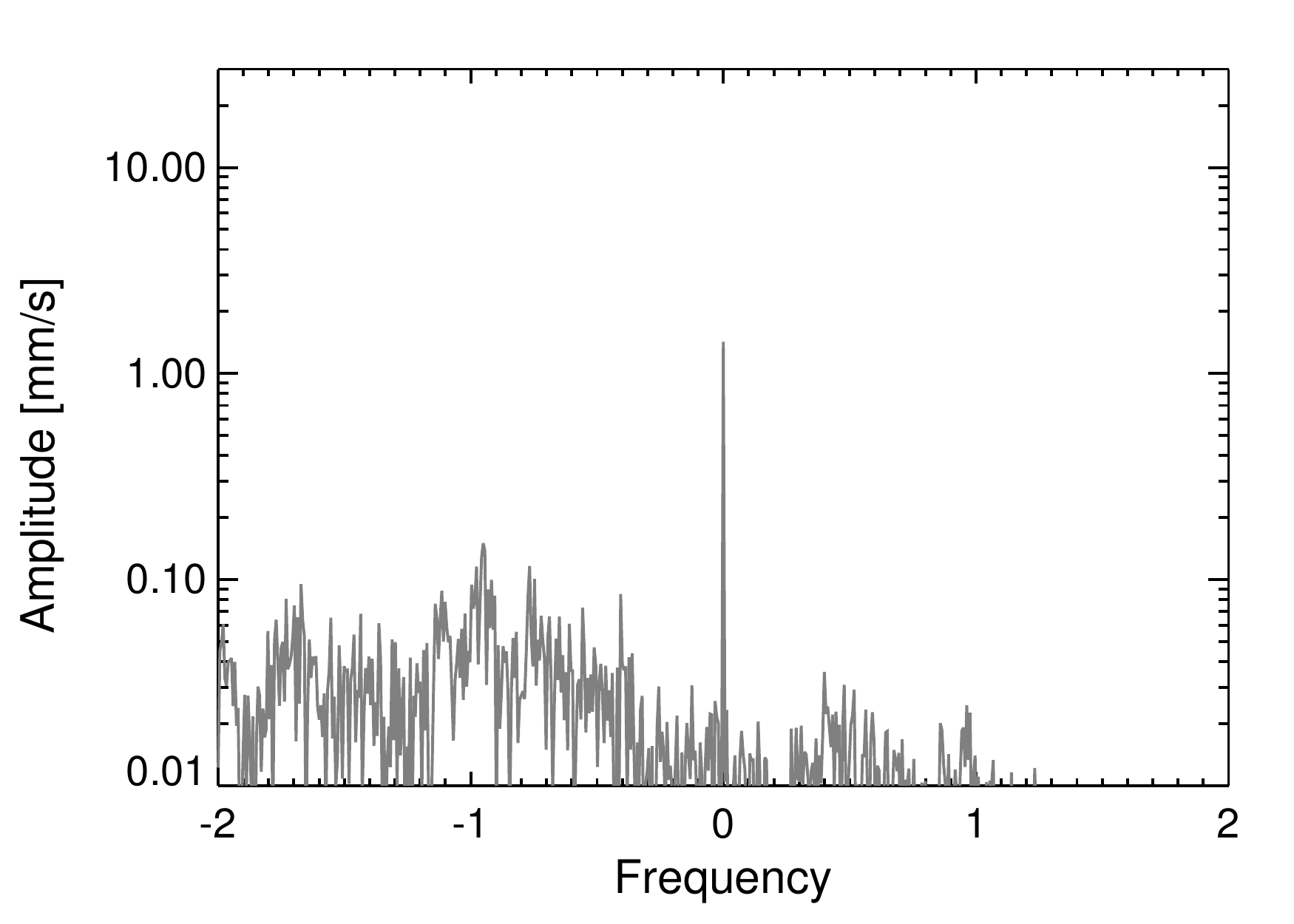}
\includegraphics[width=\sizfigsix]{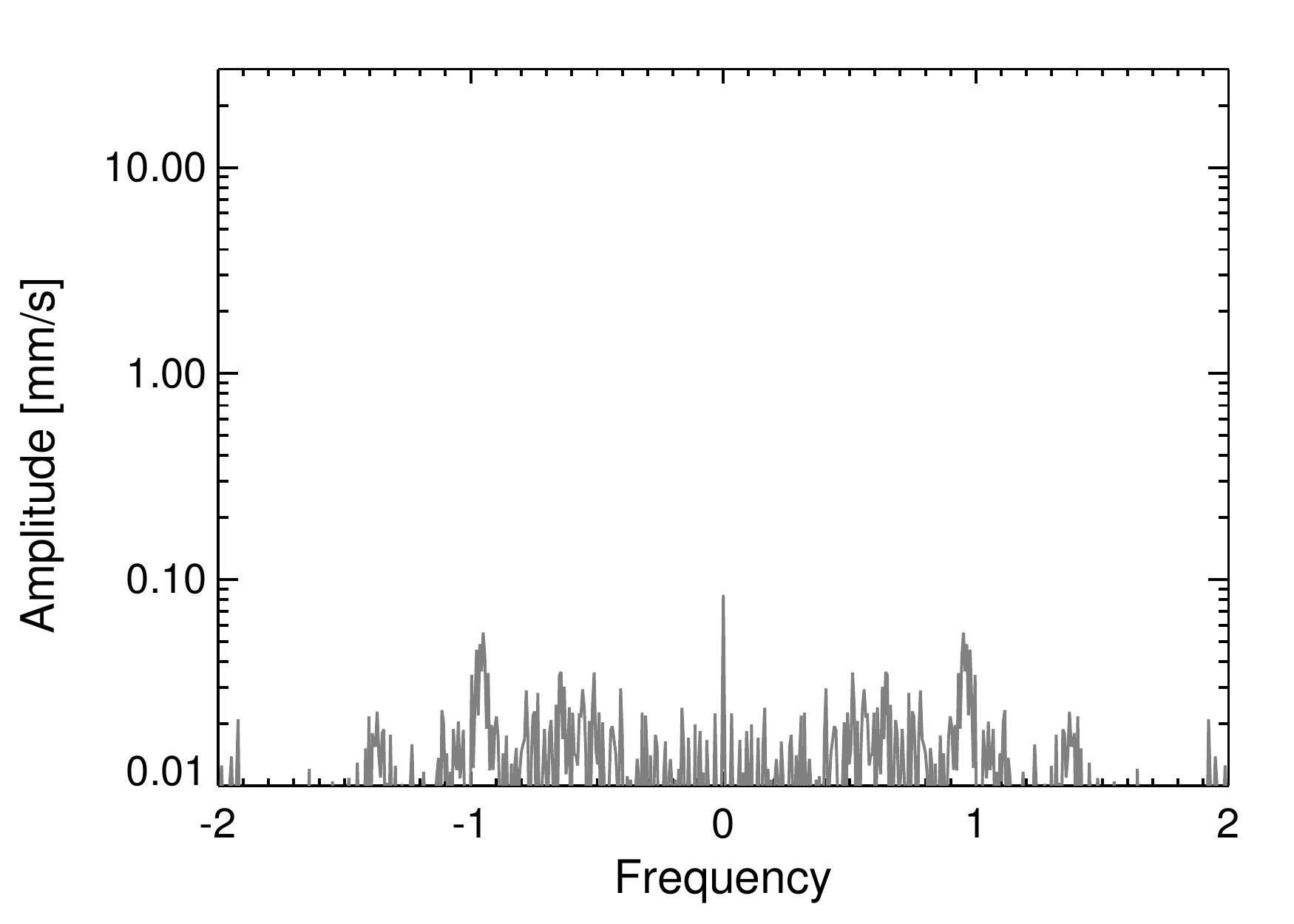}
\\
\includegraphics[width=\sizfigsix]{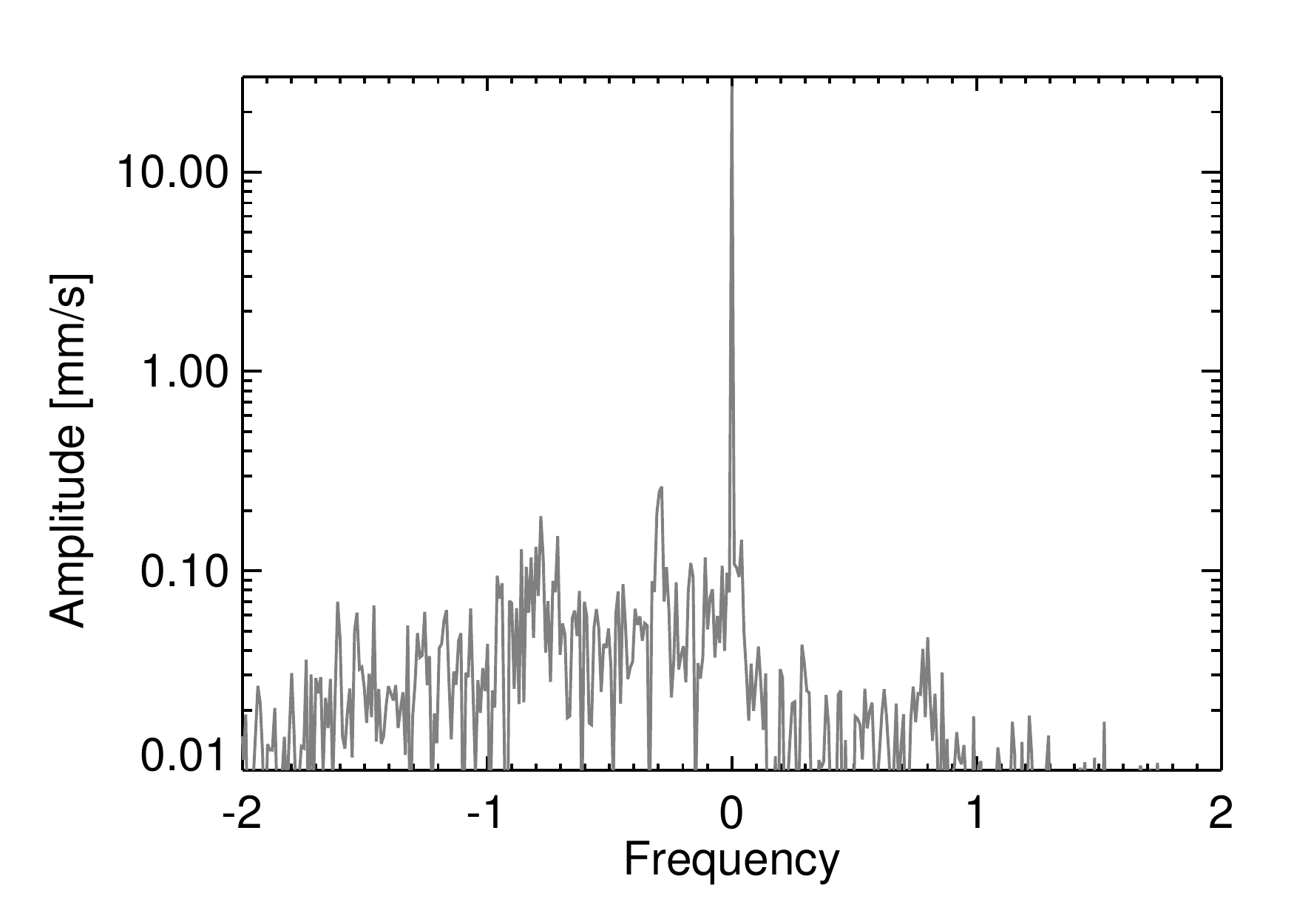}
\includegraphics[width=\sizfigsix]{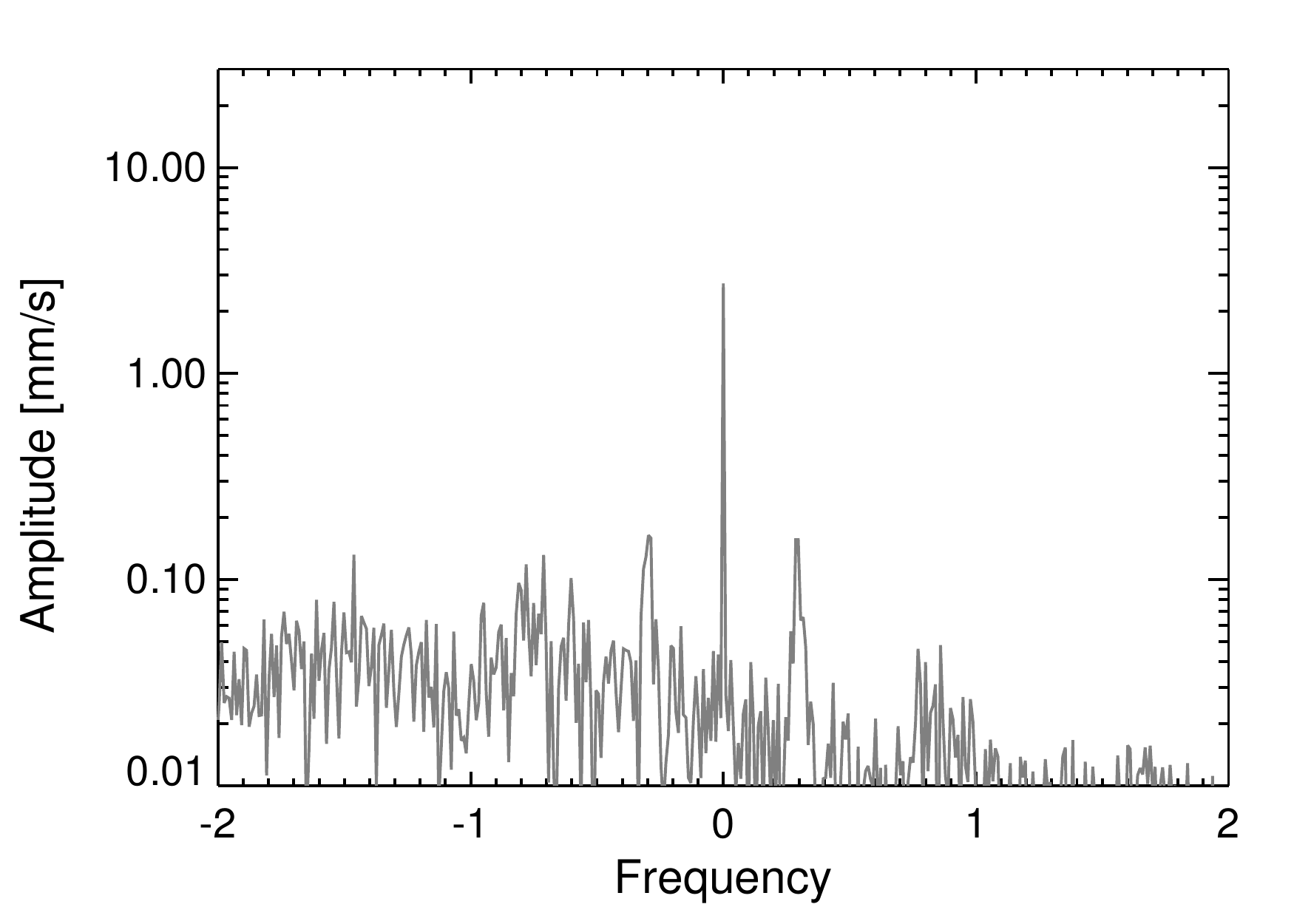}
\includegraphics[width=\sizfigsix]{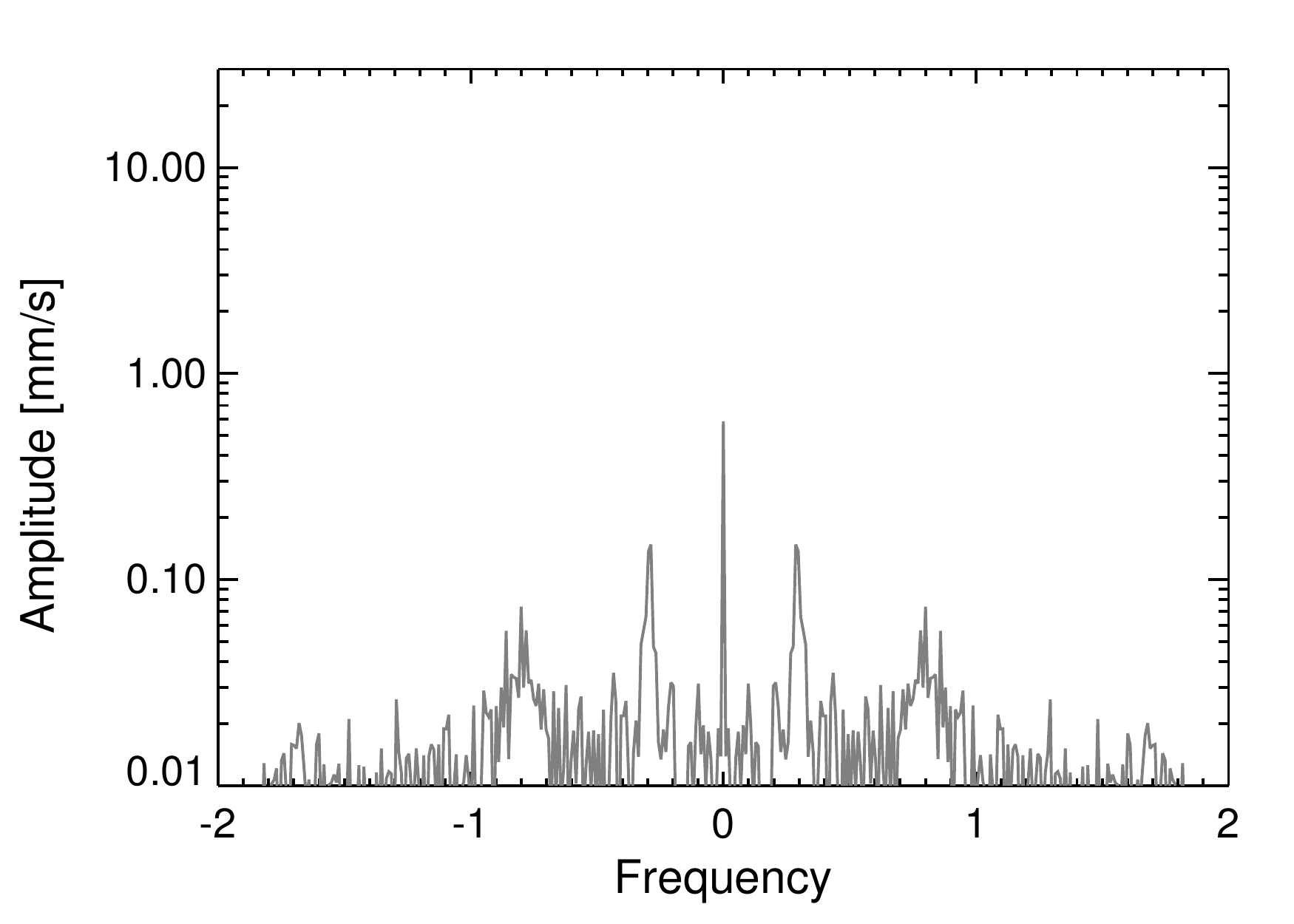}
\\
\includegraphics[width=\sizfigsix]{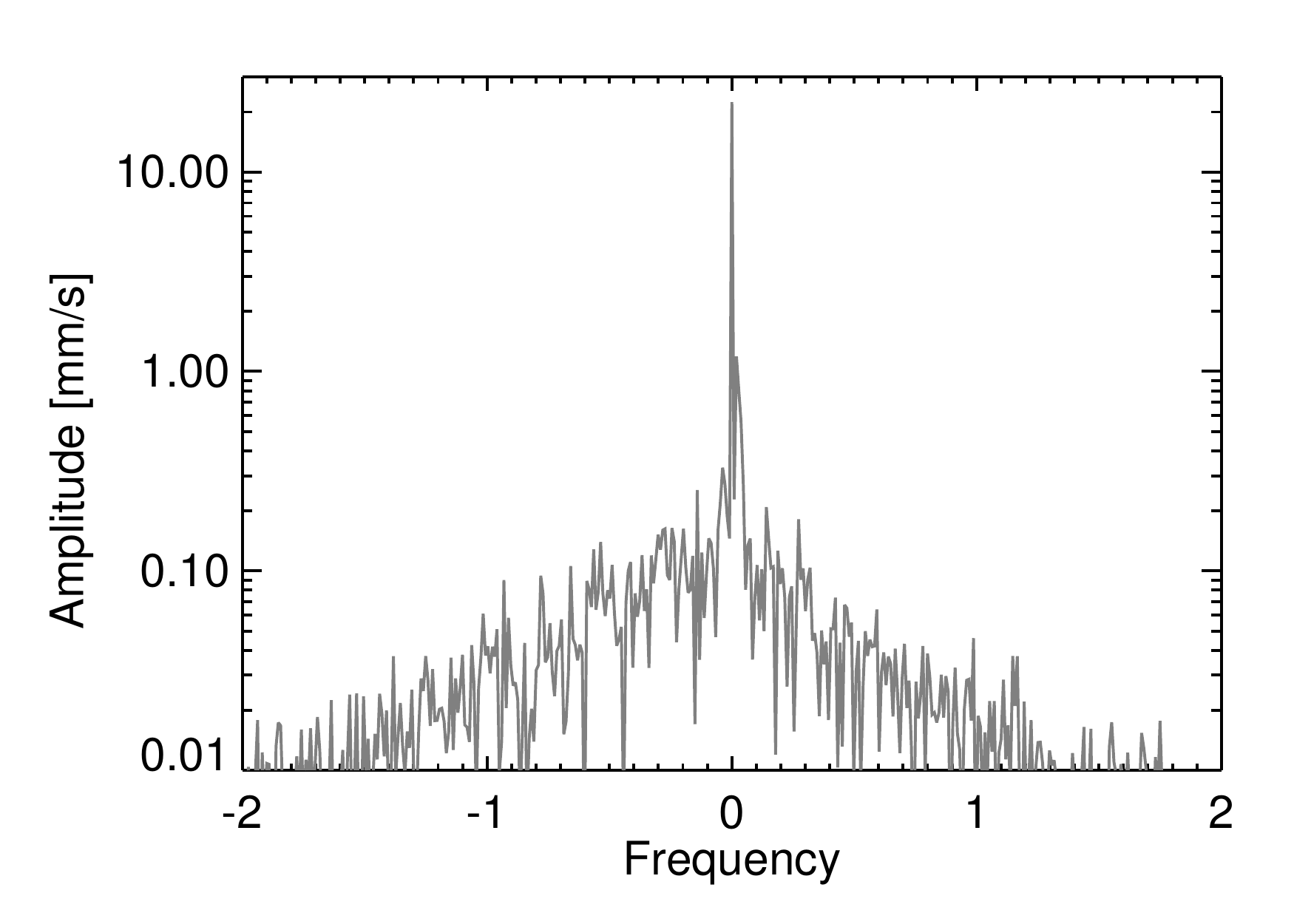}
\includegraphics[width=\sizfigsix]{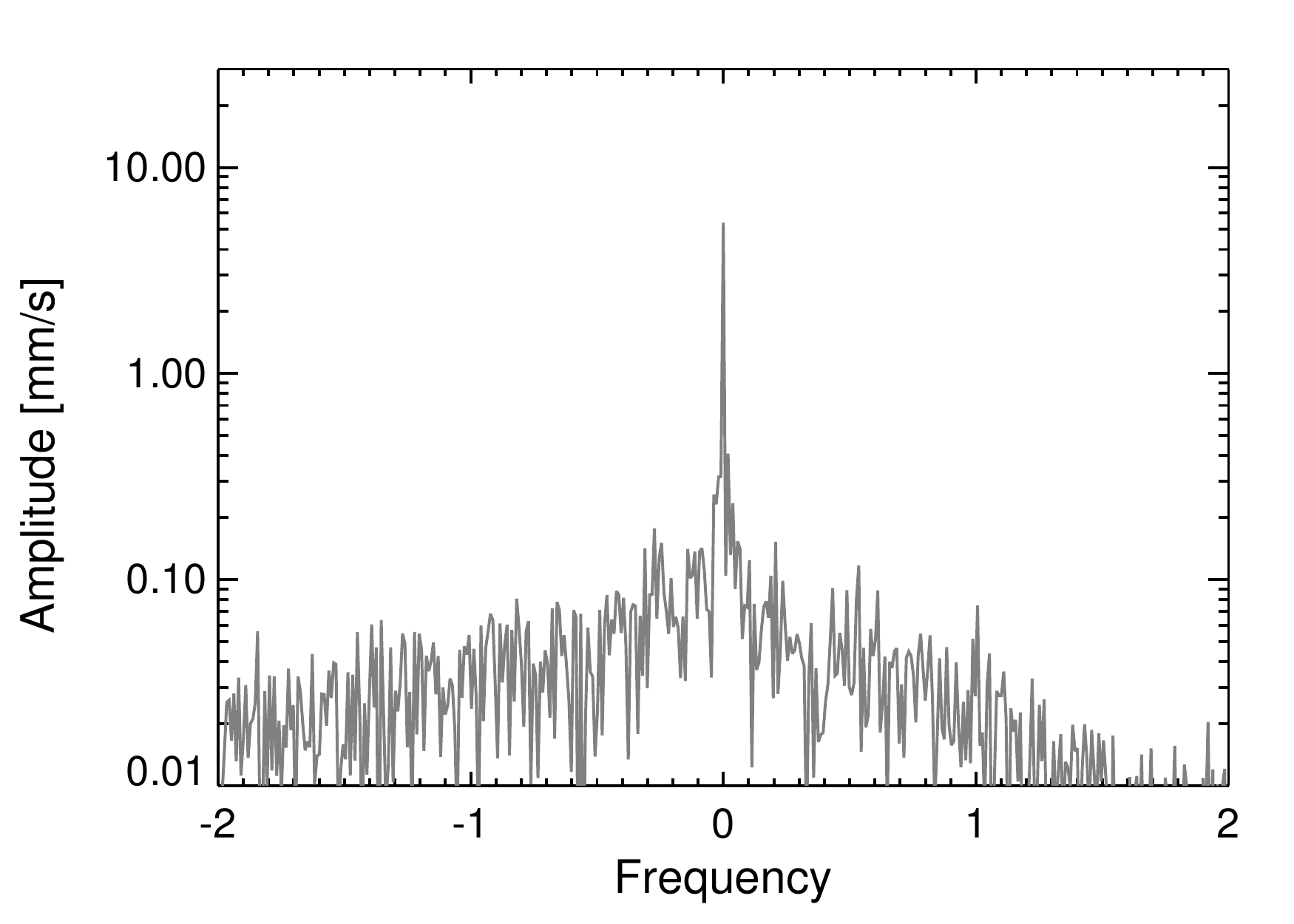}
\includegraphics[width=\sizfigsix]{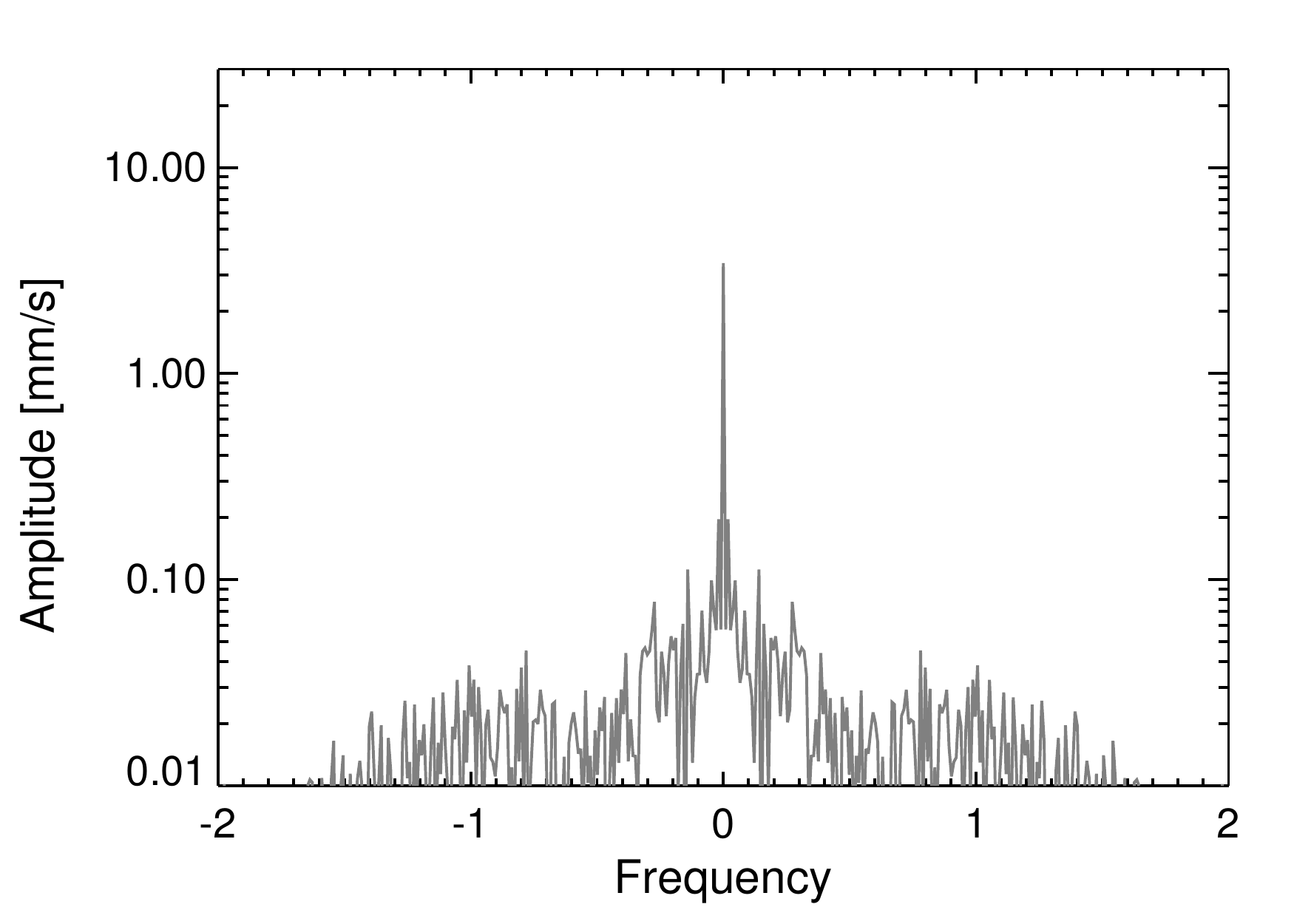}
\\
\includegraphics[width=\sizfigsix]{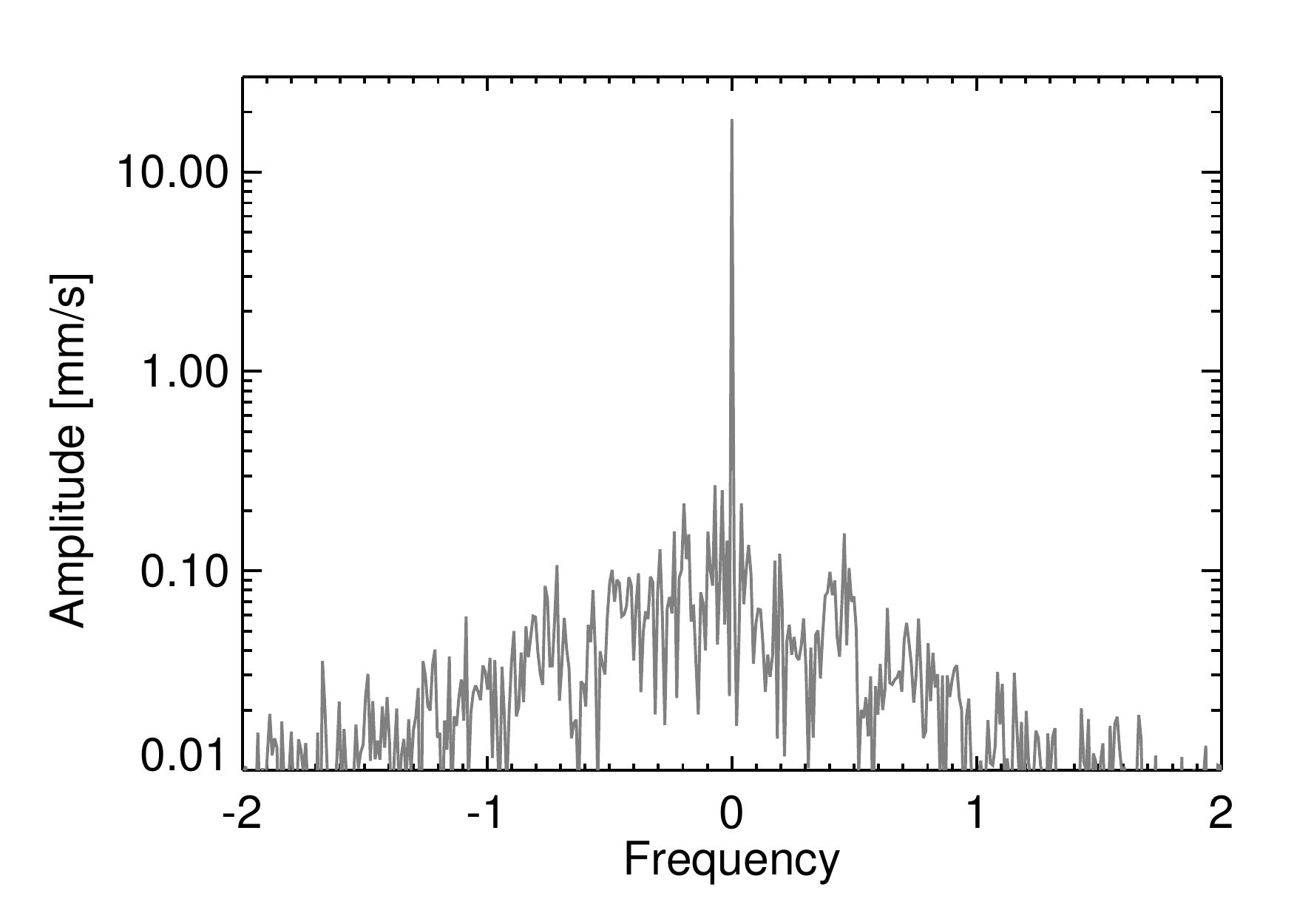}
\includegraphics[width=\sizfigsix]{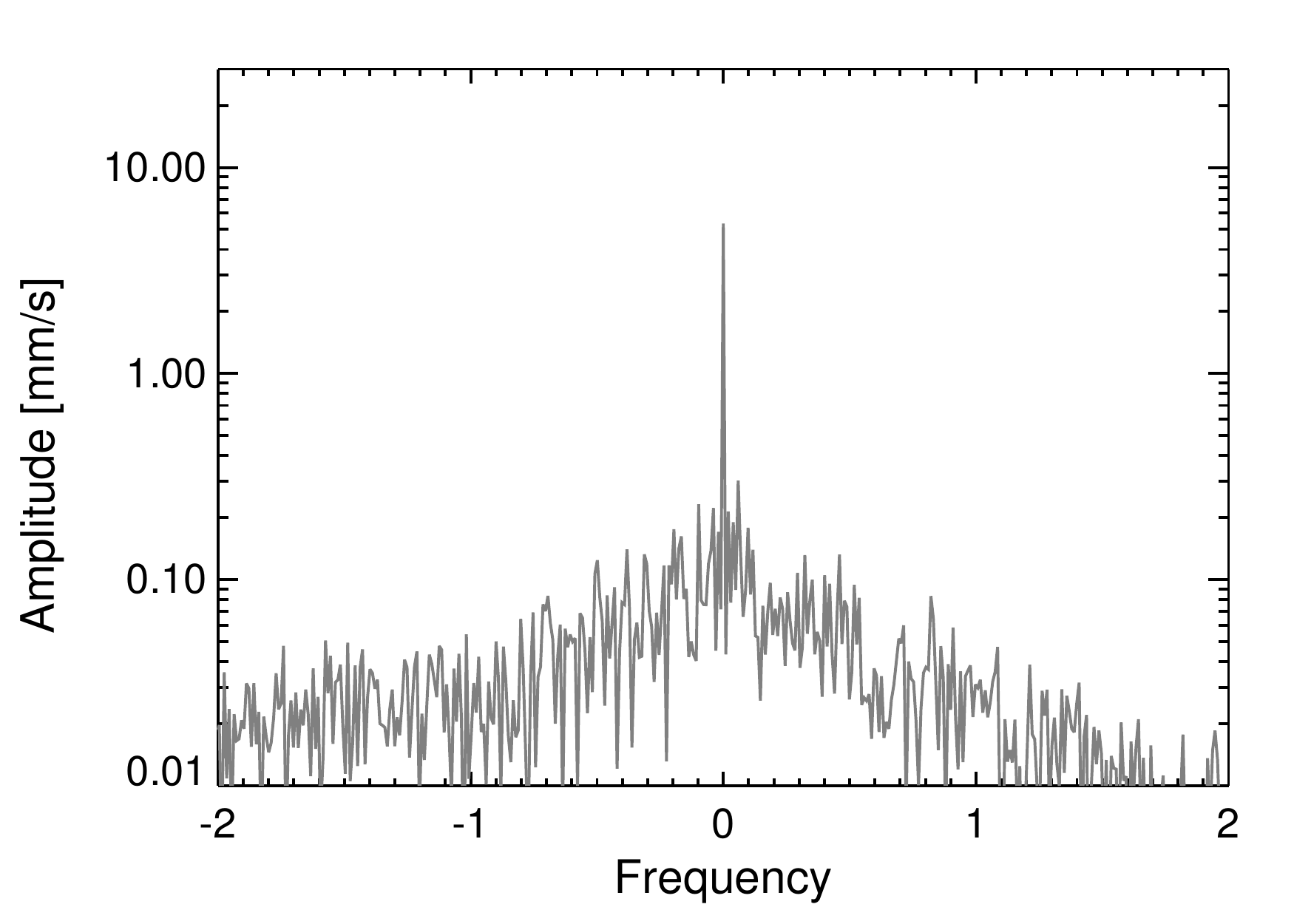}
\includegraphics[width=\sizfigsix]{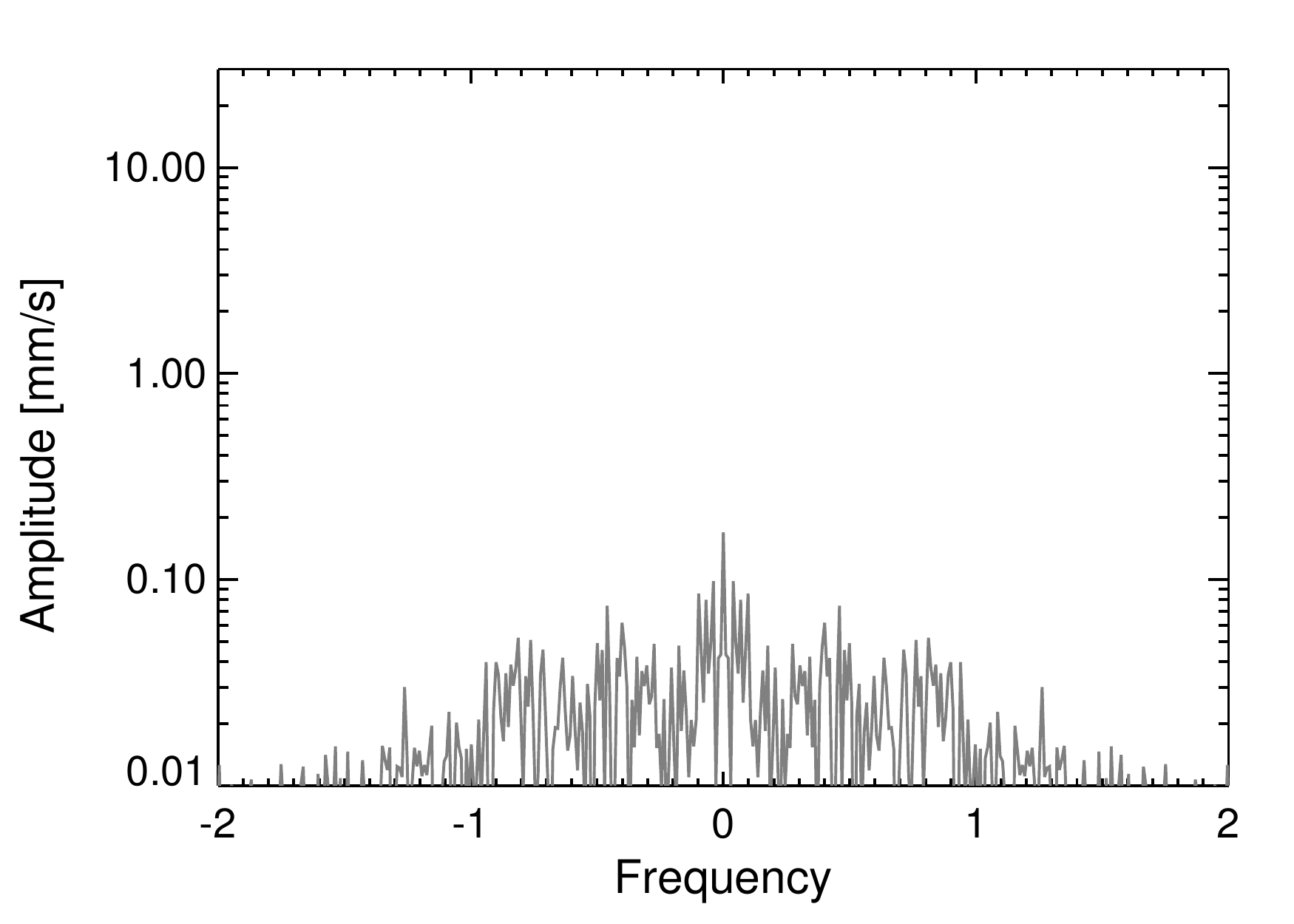}
\caption{Spectra for ${\rm{Re}}=10^4$ from simulations at
$r=0.92R$. From top to bottom: ${\rm{Po}}=0.01, 0.03, 0.05, 0.075,
0.10, 0.125$. From left to right: $(m,k)=(1,1),(2,2),(0,2)$
\label{fig::spec}}
\end{center}
\end{figure}

The flow field is always dominated by the primary forced mode $(m, k)
= (1,1)$ which is directly excited by the Poincar{\'e} force. This is
the directly forced standing inertial mode with $(m,k)=(1,1)$.  Higher
modes, i.e., the modes with $(m,k)=(2,2)$ (second column) result from
nonlinear self-interaction of the direct forced mode. In principle,
other modes with $(m,k) = (3,3), (4,4), \dots $ (and so on) or Kelvin
modes that may be directly forced (e.g. $(m,k)=(1,3), (1,5),...$) can
also be observed. However, their amplitudes are smaller and the most
dominant modes are those presented in figure~\ref{fig::spec}.

For small ${\rm{Po}}$ (i.e. ${\rm{Po}}=0.01$ and ${\rm{Po}}=0.03$)
weak broad peaks with $\omega \neq 0$ can be recognized. In these
regimes temporal variations in terms of irregularly appearing
azimuthally drifting vortices emerge parallel to the cylinder
axis. However, these features remain intermittent and do not show a
strong amplitude. Similar features have been found experimentally
\citep{mouhali2012} and numerically \citep{lin2016}.  For small
precession ratios we also see weak broad maxima in the negative part
of the spectrum which may indicate inertial waves propagating in the
retrograde direction. However, these peaks are too weak and the maxima
remain too broad so that so far no conclusion is possible about the
nature of these features.  Another less striking feature are the
little peaks located symmetrically around the central peak in the runs
at ${\rm{Po}}=0.075$ (fourth row). These peaks result from a periodic
variation of the amplitude of the dominant mode. Again the
corresponding amplitudes are weak and we will ignore these
contributions in the following.

A much more noticeable peculiarity, which turns out to be important
for the dynamo experiment, appears in the spectra shown in the right
column that corresponds to an axisymmetric mode with $(m,k)=(0,2)$.
The amplitude of this mode is mostly small, except around
${\rm{Po}}\approx 0.100$ where it becomes a significant contribution
to the total flow (right plot in figure~\ref{fig::amp2}).  In summary,
the inspection of the spectra demonstrates that the precession driven
flow in the turntable system is essentially dominated by standing
inertial waves (Kelvin modes) and fluctuations do not play an
important role. This is in particular valid for strongly precessing
fluids. In the following, we ignore the time-dependent contributions
and only discuss the amplitudes of modes that are standing in the
turntable reference frame and have the frequency $\omega=0$.
\newcommand{\abstand}{-0.64cm}
\begin{figure}[t!]
\begin{center}
{
\includegraphics[width=0.35\textwidth]{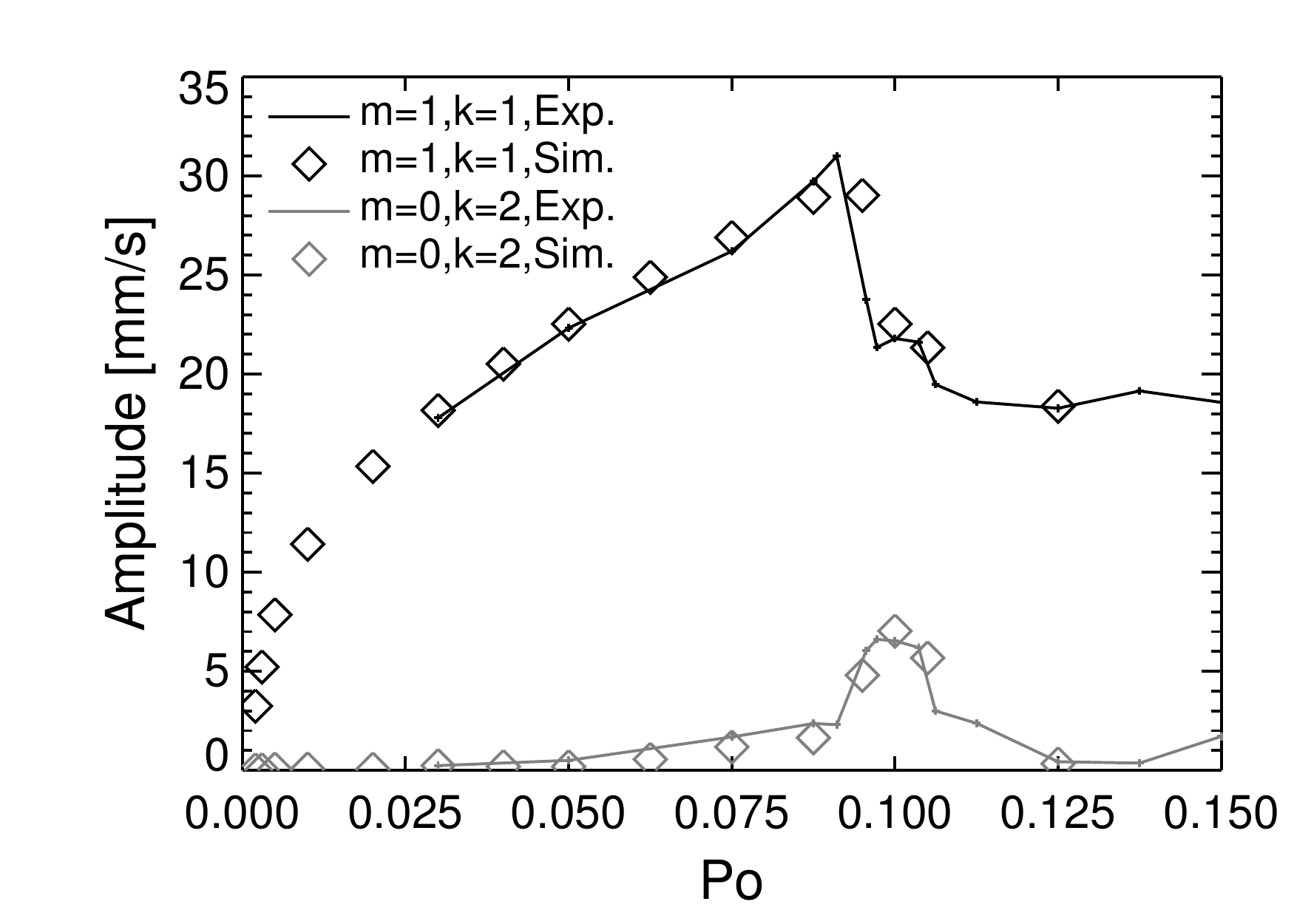}
\hspace*{\abstand}
\includegraphics[width=0.35\textwidth]{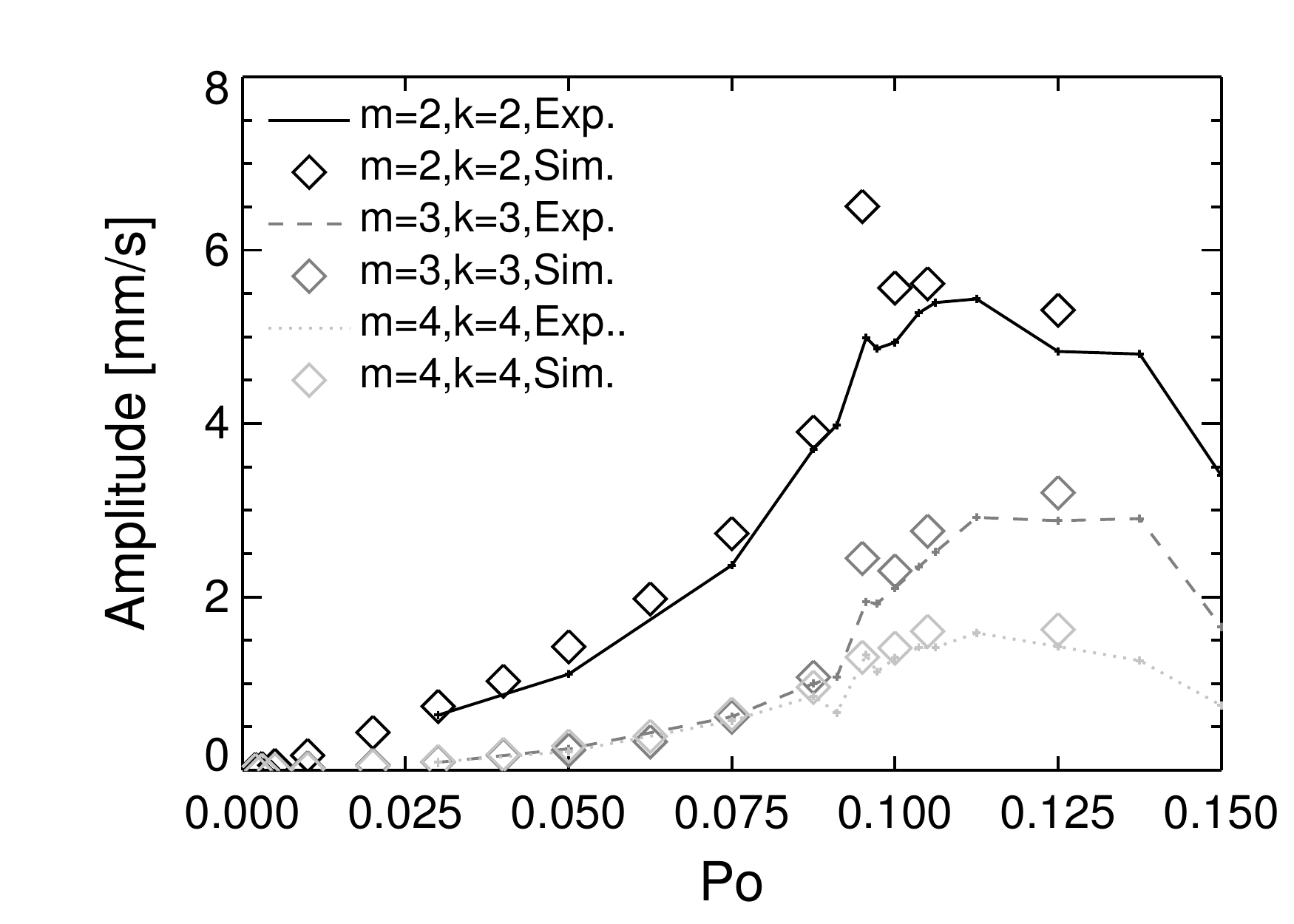}
\hspace*{\abstand}
\includegraphics[width=0.35\textwidth]{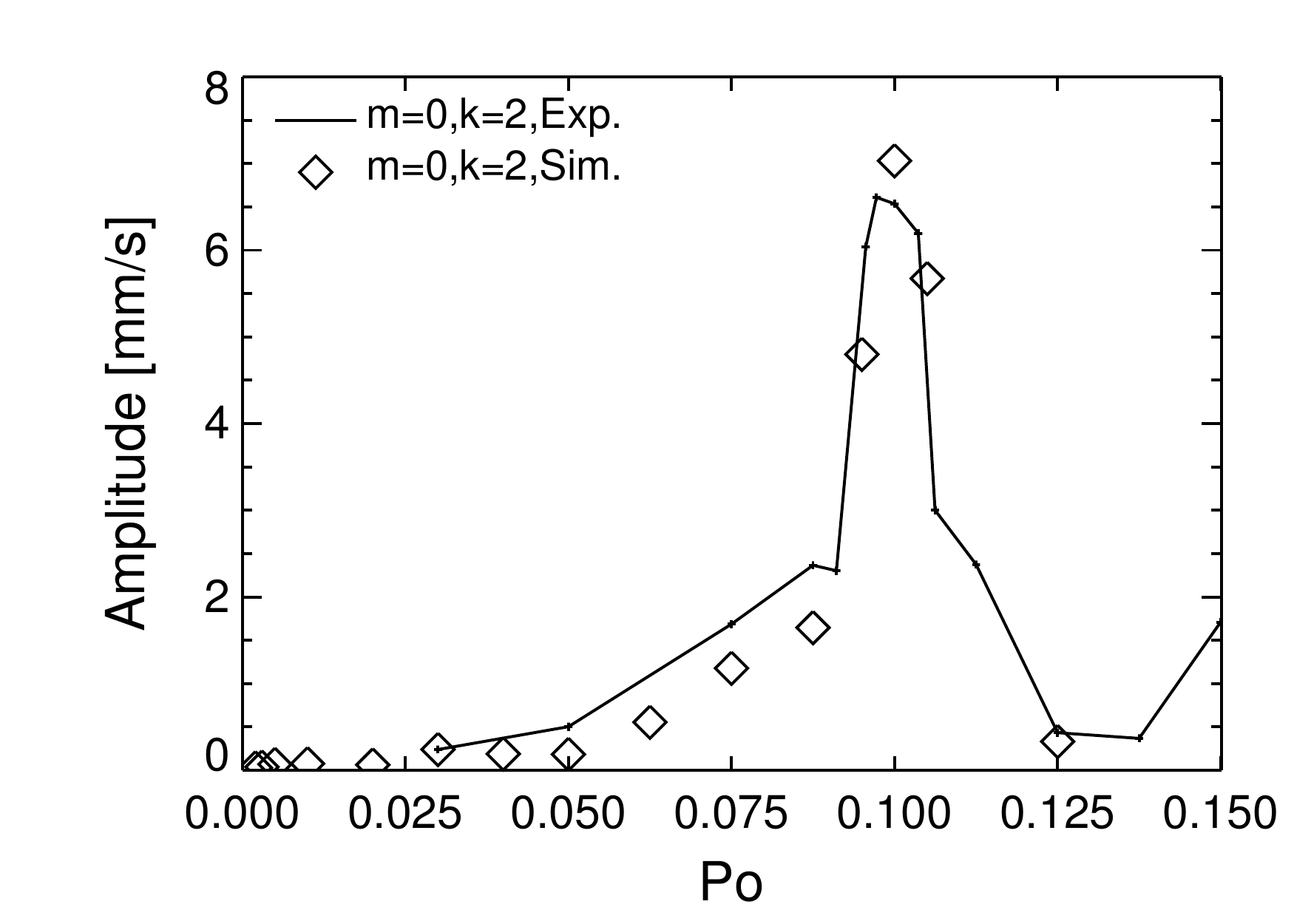}
}
\caption{\label{fig::amp2}
Amplitudes of the most dominant inertial modes from simulations
(diamonds) and UDV measurements (solid curves) versus precession
ratio.  (a) $(m,k)=(1,1)$ (black curve) and $(m,k)=(0,2)$ (grey curve)
(b) $(m,k)=(2,2)$ (black curve), $(m,k)=(3,3)$ (dark dashed grey
curve) and $(m,k)=(4,4)$ (light dotted grey curve) (c) close up of
$(m,k)=(0,2)$
}
\end{center}
\end{figure}
\renewcommand{\siz}{0.34\textwidth}
\renewcommand{\abstand}{-0.5cm}
\begin{figure}[t!]
\begin{center}
\includegraphics[width=\siz]{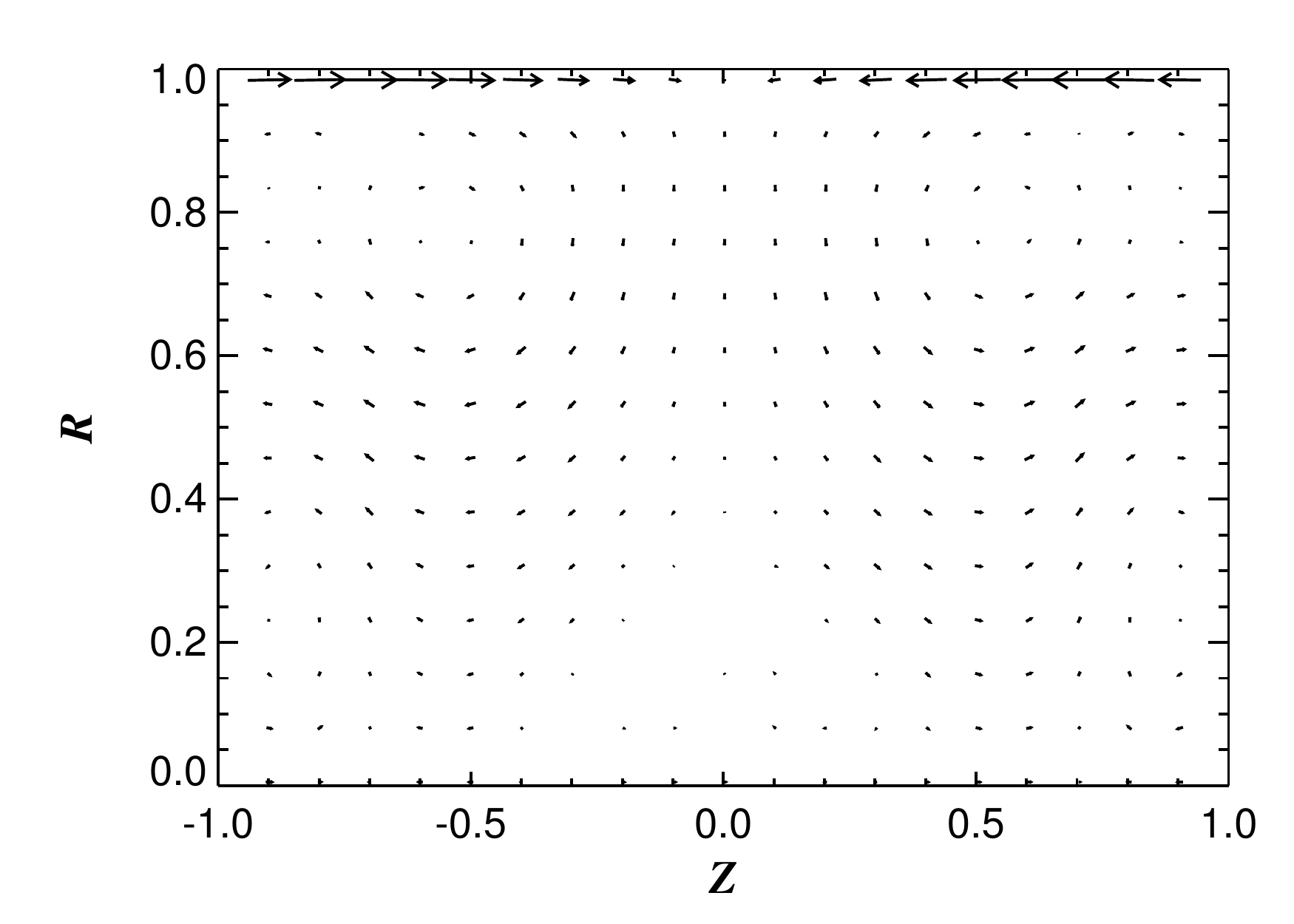}
\hspace*{\abstand}
\includegraphics[width=\siz]{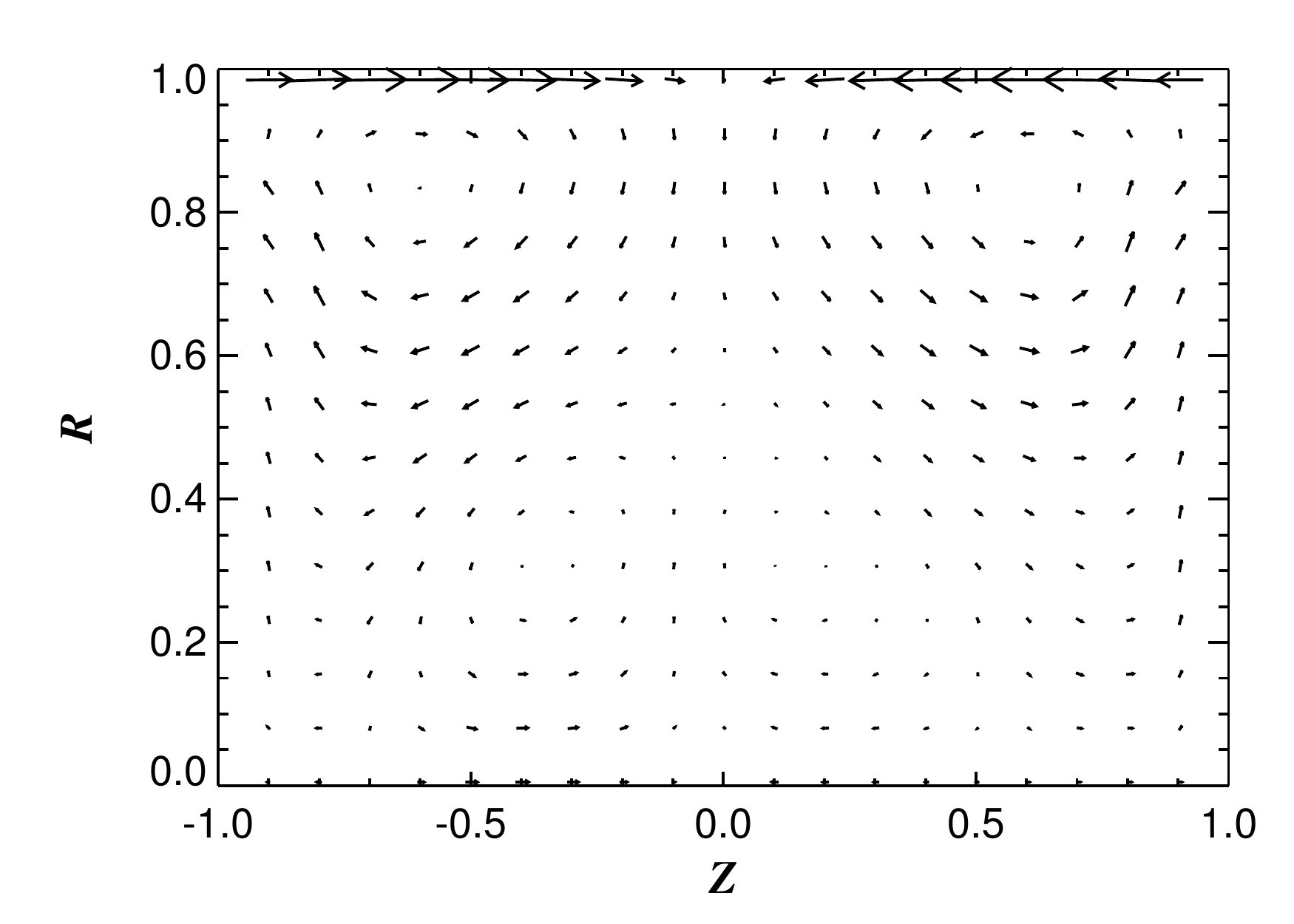}
\hspace*{\abstand}
\includegraphics[width=\siz]{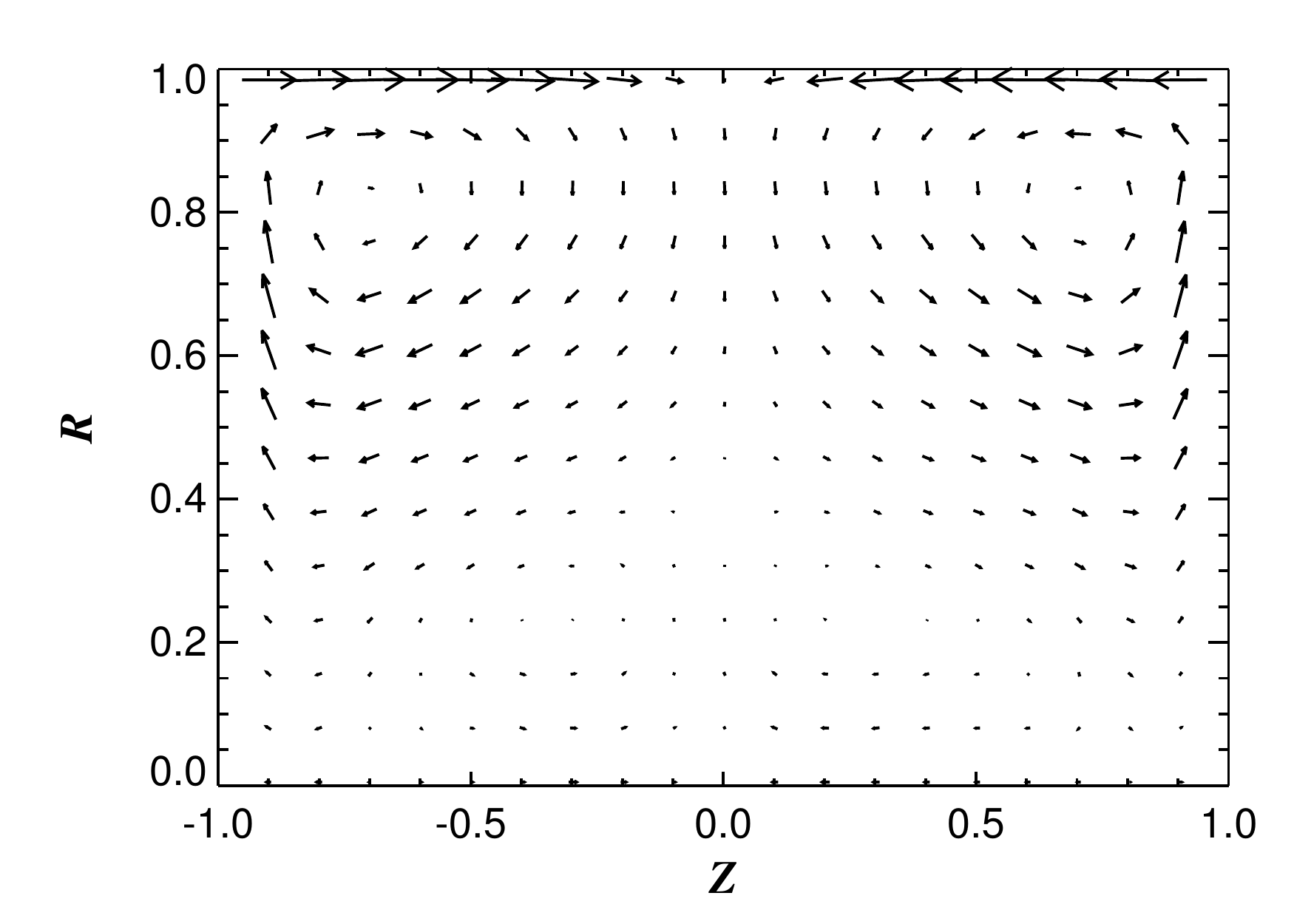}
\\[-0.3cm]
\includegraphics[width=\siz]{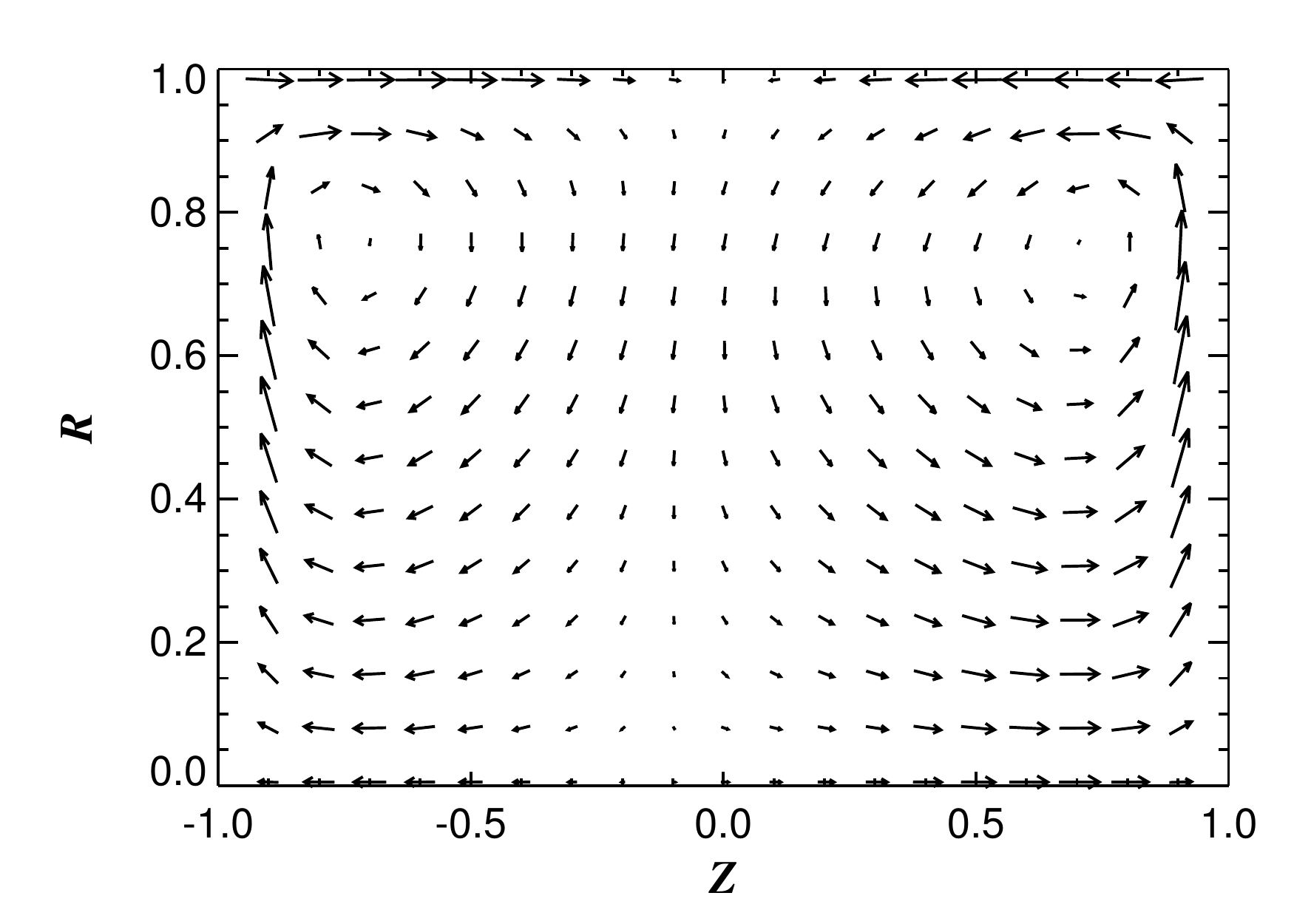}
\hspace*{\abstand}
\includegraphics[width=\siz]{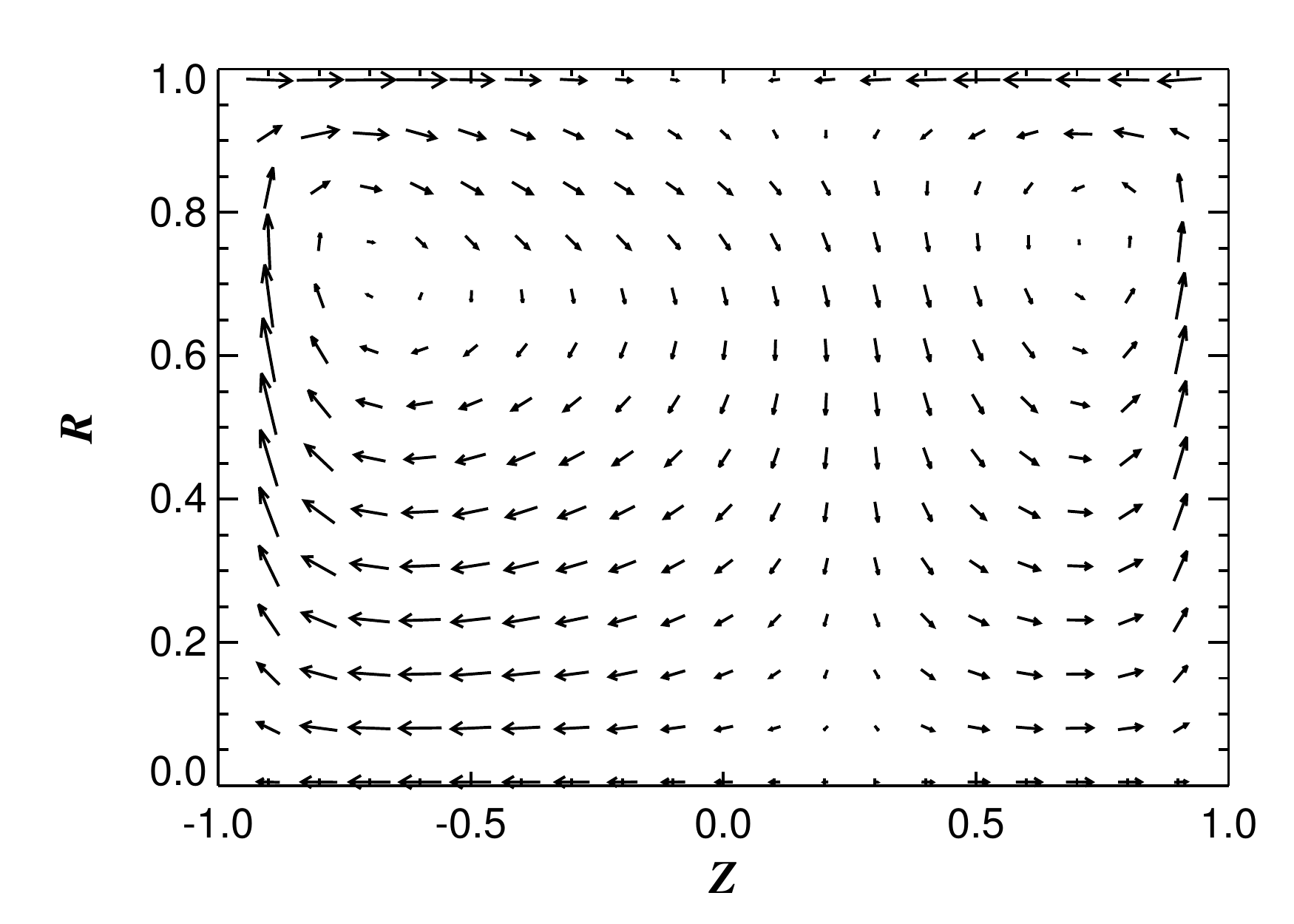}
\hspace*{\abstand}
\includegraphics[width=\siz]{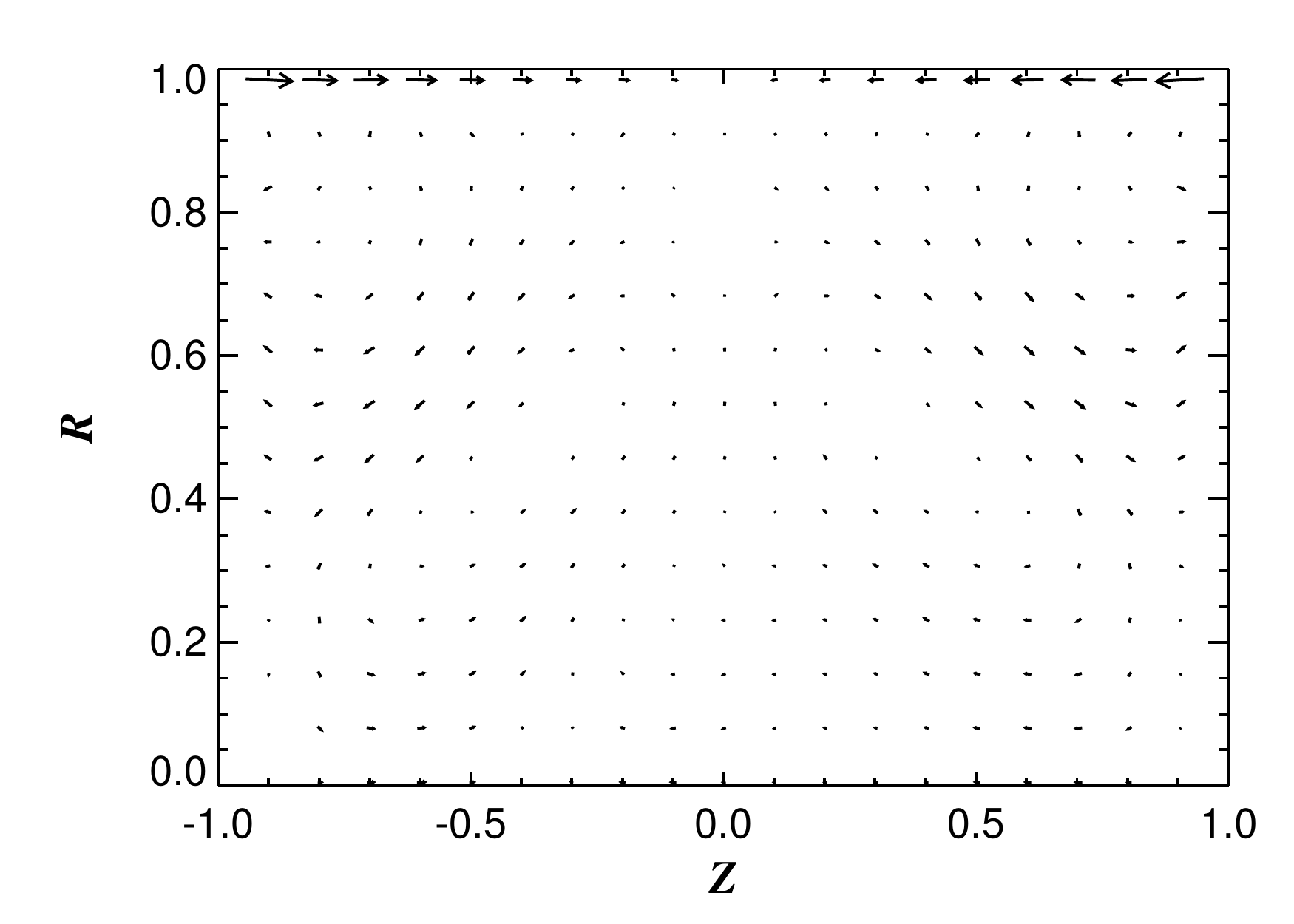}
\caption{\label{fig::axisymstruc}
Structure of the poloidal components $u_r$ and $u_z$ of the
time-averaged axisymmetric velocity field for various precession
ratios. From top left to bottom right:
${\rm{Po}}=0.05,0.0875,0.095,0.100,0.105, 0.125$.}
\end{center}
\end{figure}
The amplitudes of the most important inertial modes are shown in
figure~\ref{fig::amp2}.  The plots show the data obtained from
simulations and experiments at ${\rm{Re}}=10^4$.  The left plot
compares the directly forced mode $(m,k)=(1,1)$ (black curve) and the
axisymmetric mode $(m,k)=(0,2)$ (grey curve), the central plot shows
the corresponding behavior of the most important multiples excited
from nonlinear self-interactions, $(m,k)=(2,2)$ (black curve),
$(m,k)=(3,3)$ (grey curve), and $(m,k)=(4,4)$ (light grey curve), and
the right panel shows a close up of the behavior of the axisymmetric
mode $(m,k)=(0,2)$.  In all cases simulations and experiments agree
reasonable well (the diamonds denote simulations and the solid curves
denote experimental data).

The plots in figure~\ref{fig::amp2} illustrate the two essential
characteristics of the flow amplitudes around ${\rm{Po}}=0.1$, which
are the abrupt decrease of the amplitude of the $m=1$ mode and the
simultaneous emergence of the axially symmetric mode with $(m,k)=
(0,2)$ in the range $0.095 \lesssim {\rm{Po}} \lesssim 0.105$.  In the
meridional plane, the structure of the axially symmetric mode
corresponds to a double roll pattern as it is shown in
figure~\ref{fig::axisymstruc} which presents the poloidal flow
components ($u_r$ and $u_z$) for different precession ratios.  The
double roll pattern only appears in the range ${\rm{Po}}\in [0.095,
0.105]$ (figure~\ref{fig::axisymstruc}(c), (d), (e)) with the clearest
formation at ${\rm{Po}}=0.100$ where the rolls occupy almost the
entire volume.

\begin{figure}[b!]
\begin{center}
\includegraphics[width=\siz]{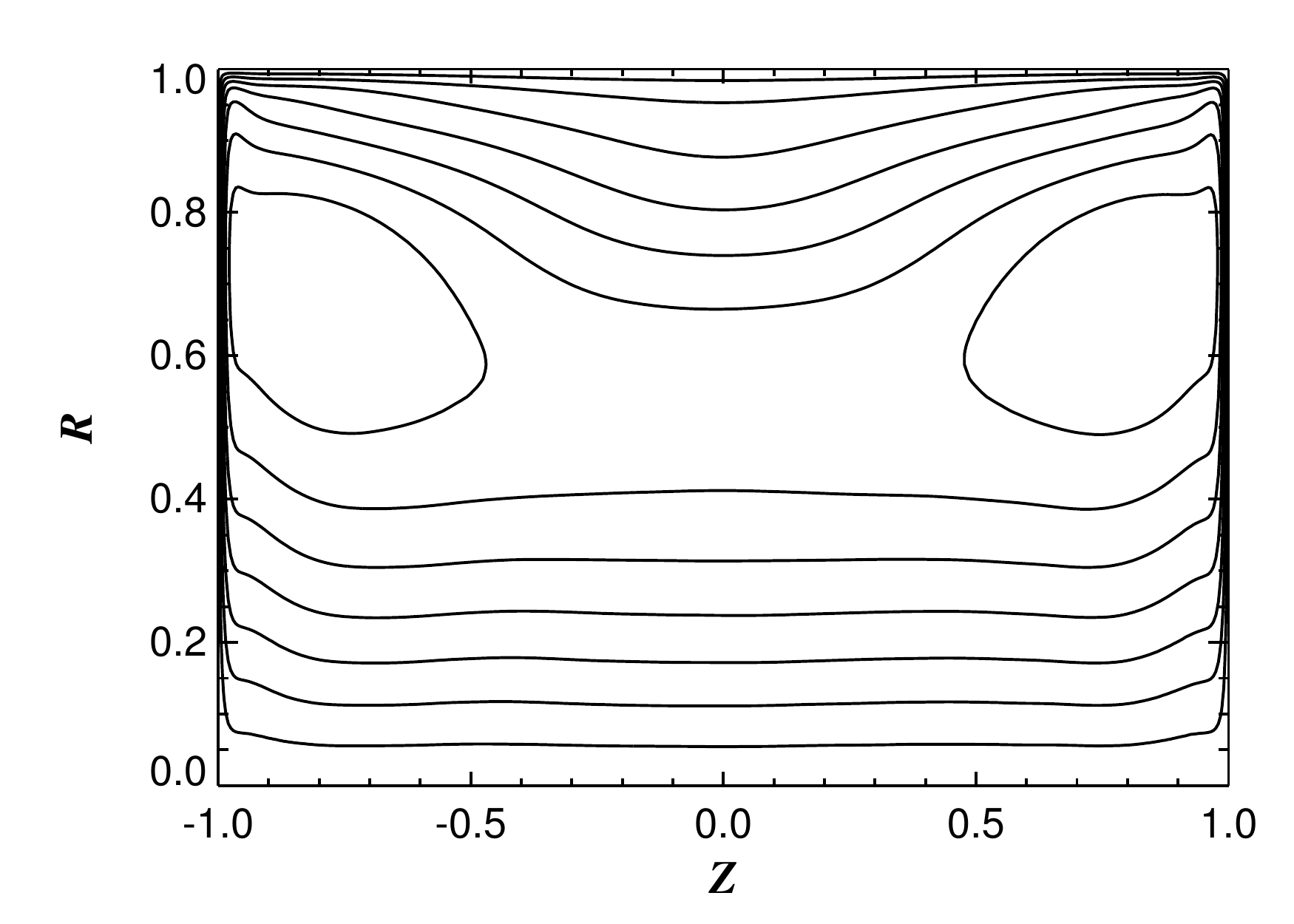}
\hspace*{\abstand}
\includegraphics[width=\siz]{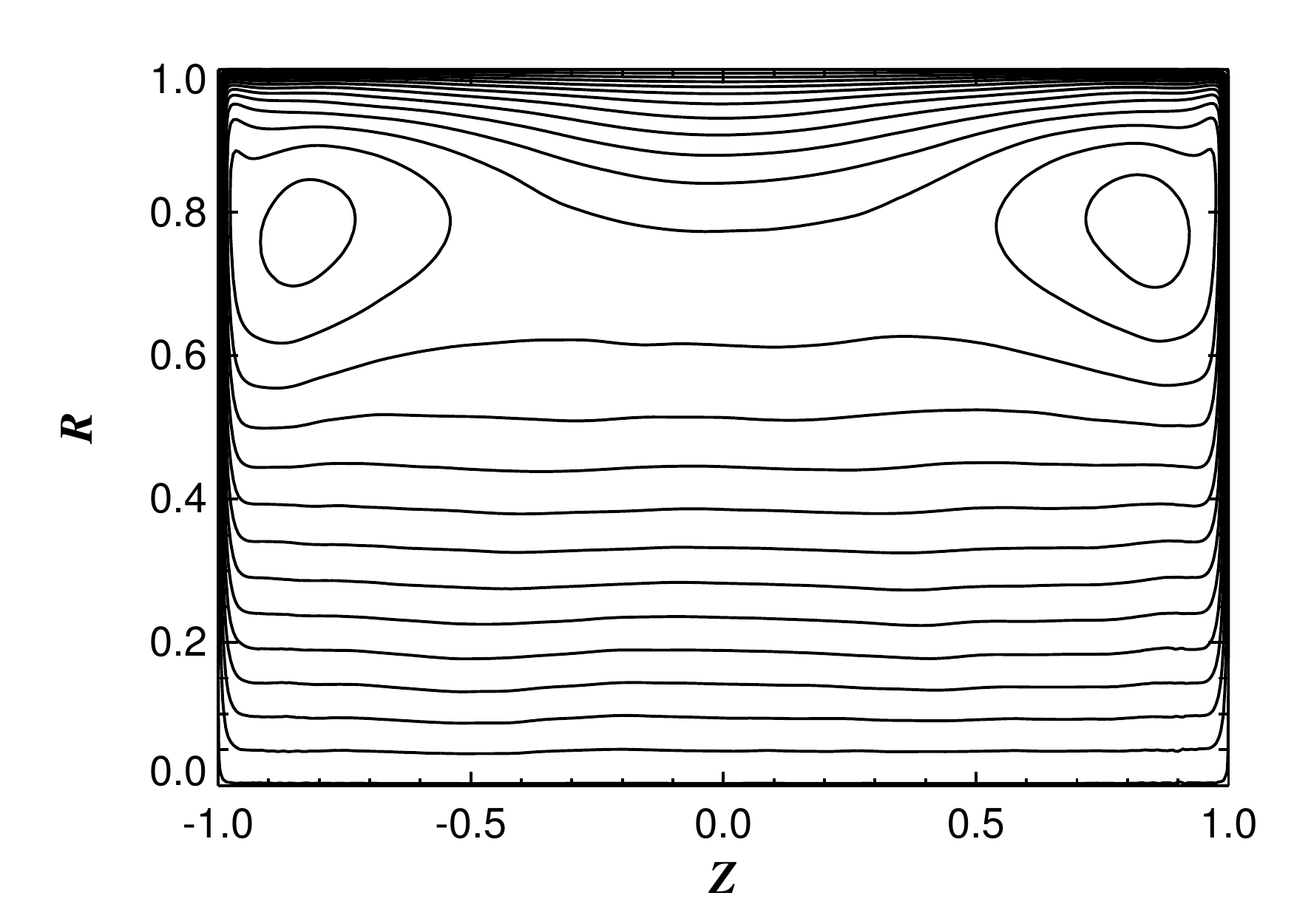}
\hspace*{\abstand}
\includegraphics[width=\siz]{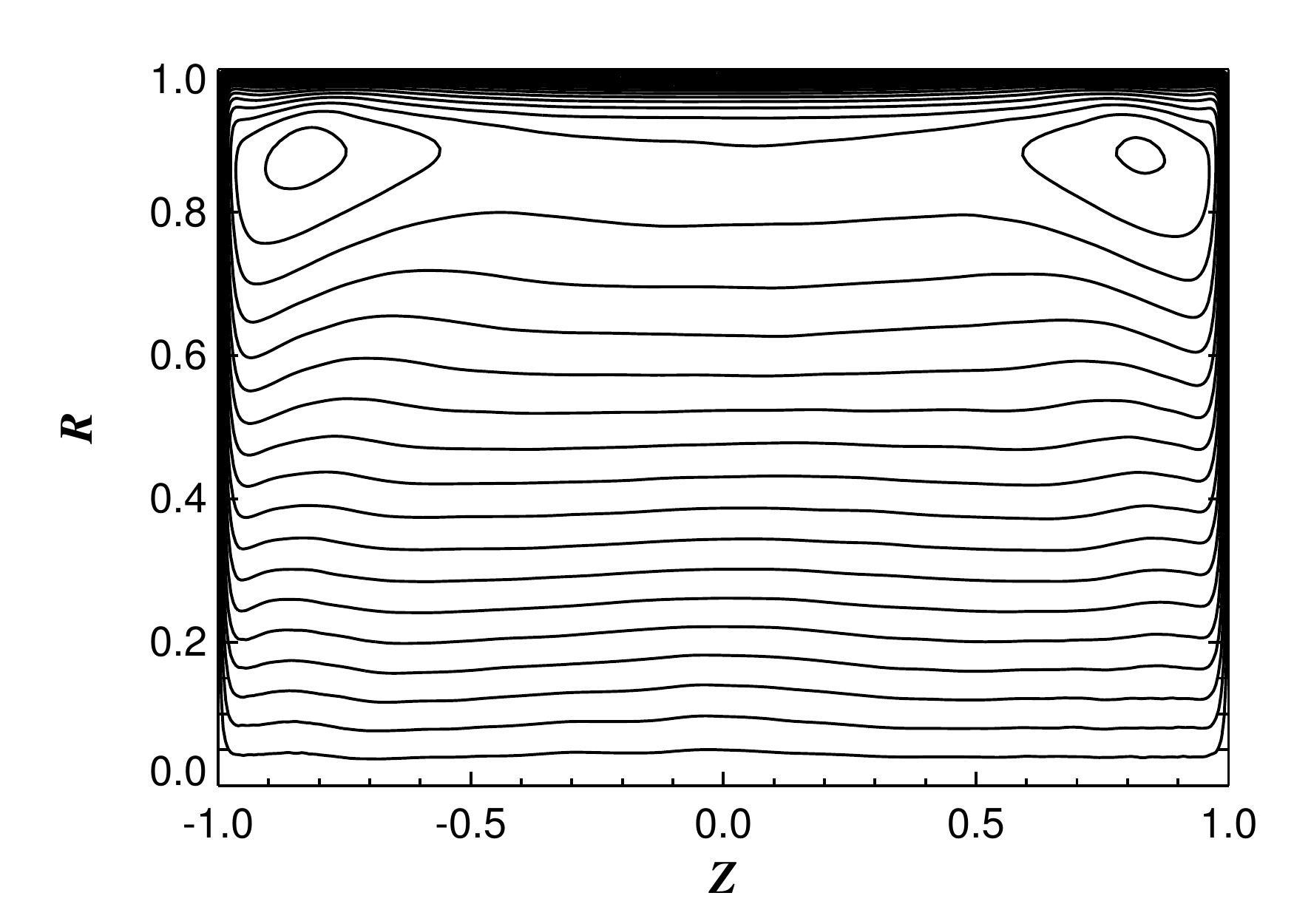}
\\[-0.3cm]
\includegraphics[width=\siz]{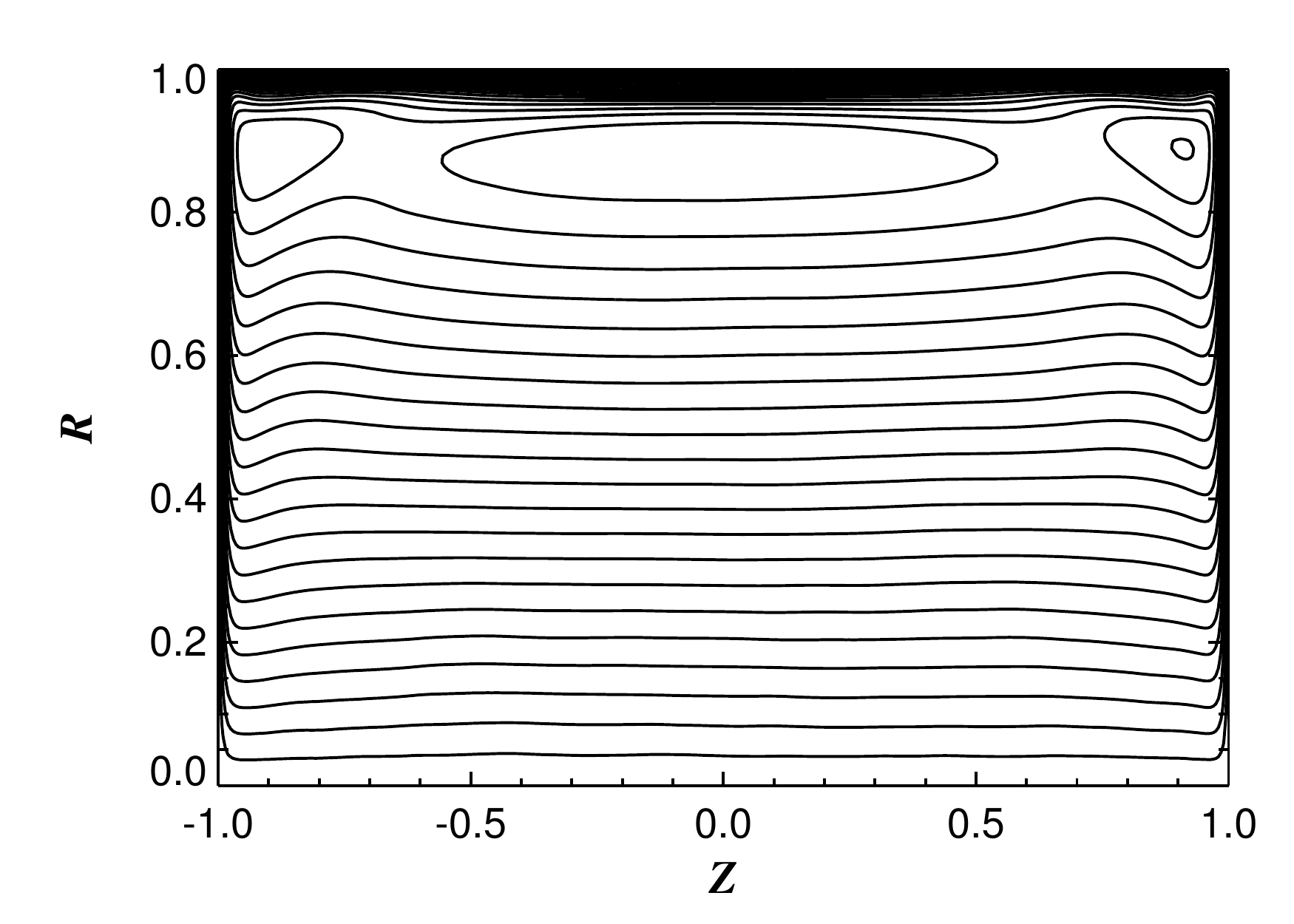}
\hspace*{\abstand}
\includegraphics[width=\siz]{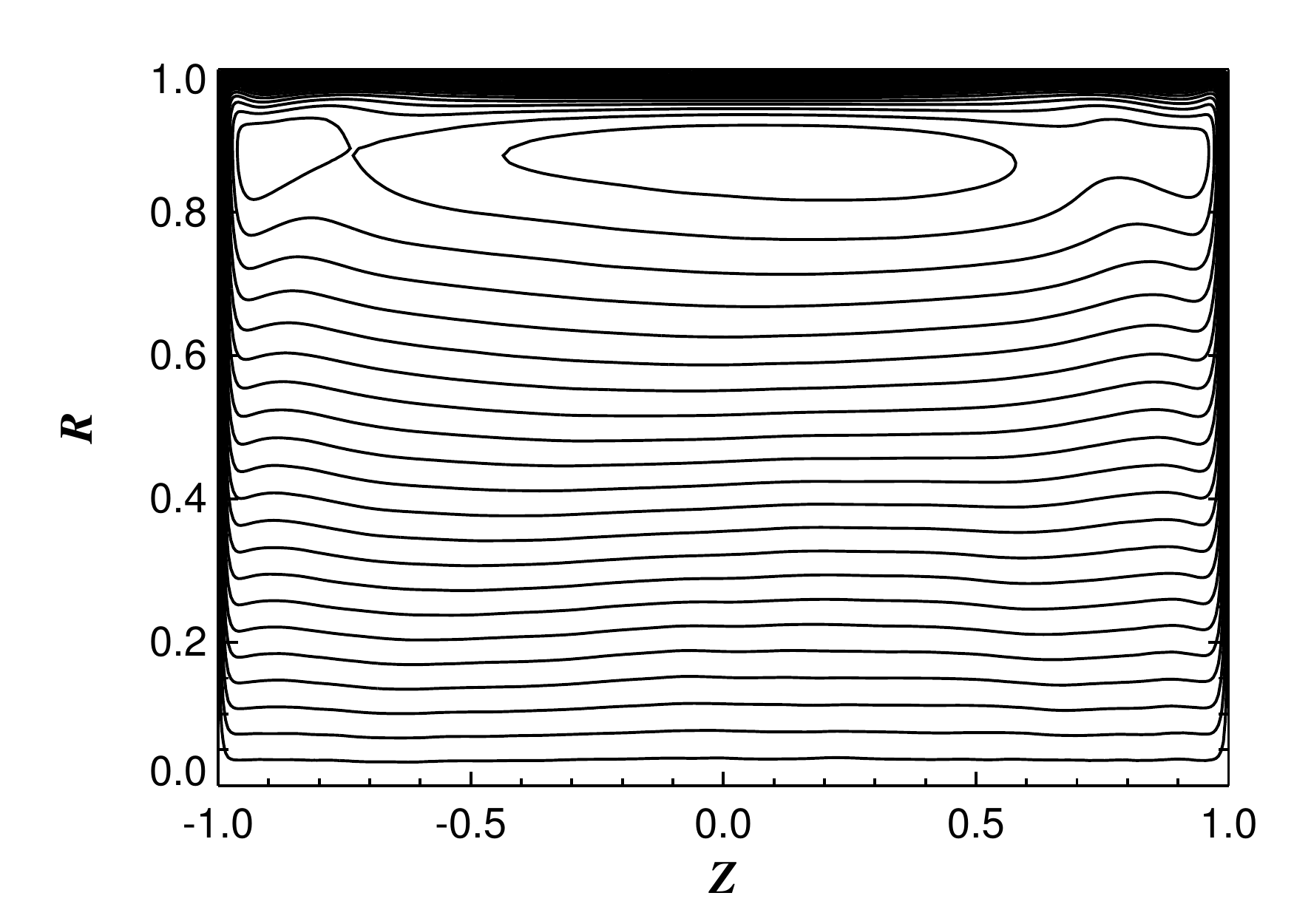}
\hspace*{\abstand}
\includegraphics[width=\siz]{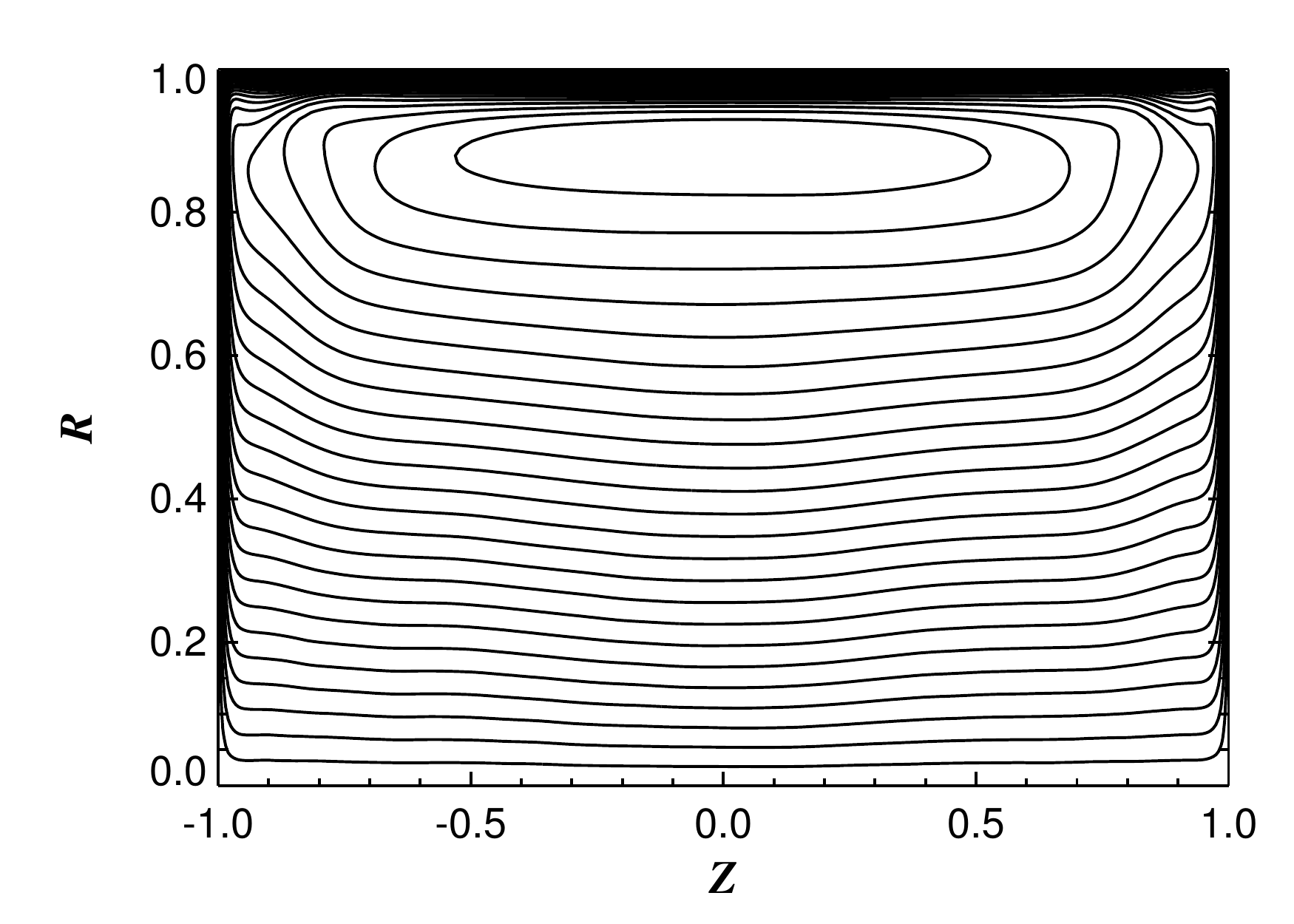}
\caption{\label{fig::axisymstruc_uphi}
Structure of the time-averaged axisymmetric azimuthal velocity
$u_{\varphi}$ without solid body rotation for various
precession ratios. From top left to bottom right:
${\rm{Po}}=0.05,0.0875,0.095,0.100,0.105,0.125$.} 
\end{center}
\end{figure}

Regarding the mean azimuthal axisymmetric flow, we find that the
modification of the initial solid body rotation continuously increases
with ${\rm{Po}}$.  This change, shown in
figure~\ref{fig::axisymstruc_uphi}, is described by a circulation flow
that is largely independent of $z$ and oriented opposite to the
rotation of the container thus corresponding to a slowdown of the --
initial -- solid body rotation.  Note also the change in the structure
of this induced azimuthal flow. While there are two maxima close to
the end caps for smaller ${\rm{Po}}$, there is only one maximum in the
equatorial plane for larger precession ratios.

\subsection{A centrifugal instability}
\begin{figure}[t!]
\begin{center}
\subfigure[]
{\includegraphics[width=0.49\textwidth]{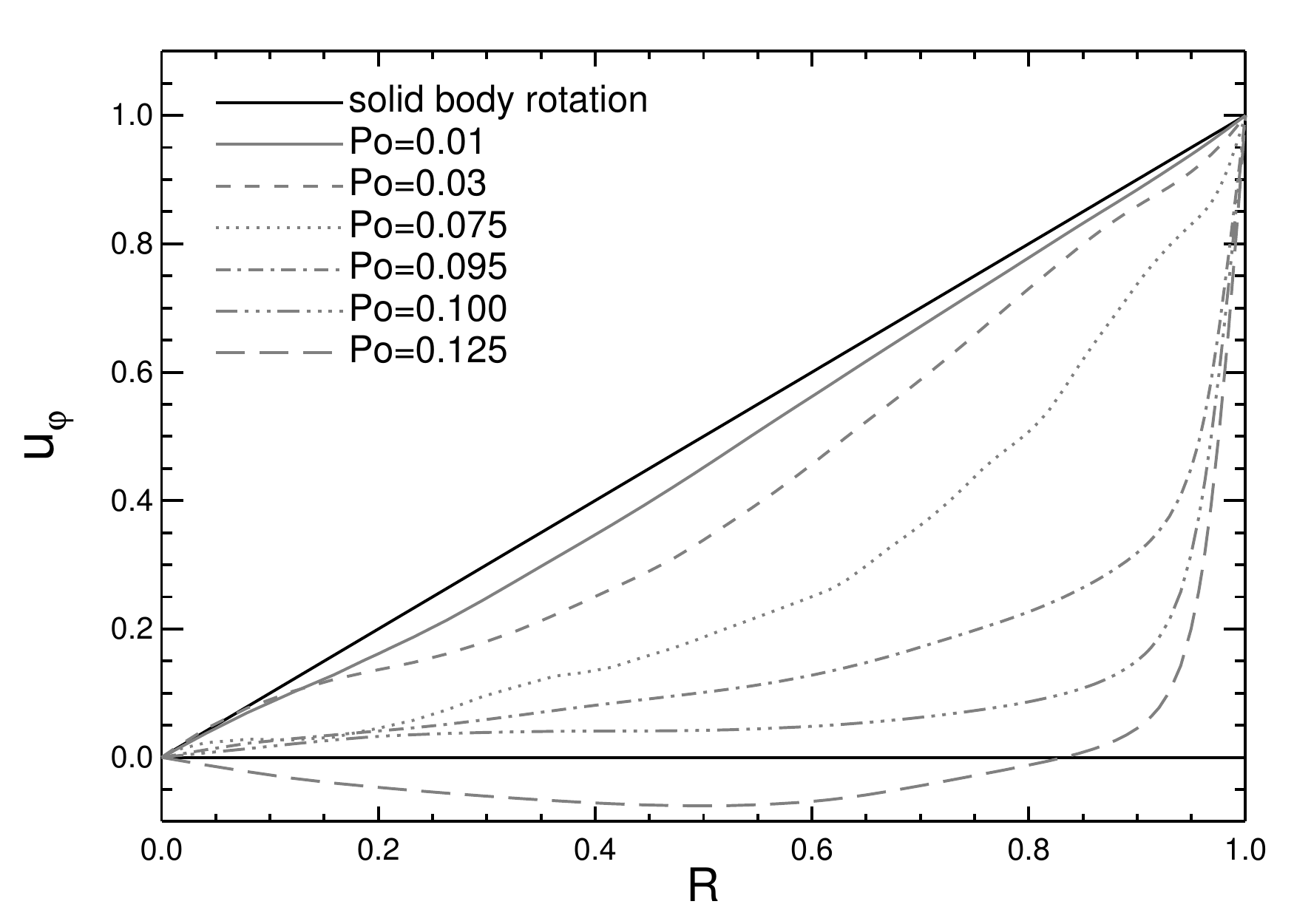}}
\subfigure[]{
\includegraphics[width=0.49\textwidth]{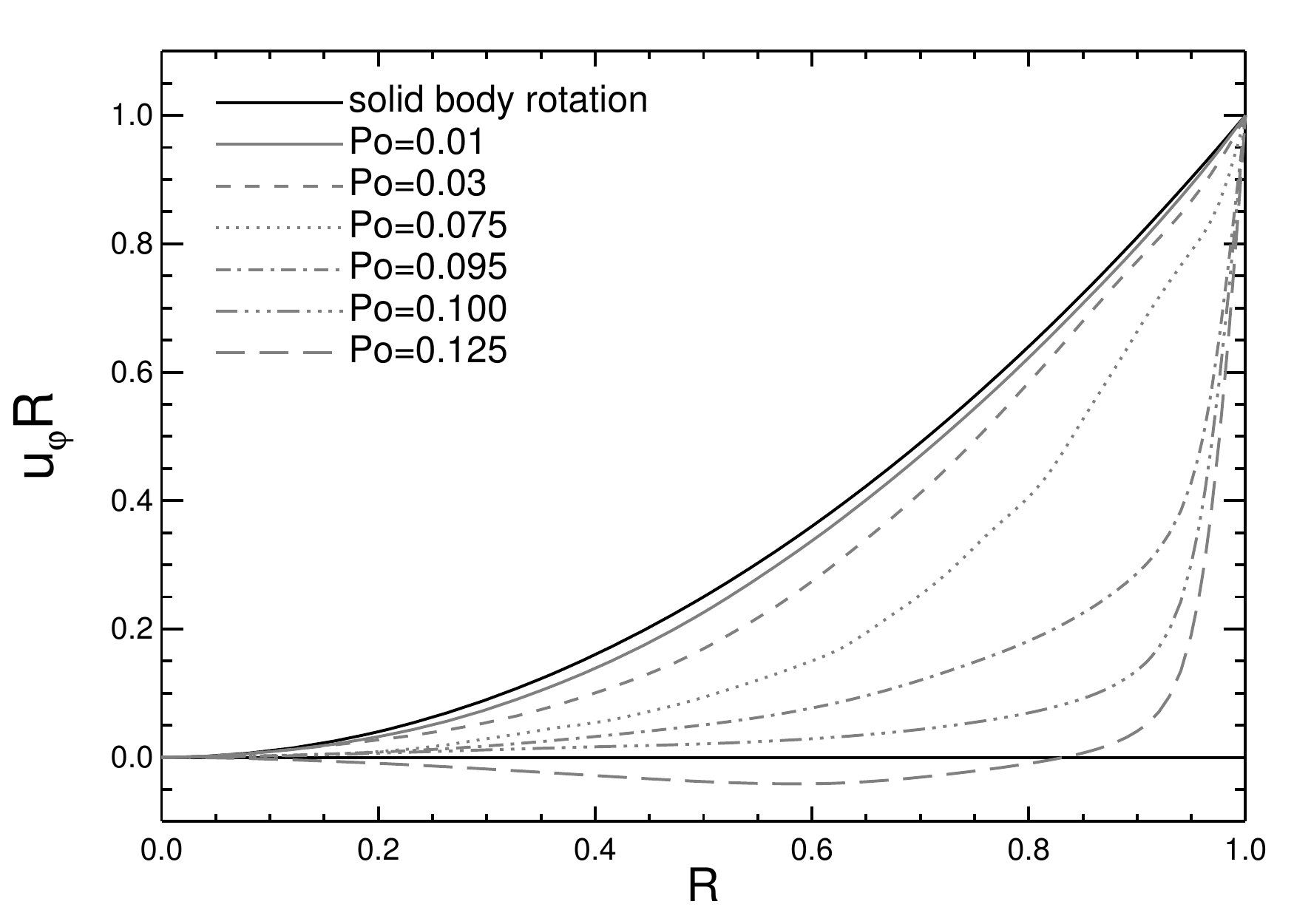}}
\caption{\label{fig::radial_profile_uphi}
(a) Radial profile of the azimuthal axisymmetric velocity
$u_{\varphi}$ including the solid body rotation from simulations with
various precession ratios at ${\rm{Re}}=10^4$.  (b) Corresponding
profiles of the angular momentum.}
\end{center}
\end{figure}

The change in the azimuthal circulation described in the previous
section is surprisingly strong and is supposed to be the reason for
the occurrence of the double roll flow which is reminiscent of the
vortices arising in a Taylor Couette setup.  The {\it{induced}}
azimuthal velocity is nearly independent of the axial coordinate and
oriented opposite to the solid body rotation that would exist in a
pure rotating system without precessional forcing. This slowdown of
the solid body rotation increases with the precession ratio
${\rm{Po}}$ such that nearly exactly at the critical point, i.e. at
${\rm{Po}}\sim 0.1$, the Rayleigh criterion for the stability of a
rotating fluid becomes violated. This can be seen in
figure~\ref{fig::radial_profile_uphi} which shows the radial profile
of $u_{\varphi}$ (left) and the corresponding angular momentum
$ru_{\varphi}$ (right) for increasing ${\rm{Po}}$ including the solid
body rotation. The slope of the angular momentum is indeed negative in
the bulk when ${\rm{Po}}\gtrsim 0.1$ which is the criterion for the
occurrence of a centrifugal instability.
In the specific case of the azimuthal velocity profiles as they are
generated by precession, the axisymmetric double roll structure occurs
shortly before reaching the Rayleigh line, which can be explained by
the presence of the strong $m = 1$ mode due to the direct precessional
forcing. A more detailed investigation would therefore require a
stability analysis of the superposition of the induced circulation and
the directly driven Kelvin mode. Since this is not straightforward --
and probably requires the additional consideration of boundary layers
that are not well known, we postpone such an calculation to a future
study.

\subsection{Dependence on Reynolds number}

\begin{figure}[t!]
\begin{center}
\subfigure[]
{\includegraphics[width=0.5\textwidth]{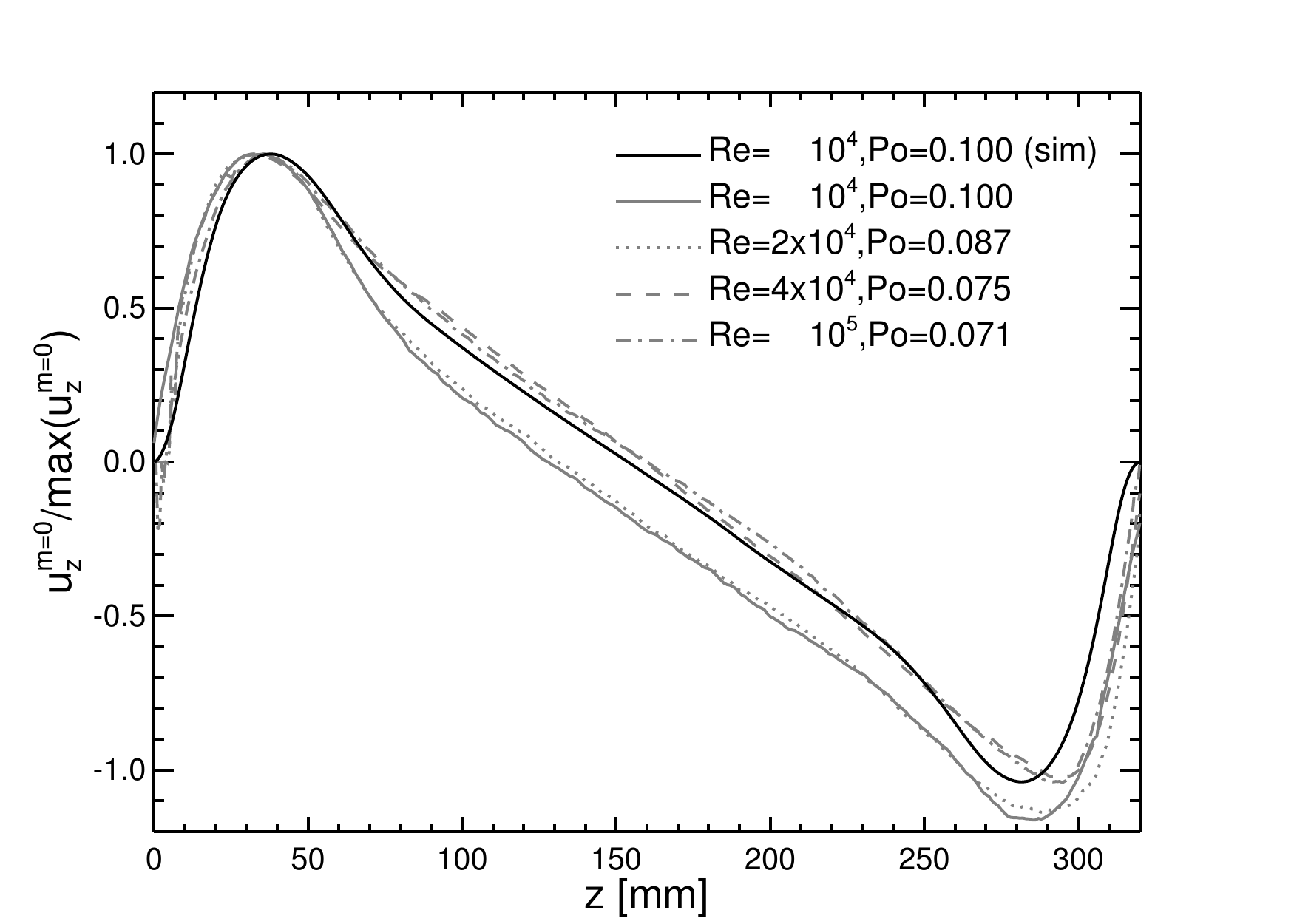}}
\subfigure[]
{\includegraphics[width=0.475\textwidth]{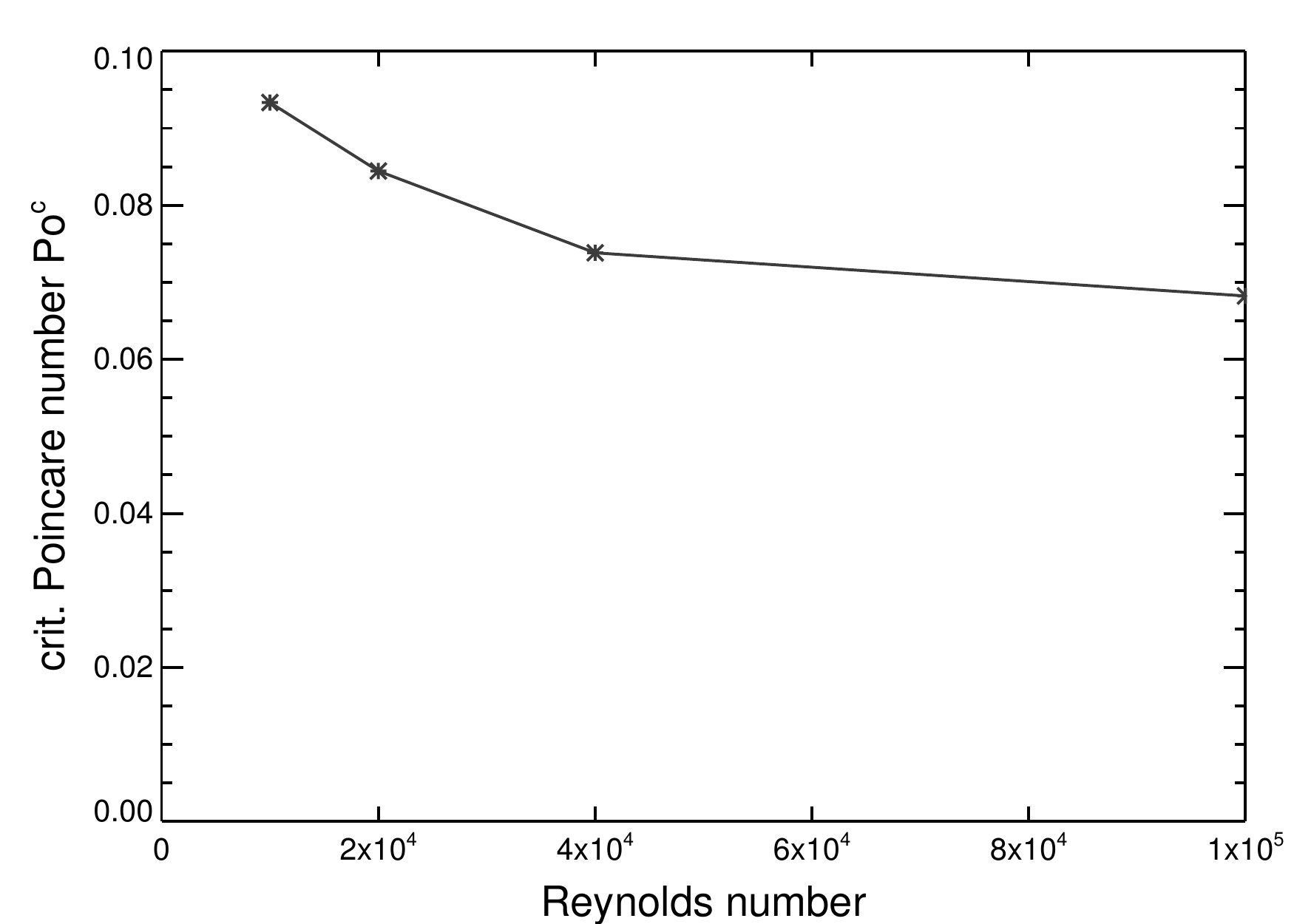}}
\caption{\label{fig::axisymprof} %
(a) Axial profile of the axisymmetric part of $u_z$ for various
${\rm{Re}}$ scaled by the respective maximum.  (b) Critical value for
the precession ratio ${\rm{Po}}^{\rm{c}}$ at which the axisymmetric
mode $(m,k)=(0,2)$ emerges.}%
\end{center}
\end{figure}

Whereas numerical simulations are limited to ${\rm{Re}} \lesssim
10^4$, UDV measurements in the water experiment are possible up to
${\rm{Re}}=10^5$.  The results show that the essential characteristics
of the flow behavior in dependence of the precession ratio hardly
change.  This can be seen in figure~\ref{fig::pattern} for the total
flow, and for the axisymmetric flow in figure~\ref{fig::axisymprof}(a)
which shows the axial profile of the time-average of $u_z$ versus $z$
measured with UDV at $r=0.9 R$.  The flow profiles are scaled by the
respective maximum value so that all curves roughly collapse,
indicating that only minor changes of the flow structure appear when
increasing ${\rm{Re}}$.

An exception is the critical value for the precession ratio
${\rm{Po}}^{\rm{c}}$ at which the axisymmetric mode emerges and the
amplitude of the directly forced flow drops (left plot in
figure~\ref{fig::axisymprof}). ${\rm{Po}}^{\rm{c}}$ decreases with
increasing ${\rm{Re}}$, however, the reduction of ${\rm{Po}}^{\rm{c}}$
mainly takes place for ${\rm{Re}} \lesssim 4\times 10^4$ (from
${\rm{Po}}^{\rm{c}}\approx0.100$ at ${\rm{Re}}=10^4$ to
${\rm{Po}}^{\rm{c}}\approx0.075$ at ${\rm{Re}}=4\times 10^4$). Further
increasing ${\rm{Re}}$, the measurements indicate an asymptotic
behavior with a saturation around ${\rm{Po}}^{\rm{c}}\approx
0.07$. This is supported by measurements of the pressure fluctuations
at the end caps presented in \cite{herault2015}.  These pressure
measurements provide the precession ratio at which the amplitude of
the $m=1$ mode drops abruptly but give only little information on the
very flow structure or the contributing modes.  The pressure data is
available up to ${\rm{Re}}\approx 2\times 10^6$ and shows that nearly
no further change of ${\rm{Po}}^{\rm{c}}$ occurs up to this value of
the Reynolds number.  However, these measurements can only be regarded
as a hint for the behavior of ${\rm{Po}}^{\rm{c}}$, and further
experiments in the intermediate regime -- in particular with better
information on the flow structure -- are required.

%%%%%%%%%%%%%%%%%%%%%%%%%%%%%%%%%%%%%%%%%%%%%%%%%%%%%%%%%%%%%%%%%%%%%%%%%%%%%%%%%%%%

\section{The kinematic dynamo problem}

\subsection{Growth rates}

In order to deduce whether a precessing conducting liquid is capable
to sustain a magnetic field we run simulations of the magnetic
induction equation which describes the temporal evolution of the
magnetic field $\vec{B}$ and reads
\begin{equation}
\frac{\upartial\vec{B}}{\upartial t}\,
=\,{\bm \nabla}\times\bigl(\vec{u}\times\vec{B}
\,-\,\eta{\bm \nabla}\times\vec{B}\bigr).
\label{eq::magind}
\end{equation}
Since we have seen in the previous sections that the dominant
contribution of the fluid flow is essentially stationary in the
turntable reference frame, we revert to the kinematic approach and
apply the time-averaged velocity field obtained from our hydrodynamic
simulations.  An advanced model would include turbulent fluctuations
on top of the stationary flow. It is well known that such
contributions may inhibit dynamo action
\citep{petrelis2006,petrelis2007} because these fluctuations cause an
decrease of the effective electrical conductivity
($\beta$-effect). However, experiments at the Riga dynamo facility
where a turbulence level of roughly $7\%$ on top of the flow was
observed, showed that the restriction to a simple laminar flow model
is sufficient to describe the onset of dynamo action in this
experimental system \citep{gailitis2000,gailitis2004,gailitis2008}. We
thus suppose that temporal fluctuations of the flow field only have a
minor impact for the onset of dynamo action, at least in the regime
that is relevant for the present study.

\begin{table}[b!]
\begin{center}
\begin{tabular}{c||c|c|c|c|c|c|c}
${\rm{Po}}$ & $0.05$ & $0.075$ & $0.0875$ & $\boldsymbol{0.095}$ &
$\boldsymbol{0.100}$ & $\boldsymbol{0.105}$ & $0.125$\\ 
\hline
${\rm{Rm}}^{\rm{c}}$ & $> 5000$ & $> 5000$ & $\sim 4900$ & $\boldsymbol{1045}$ &
$\boldsymbol{430}$ & $\boldsymbol{530}$ & $1930$\\
\end{tabular}
\vspace*{0.3cm}
\caption{\label{tab::rmc}Critical magnetic Reynolds number for various
precession ratios. For ${\rm{Po}} =0.05$ and ${\rm{Po}}=0.075$ we did
not find dynamos up to ${\rm{Rm}}=5000$ and the development of the
growthrates suggest that these flows do not show dynamo action at
all. The bold numbers denote the cases with ${\rm{Rm}}^{\rm{c}}$ in
the range of experimentally achievable values.}
\end{center}
\end{table}
We utilize different time-averaged velocity fields obtained from
hydrodynamic simulations at ${\rm{Re}}=10^4$ and various ${\rm{Po}}$.
The magnetic induction equation~(\ref{eq::magind}) is solved
numerically using a finite volume scheme with constraint transport in
order to ensure ${\bm \nabla}{\bm \cdot}\vec{B}=0$ \citep{giesecke2008}.  For all
simulations we apply pseudo-vacuum boundary conditions for the
magnetic field, i.e. at the boundary the tangential field components
vanish.  This behavior mimics an exterior domain with infinite
permeability and it is known from numerical simulations for the
von-K{\'a}rm{\'a}n-Sodium dynamo that these boundary conditions may
have a beneficial impact for dynamo action \citep[see also
section~\ref{sec::4p2} below]{giesecke2010b}.

\begin{figure}[t!]
\subfigure[]
{\includegraphics[width=0.5\textwidth]{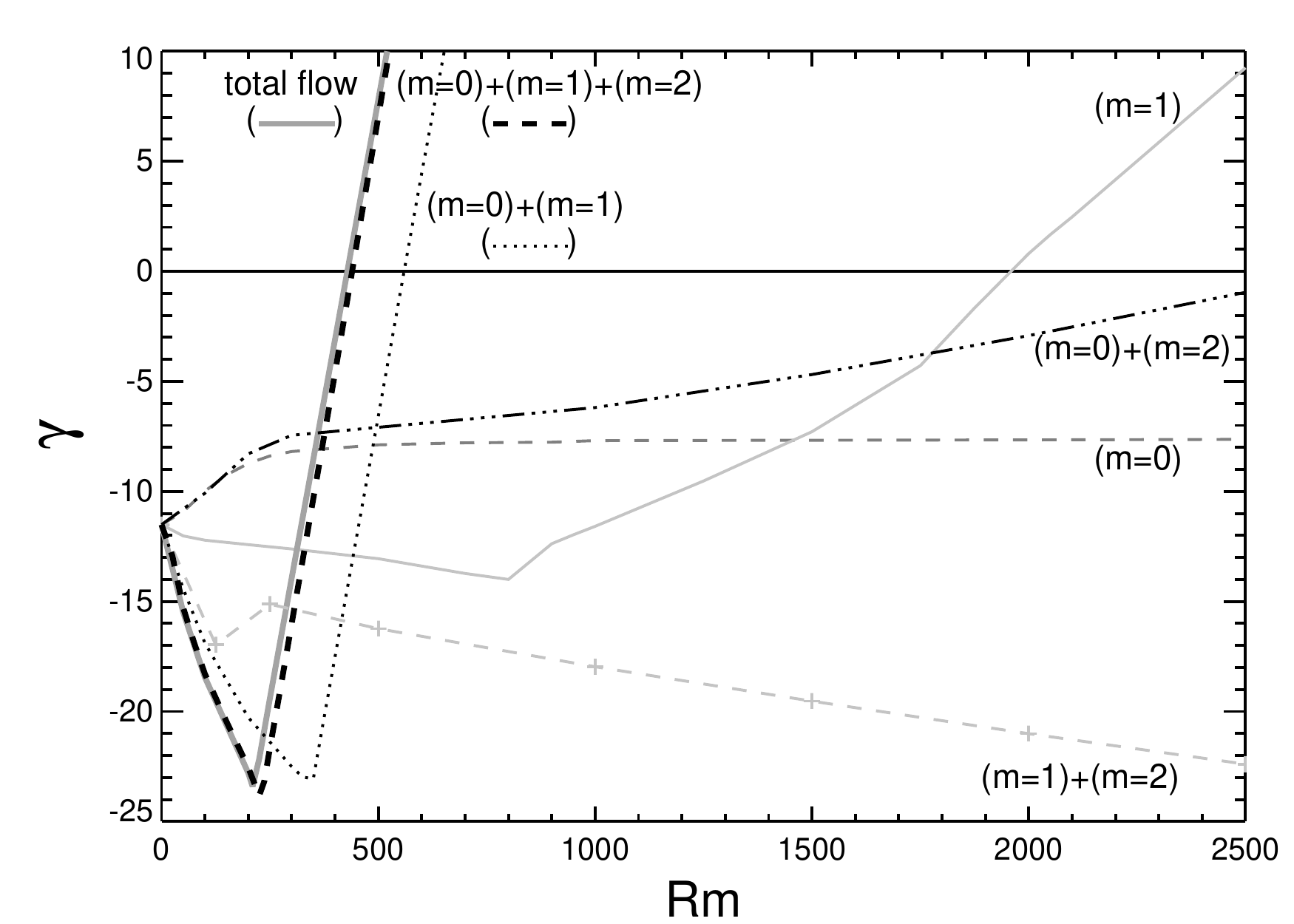}}
\subfigure[]
{\includegraphics[width=0.5\textwidth]{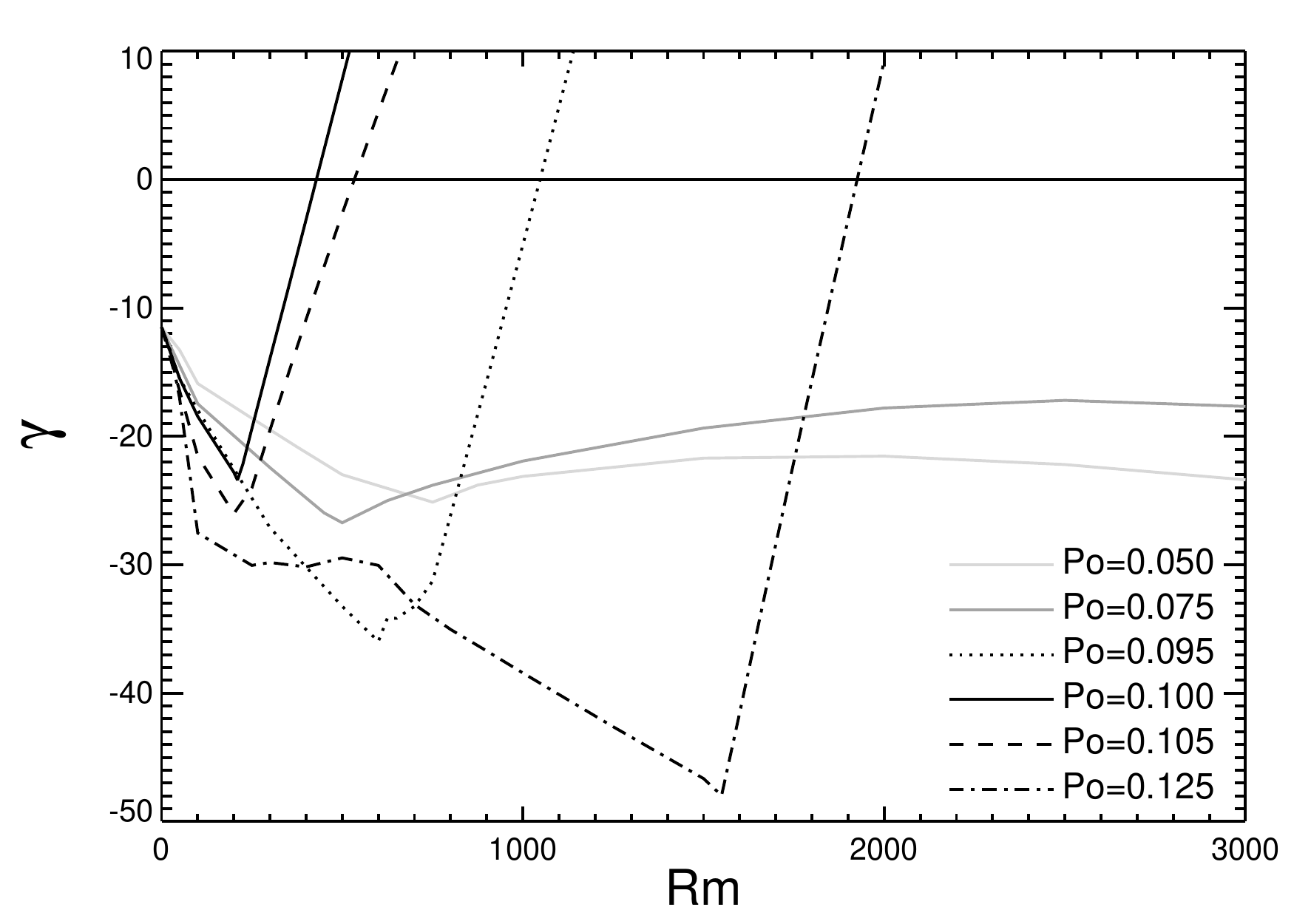}}
\caption{\label{fig::gr} %
(a) Magnetic field growth rates versus ${\rm{Rm}}$ induced by various
combinations of individual azimuthal flow modes of the hydrodynamic
flow obtained at ${\rm{Re}}=10^4$ and ${\rm{Po}}=0.1$. (b) Magnetic
field growth rates versus ${\rm{Rm}}$ induced by the time-averaged
total flow obtained at ${\rm{Re}}=10^4$ and various precession ratios
${\rm{Po}}$.
}
\end{figure}

For the particular case ${\rm{Re}}=10^4$ and ${\rm{Po}}=0.1$ we
decompose the flow field into different azimuthal modes with $m=0,
m=1, m=2$ and compute solutions of multiple combinations of these
individual contributions.  Figure~\ref{fig::gr}(a) shows that the
axisymmetric flow alone ($m=0$) is not able to drive a dynamo, and
although the cases $(m=1)$ and $(m=0)+(m=2)$ show a positive growth
rate -- and thus provide a dynamo -- this only happens at relatively
large magnetic Reynolds numbers ${\rm{Rm}}=\varOmega_{\rm{c}}R^2/\eta$,
which will not be achievable in the experiment.  A critical magnetic
Reynolds numbers sufficiently low for the planned experiment is only
obtained when considering at least the directly forced flow and the
axisymmetric flow, i.e., $(m=0)+(m=1)$ (dotted curve in
figure~\ref{fig::gr}(a)).

\begin{figure}[b!]
\begin{center}
\subfigure[$t=0.000$]{
\resizebox*{3.5cm}{!}%
{\includegraphics{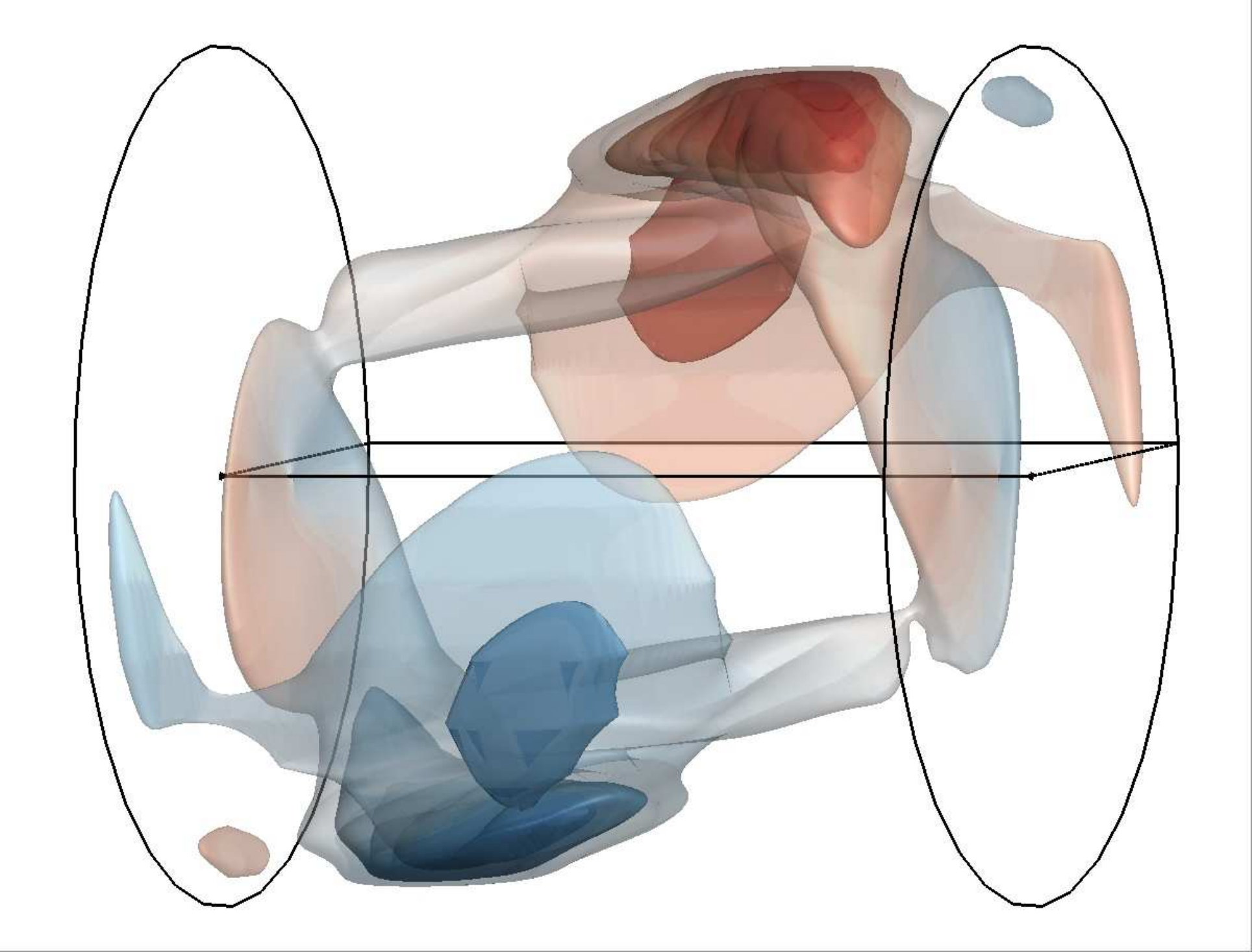}}}%
\subfigure[$t=0.2198$]{
\resizebox*{3.5cm}{!}%
{\includegraphics{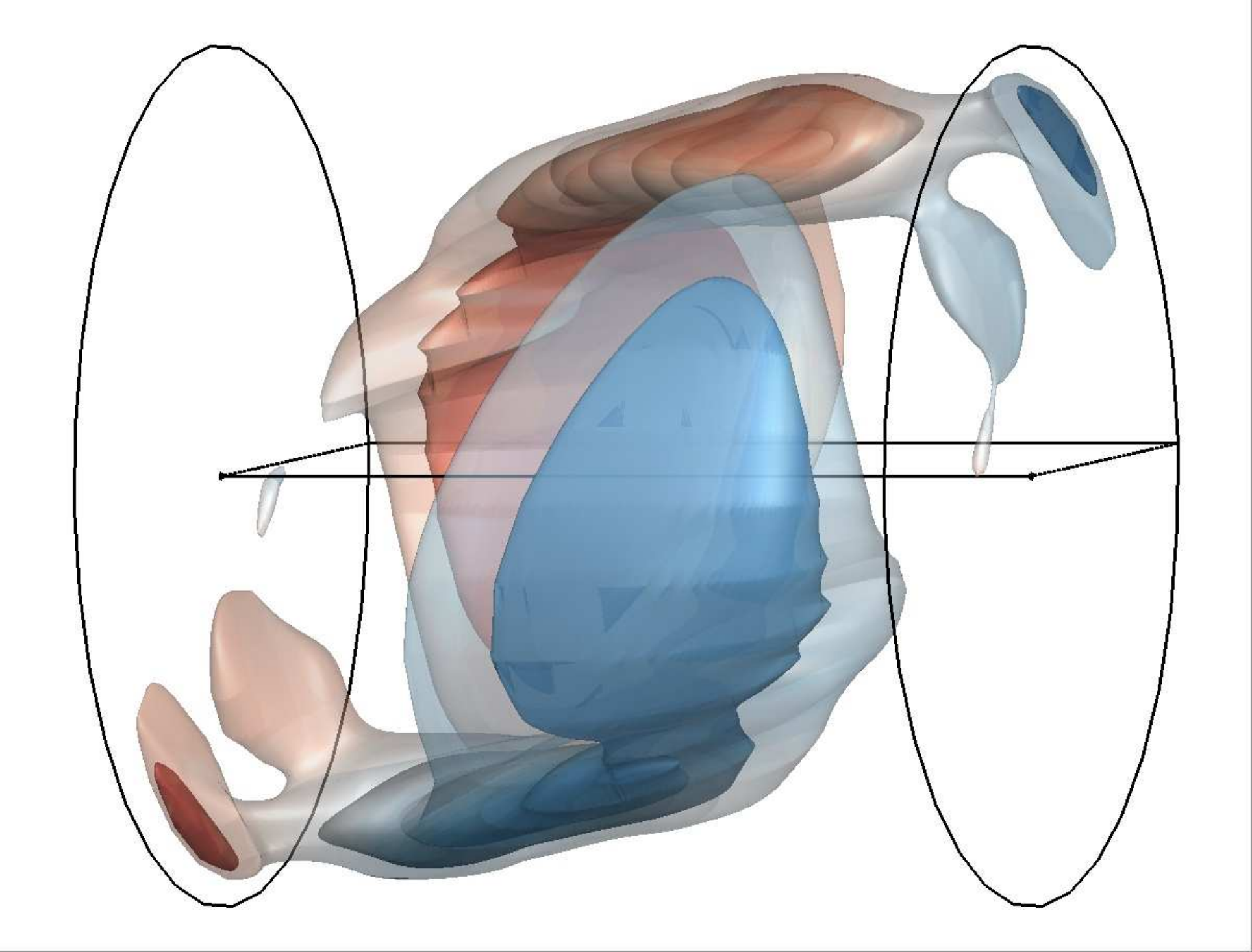}}}%
\subfigure[$t=0.4396$]{
\resizebox*{3.5cm}{!}%
{\includegraphics{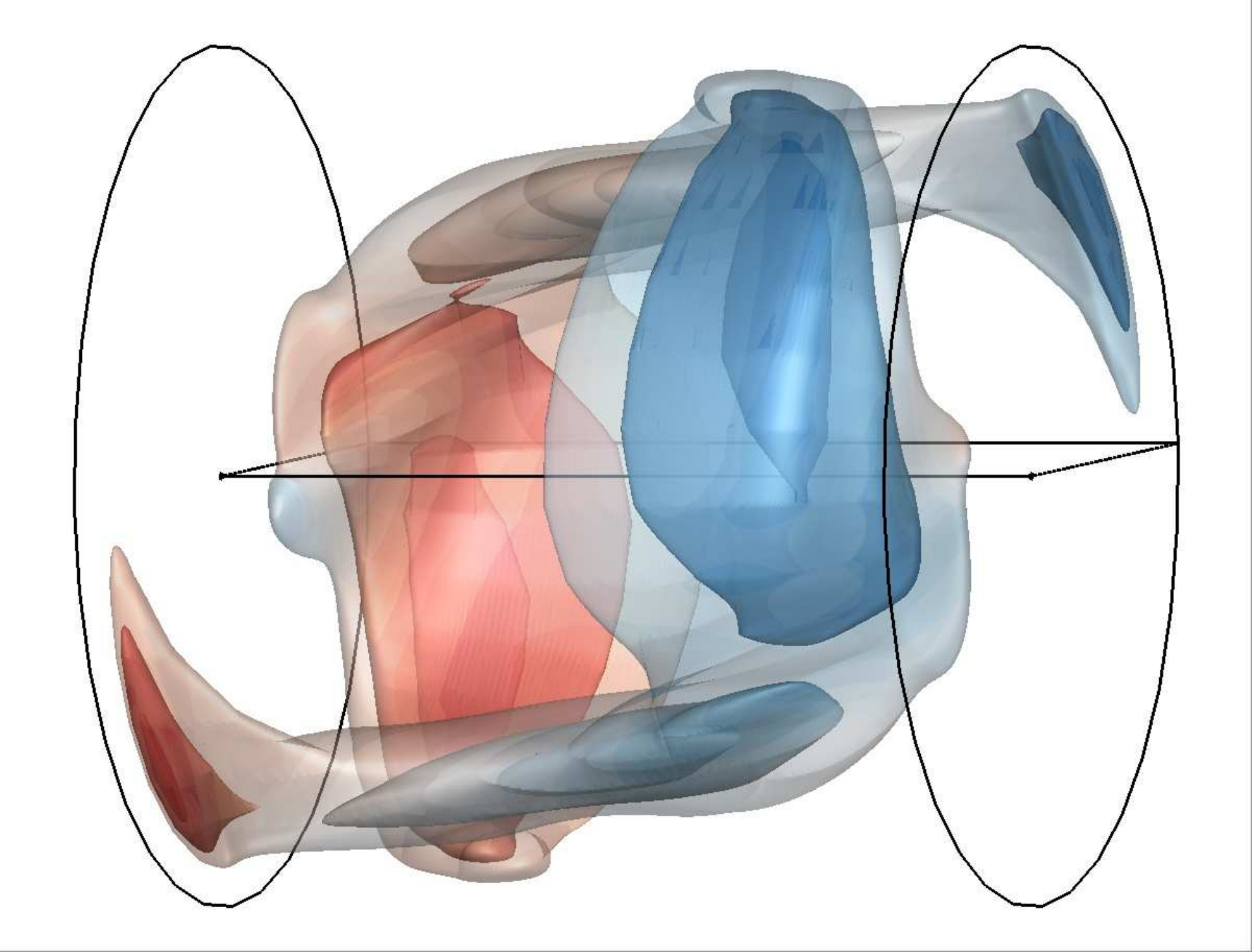}}}%
\subfigure[$t=0.6594$]{
\resizebox*{3.5cm}{!}%
{\includegraphics{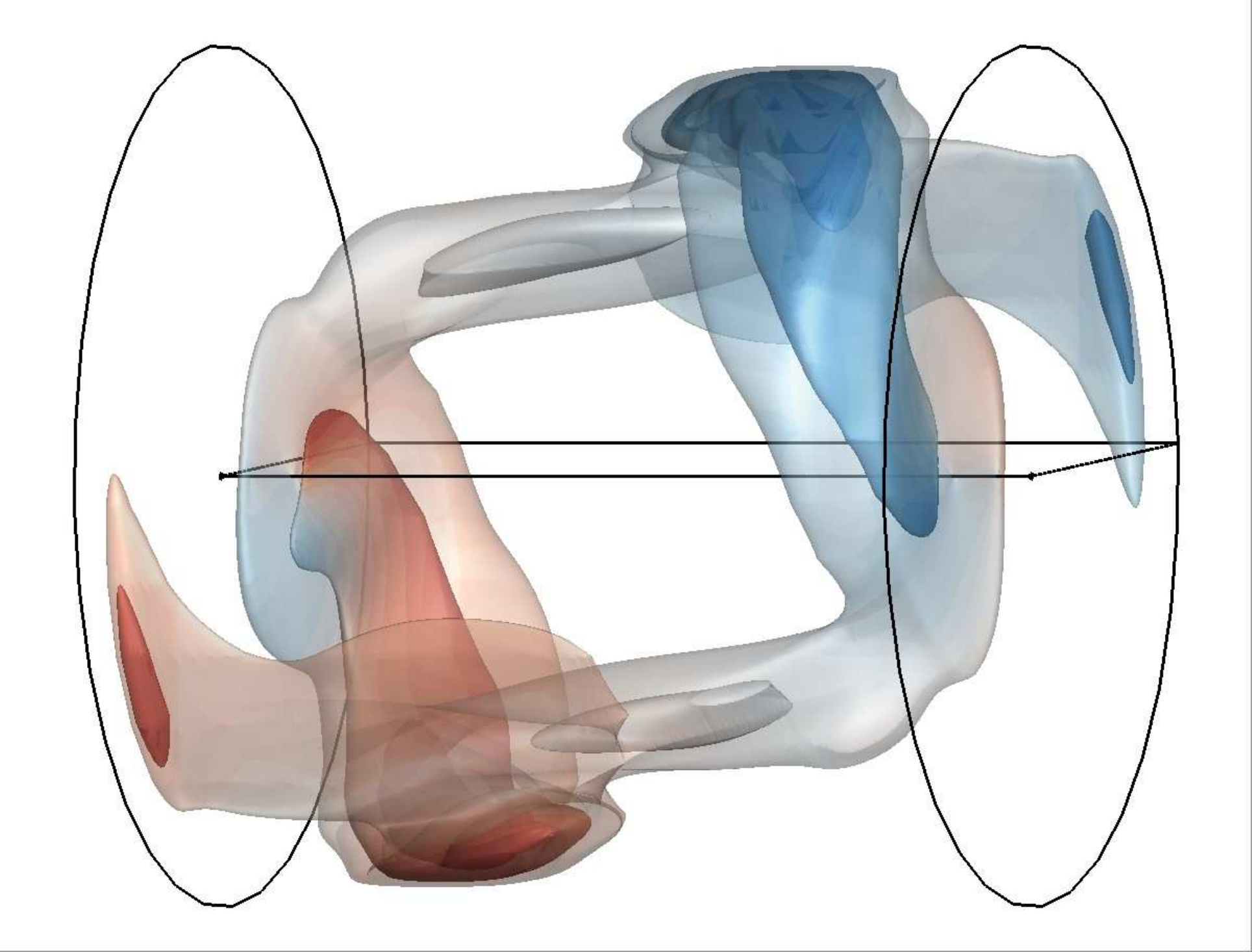}}}%
\\
\subfigure[$t=0.8792$]{
\resizebox*{3.5cm}{!}%
{\includegraphics{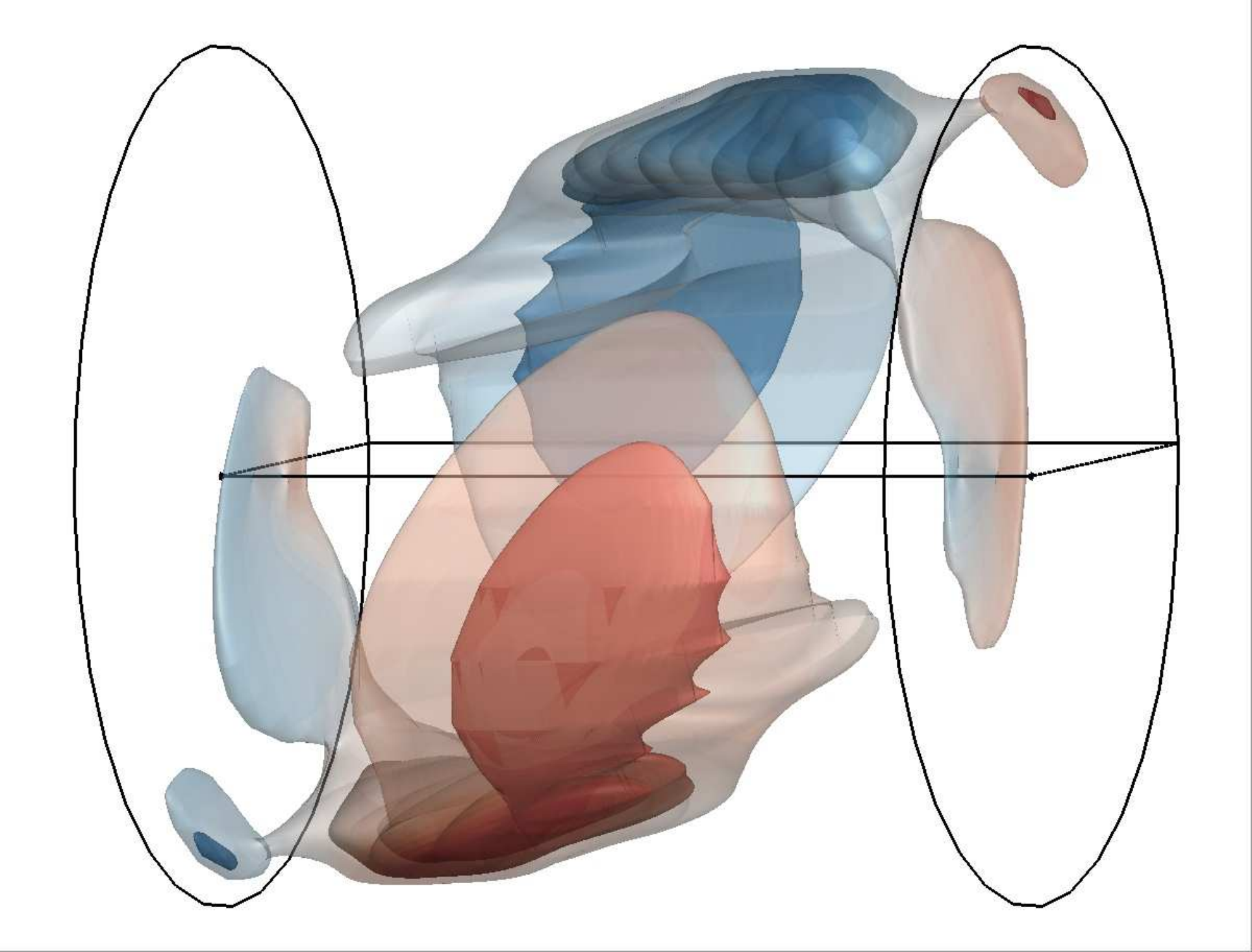}}}%
\subfigure[$t=1.0990$]{
\resizebox*{3.5cm}{!}%
{\includegraphics{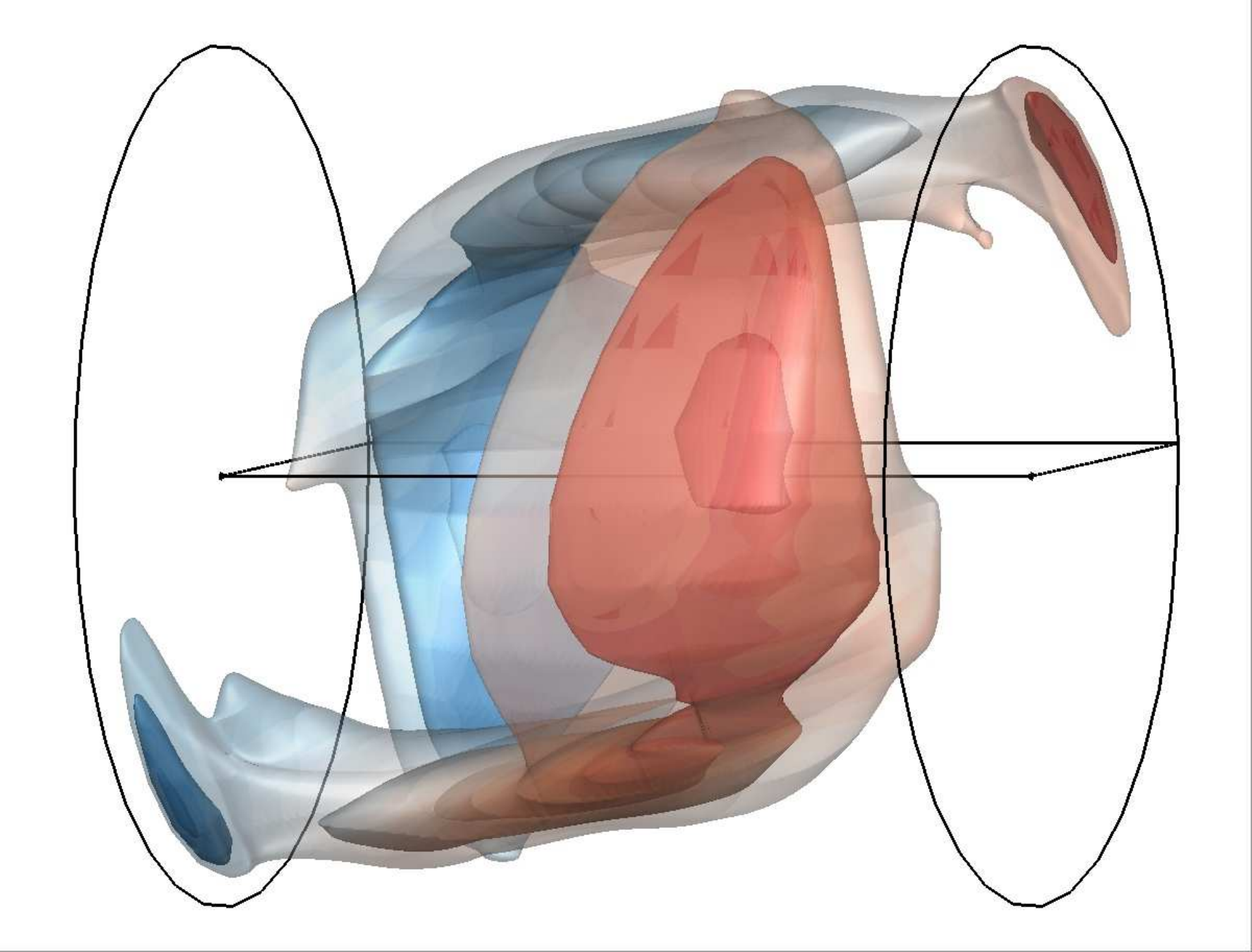}}}%
\subfigure[$t=1.3187$]{
\resizebox*{3.5cm}{!}%
{\includegraphics{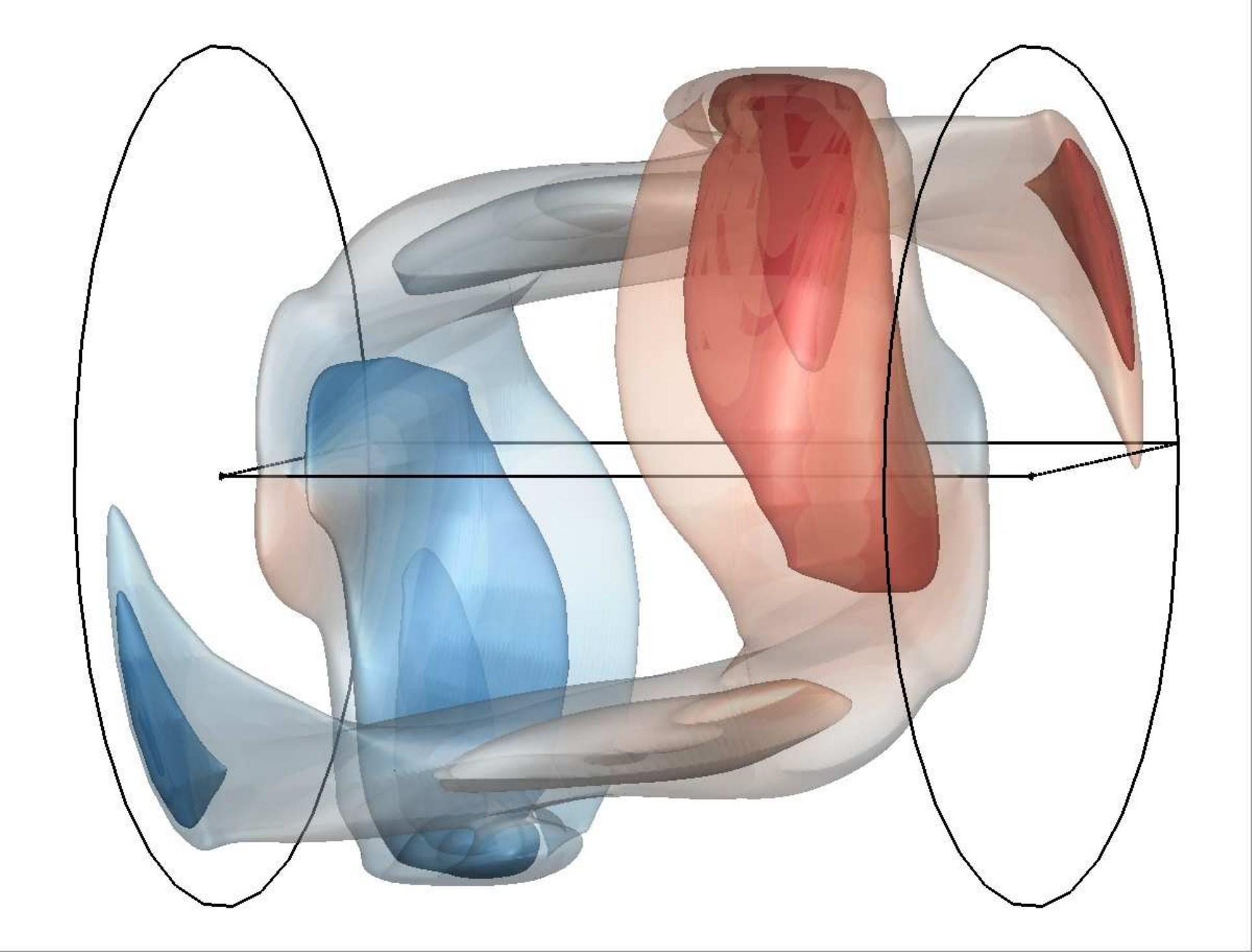}}}%
\subfigure[$t=1.5385$]{
\resizebox*{3.5cm}{!}%
{\includegraphics{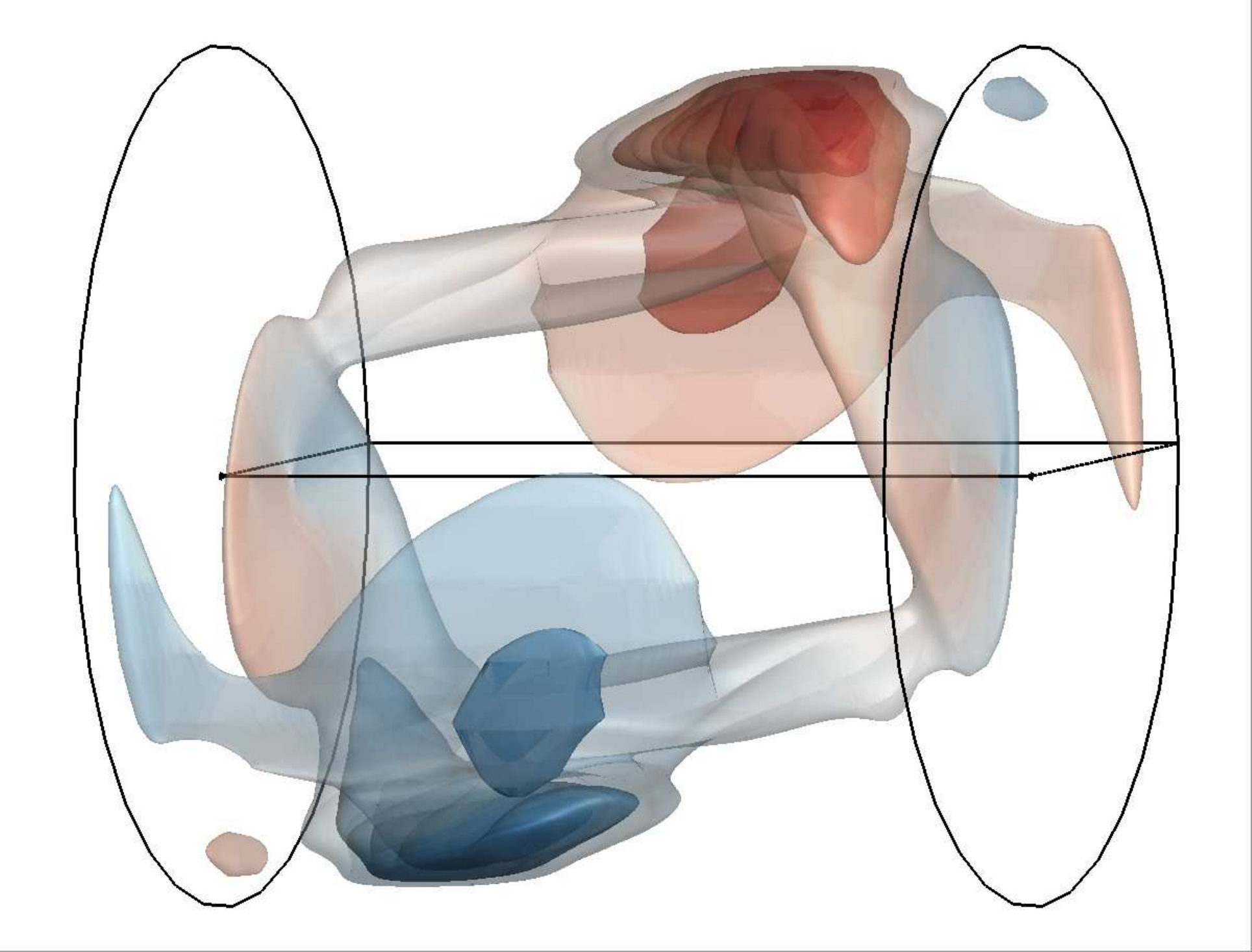}}}%
\caption{\label{fig::flowstruc} %
Magnetic energy density at $10,25,50 \%$ of the maximum value mapped
with the azimuthal field $B_{\rm{\varphi}}$. ${\rm{Rm}}=450$.  The
time-averaged flow field results from hydrodynamic simulations at
${\rm{Re}}=10^4$ and ${\rm{Po}}=0.1$. The time is denoted in units of
the rotation time (colour online).
}
\end{center}
\end{figure}
We conclude that the best way to drive the dynamo is attained from the
largest scales of the velocity field and there is nearly no difference
in the growth rates between dynamos driven by a velocity field
consisting only of the $m=0, m=1$ and $m=2$ mode (dashed black curve)
and dynamos that are driven by the full (time-averaged) velocity field
(solid grey curve).

Finally, we briefly discuss the results obtained with the total flow
field obtained from the hydrodynamic simulations at different
precession ratios ${\rm{Po}}$ with the Reynolds number fixed at
${\rm{Re}}=10^4$. Figure~\ref{fig::gr}(b) shows the magnetic energy
growth rates for the cases ${\rm{Po}}=0.05, 0.075, 0.095, 0.100,
0.105$, and ${\rm{Po}}=0.125$.  We find dynamos only for flows
obtained for ${\rm{Po}}>0.075$ (see table~\ref{tab::rmc}).  However,
the critical value for ${\rm{Rm}}^{\rm{c}}$ is only within the regime
that will be achievable in the dynamo experiment when using velocity
fields obtained around ${\rm{Po}}\approx 0.100$, i.e. when the
axisymmetric double roll structure is present (bold numbers in
table~\ref{tab::rmc}).  The minimum value for ${\rm{Rm}}^{\rm{c}}$ is
found with the velocity field obtained at ${\rm{Re}}=10^4$ and
${\rm{Po}}=0.100$ where the relation of the axisymmetric mode
$(m,k)=(0,2)$ to the directly forced mode $(m,k)=(1,1)$ has its
maximum value.  A common feature for all cases is the strong initial
decrease of the growth rates when increasing ${\rm{Re}}$.  This
behavior changes abruptly at a certain value of ${\rm{Rm}}$ at which
the growth rates begin to increase with a rather steep slope. The
sudden change goes along with a change in the temporal behavior of the
magnetic field. Whereas the solution is stationary in the regime with
decreasing growth rate, behind the sudden bent the magnetic field has
an oscillatory contribution.  However, the main part of the magnetic
field is constant in time with the eigenmode performing a rotation
around the symmetry axis of the cylinder
(figure~\ref{fig::flowstruc}).

\subsection{Magnetic boundary conditions\label{sec::4p2}}

In the following, we will concentrate on the behavior of the magnetic
field for the case ${\rm{Po}}=0.100$ for which we obtain the smallest
${\rm{Rm}}^{\rm{c}}$. An important issue that determines
self-excitation of magnetic fields is the impact of the electrical
boundary conditions for which pseudo vacuum conditions (i.e. vanishing
tangential fields) have been utilized.  In the planned experiment the
container will be made of stainless steel with a magnetic permeability
very close to $\mu_{\rm{r}}=1$ but with an magnetic diffusivity that
at a typical operating temperature of $T \approx 120^{\circ}\mbox{C}$
is roughly a factor six larger than the diffusivity of liquid sodium.
In that case the solution for the magnetic field -- in particular its
growth rate -- is affected by the material properties of the
container, i.e., its magnetic diffusivity $\eta_{\rm{wall}}$ and its
thickness $d_{\rm{wall}}$, whereas the electric properties of the very
outer regime remain less important.  We extend our previous dynamo
models by including an outer shell with thickness
$d_{\rm{wall}}=5\mbox{ cm}$. The real value in the planned dynamo
experiment will be $d_{\rm{wall}}=3\mbox{ cm}$, but for numerical
reasons we seek to consider a sufficient number of grid cells that
represent the container wall. Within the wall the ``flow'' is
prescribed by a solid body rotation, and the magnetic diffusivity
$\eta_{\rm{wall}}$ may differ from the diffusivity in the fluid
domain.
\begin{figure}[b!]
\begin{center}
\includegraphics[width=0.55\textwidth]{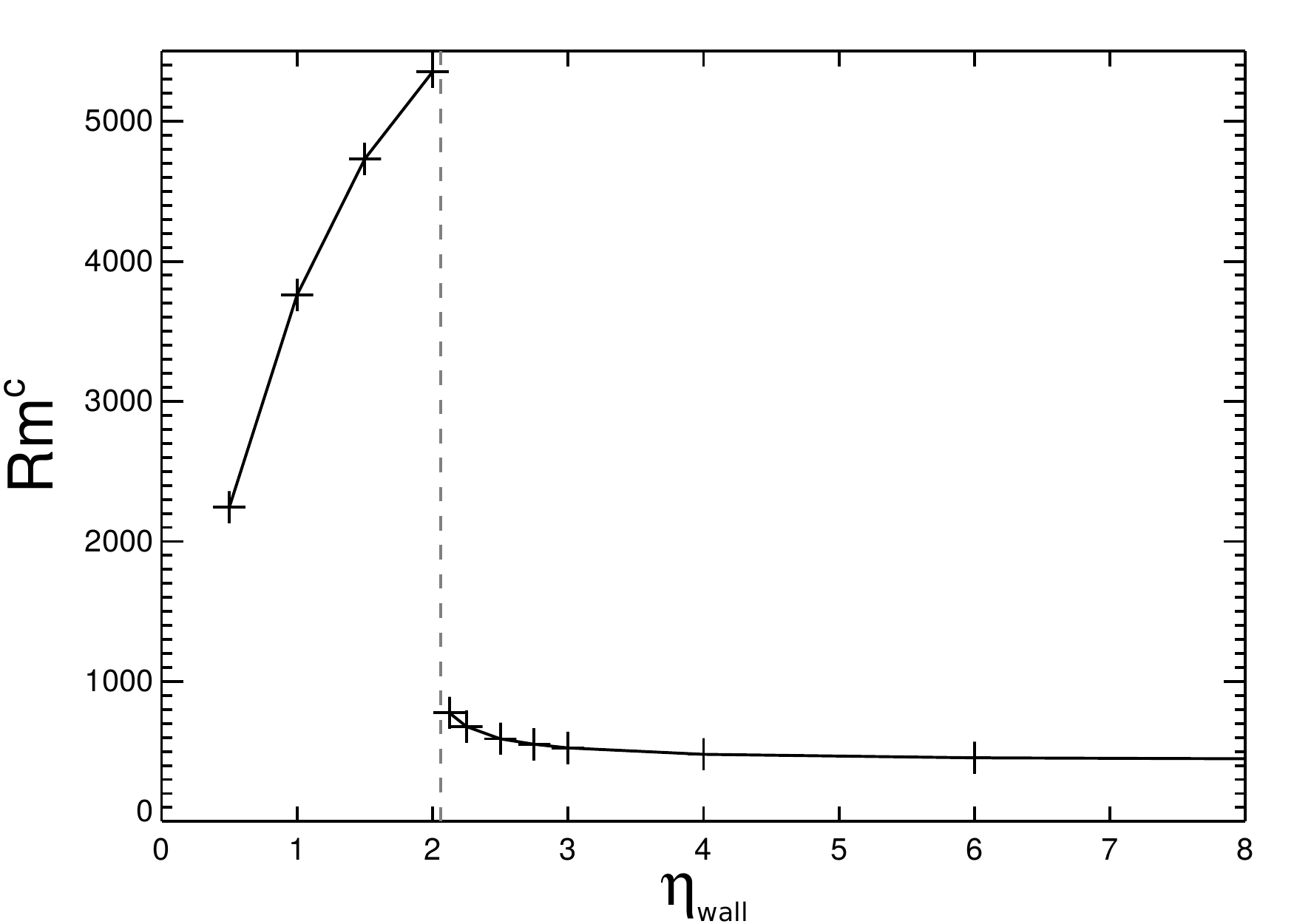}
\caption{\label{fig::rmcrit} %
Critical magnetic Reynolds number versus the diffusivity of the outer
layer $\eta_{\rm{wall}}$ with thickness $d_{\rm{wall}}=5\mbox{
cm}$. The flow field is the time-averaged field obtained in
hydrodynamic simulations at ${\rm{Re}}=10^4$ and ${\rm{Po}}=0.1$.}
\end{center}
\end{figure}
We find two different dynamo states that are distinguished by the
container diffusivity (figure~\ref{fig::rmcrit}). For
$\eta_{\rm{wall}}/\eta \lesssim 2$ the critical magnetic Reynolds
number is large and remains far above values that will be achievable
in the planned experiment. Above $\eta_{\rm{wall}}$, however,
${\rm{Rm}}^{\rm{c}}$ drops significantly, and remains nearly
independent of $\eta_{\rm{wall}}$ with ${\rm{Rm}}^{\rm{c}}\approx 450$
for the realistic case $\eta_{\rm{wall}}=6\eta$. It is surprising that
the structure of the magnetic field in both regimes is quite similar,
and that without an outer shell the simplifying pseudo-vacuum boundary
conditions emulate the impact of container walls with increased
diffusivity.

Less unusual is the sensitivity of the solution to small variations in
the diffusivity of this outer shell, which is a typical property of
boundary eigenvalue problems of simple dynamo models
\citep{kirillov2009}. In connection with the abrupt and sharp change
of the growth rates and the transition from time-independent to
oscillating solutions, this behavior suggests the presence of
exceptional points. These points represent distinguished locations in
the spectrum at which two degenerated solutions occur with equal
eigenvalues and identical eigenvectors and are important ingredients
for example in models of reversals of the geomagnetic field.

\section{Conclusions}

Our kinematic dynamo models indicate that magnetic self-excitation by
a precession driven flow in a cylindrical container may be possible in
a narrow parameter regime that indeed will be achievable in the
planned precession dynamo experiment at HZDR. This regime is
essentially characterized by the presence of an axisymmetric double
roll structure in the meridional plane.  For small ${\rm{Re}}$ the
critical precession ratio ${\rm{Po}}^{\rm{c}}$, at which the
axisymmetric flow mode emerges, decreases with increasing ${\rm{Re}}$.
However, the UDV measurements at larger ${\rm{Re}}$ as well as the
behavior of the pressure and the hysteresis deduced from the power
consumption measured by \cite{herault2015} suggest that for
sufficiently large ${\rm{Re}}$ the critical precession ratio becomes
largely independent of ${\rm{Re}}$ with values around
${\rm{Po}}^{\rm{c}} \approx 0.07$ when ${\rm{Re}}\gtrsim 10^6$.

Nevertheless, it remains quite ambitious to scale the results obtained
at comparable small Reynolds number ${\rm{Re}}\sim 10^4$ to the large
device, where, in order to reach the predicted critical values
${\rm{Rm}}^{\rm{c}}\approx 400\dots 500$, the fluid flow will
correspond to Reynolds numbers of the order ${\mathrm O}(10^7)$.  Our
investigations indicate that for ${\rm{Po}} \lesssim
{\rm{Po}}^{\rm{c}}$, i.e. before the transition into the turbulent
regime, the flow is essentially quasi-laminar and does hardly change
its structure when increasing ${\rm{Re}}$. In particular the
axisymmetric mode, which emerges immediately before the transition to
the turbulent regime and which is a crucial ingredient for dynamo
action at low ${\rm{Rm}}$, does not vanish in the fast rotating case
(i.e. large ${\rm{Re}}$) so that there we have some basic confidence
in the transferability of our results to the large dynamo experiment.
The minimum value of the critical magnetic Reynolds number
${\rm{Rm}}^{\rm{c}}$ for the onset of dynamo action found in our
dynamo models is ${\rm{Rm}}^{\rm{c}}\approx 430$. This value is
obtained from kinematic dynamo simulations using the time-averaged
velocity field computed in hydrodynamical simulations at
${\rm{Re}}=10^4$ and ${\rm{Po}}=0.100$.  Assuming that the flow
structure -- and therefore the dynamo capability -- remains the same
even for the large ${\rm{Re}}$ that will occur in the real experiment
and that the critical value for the precession ratio
${\rm{Po}}^{\rm{c}}$ for the emergence of the axisymmetric mode will
be around ${\rm{Po}}^{\rm{c}}\approx 0.07$ (as suggested by the
behavior shown in figure~\ref{fig::axisymprof}(b)), we end up at a
corresponding rotation frequency of $f_{\rm{c}} \approx 6.8\mbox{ Hz}$
and a precession frequency of $f_{\rm{p}}\approx 0.5\mbox{ Hz}$ that
will be required for a dynamo in the planned experiment (assuming
$\eta=0.1\mbox{ m}^2/\mbox{s}$).  In order to confirm this values,
further studies are necessary that focus on the impact of
electromagnetic boundary conditions and the influence of
time-dependent flow contributions.

we may explicitly estimate the parameters that will be required in
order to obtain magnetic field self-excitation in the liquid sodium
experiment.  This scaling would lead to a rotation frequency of
$f_{\rm{c}} \approx 6.8\mbox{ Hz}$ and a precession frequency of
$f_{\rm{p}}\approx 0.5\mbox{ Hz}$

\bigskip

This study has been conducted in the framework of the project DRESDYN
(DREsden Sodium facility for DYNamo and thermohydraulic studies),
which provides the platform for the precession dynamo experiment at
HZDR. The authors further acknowledge support by the Helmholtz Allianz
LIMTECH and thank Bernd Wustmann for the mechanical design of the
water experiment.

\bibliographystyle{gGAF}
%\bibliography{giesecke_kinematic_precession_dynamo}
\bibliography{GGAF-2018-0018-Giesecke}

\begin{thebibliography}{61}
\providecommand{\natexlab}[1]{#1}

\bibitem[\protect\citeauthoryear{Albrecht
  {\itshape{et~al.}}}{2018}]{albrecht2018}
Albrecht, T., Blackburn, H.M., Lopez, J.M., Manasseh, R. and Meunier, P., On
  triadic resonance as an instability mechanism in precessing cylinder flow.
  {\itshape \jfm}, 2018, \textbf{841}, R3.

\bibitem[\protect\citeauthoryear{{Blackburn} and
  {Sherwin}}{2004}]{blackburn2004}
{Blackburn}, H.M. and {Sherwin}, S.J., {Formulation of a Galerkin spectral
  element-Fourier method for three-dimensional incompressible flows in
  cylindrical geometries}. {\itshape J. Comp. Phys.}, 2004, \textbf{197},
  759--778.

\bibitem[\protect\citeauthoryear{{Busse}}{1968}]{busse1968}
{Busse}, F.H., {Steady fluid flow in a precessing spheroidal shell}. {\itshape
  \jfm}, 1968, \textbf{33}, 739--751.

\bibitem[\protect\citeauthoryear{Ernst-Hullermann
  {\itshape{et~al.}}}{2011}]{hullermann2011}
Ernst-Hullermann, J., Harder, H. and Hansen, U., Finite volume simulations of
  dynamos in ellipsoidal planets. {\itshape Geophys. J. Int.}, 2011,
  \textbf{195}, 1395--1405.

\bibitem[\protect\citeauthoryear{{Gailitis}
  {\itshape{et~al.}}}{2008}]{gailitis2008}
{Gailitis}, A., {Gerbeth}, G., {Gundrum}, T., {Lielausis}, O., {Platacis}, E.
  and {Stefani}, F., {History and results of the Riga dynamo experiments}.
  {\itshape C. R. Phys.}, 2008, \textbf{9}, 721--728.

\bibitem[\protect\citeauthoryear{{Gailitis}
  {\itshape{et~al.}}}{2000}]{gailitis2000}
{Gailitis}, A., {Lielausis}, O., {Dement'ev}, S., {Platacis}, E., {Cifersons},
  A., {Gerbeth}, G., {Gundrum}, T., {Stefani}, F., {Christen}, M., {H{\"a}nel},
  H. and {Will}, G., {Detection of a Flow Induced Magnetic Field Eigenmode in
  the Riga Dynamo Facility}. {\itshape \prl}, 2000, \textbf{84}, 4365--4368.

\bibitem[\protect\citeauthoryear{{Gailitis}
  {\itshape{et~al.}}}{2004}]{gailitis2004}
{Gailitis}, A., {Lielausis}, O., {Platacis}, E., {Gerbeth}, G. and {Stefani},
  F., {Riga dynamo experiment and its theoretical background}. {\itshape \pop},
  2004, \textbf{11}, 2838--2843.

\bibitem[\protect\citeauthoryear{{Gans}}{1970}]{gans1970}
{Gans}, R.F., {On the precession of a resonant cylinder}. {\itshape \jfm},
  1970, \textbf{41}, 865--872.

\bibitem[\protect\citeauthoryear{{Gans}}{1971}]{gans1971}
{Gans}, R.F., {On hydromagnetic precession in a cylinder}. {\itshape \jfm},
  1971, \textbf{45}, 111--130.

\bibitem[\protect\citeauthoryear{{Giesecke}
  {\itshape{et~al.}}}{2015}]{giesecke2015b}
{Giesecke}, A., {Albrecht}, T., {Gundrum}, T., {Herault}, J. and {Stefani}, F.,
  {Triadic resonances in nonlinear simulations of a fluid flow in a precessing
  cylinder}. {\itshape \njp}, 2015, \textbf{17}, 113044.

\bibitem[\protect\citeauthoryear{{Giesecke}
  {\itshape{et~al.}}}{2010}]{giesecke2010b}
{Giesecke}, A., {Nore}, C., {Stefani}, F., {Gerbeth}, G., {Leorat}, J.,
  {Luddens}, F. and {Guermond}, J.L., {Electromagnetic induction in non-uniform
  domains}. {\itshape \gafd}, 2010, \textbf{104}, 505--529.

\bibitem[\protect\citeauthoryear{{Giesecke}
  {\itshape{et~al.}}}{2008}]{giesecke2008}
{Giesecke}, A., {Stefani}, F. and {Gerbeth}, G., {Kinematic simulation of
  dynamo action by a hybrid boundary-element/finite-volume method}. {\itshape
  Magnetohydrodynamics}, 2008, \textbf{44}, 237--252.

\bibitem[\protect\citeauthoryear{{Giesecke}
  {\itshape{et~al.}}}{2018}]{giesecke2018}
{Giesecke}, A., {Vogt}, T., {Gundrum}, T. and {Stefani}, F., {Nonlinear Large
  Scale Flow in a Precessing Cylinder and Its Ability To Drive Dynamo Action}.
  {\itshape Phys. Rev. Lett.}, 2018, \textbf{120}, 024502.

\bibitem[\protect\citeauthoryear{{Goepfert} and {Tilgner}}{2016}]{goepfert2016}
{Goepfert}, O. and {Tilgner}, A., {Dynamos in precessing cubes}. {\itshape
  \njp}, 2016, \textbf{18}, 103019.

\bibitem[\protect\citeauthoryear{{Goto} {\itshape{et~al.}}}{2007}]{goto2007}
{Goto}, S., {Ishii}, N., {Kida}, S. and {Nishioka}, M., {Turbulence generator
  using a precessing sphere}. {\itshape \pof}, 2007, \textbf{19},
  061705--061705.

\bibitem[\protect\citeauthoryear{{Goto} {\itshape{et~al.}}}{2014}]{goto2014}
{Goto}, S., {Matsunaga}, A., {Fujiwara}, M., {Nishioka}, M., {Kida}, S.,
  {Yamato}, M. and {Tsuda}, S., {Turbulence driven by precession in spherical
  and slightly elongated spheroidal cavities}. {\itshape \pof}, 2014,
  \textbf{26}, 055107.

\bibitem[\protect\citeauthoryear{Greenspan}{1968}]{greenspan}
Greenspan, H.P., {\itshape The theory of rotating fluids},  1968 (Cambridge
  University Press).

\bibitem[\protect\citeauthoryear{{Greenspan}}{1969}]{greenspan1969}
{Greenspan}, H.P., {On the non-linear interaction of inertial modes}. {\itshape
  \jfm}, 1969, \textbf{36}, 257--264.

\bibitem[\protect\citeauthoryear{{Herault}
  {\itshape{et~al.}}}{2018}]{herault2018}
{Herault}, J., {Giesecke}, A., {Gundrum}, T. and {Stefani}, F., {Instability of
  precession driven Kelvin modes: evidence of a detuning effect}. {\itshape
  ArXiv e-prints}, 2018 Submitted to Phys. Rev. Fluids.

\bibitem[\protect\citeauthoryear{{Herault}
  {\itshape{et~al.}}}{2015}]{herault2015}
{Herault}, J., {Gundrum}, T., {Giesecke}, A. and {Stefani}, F., {Subcritical
  transition to turbulence of a precessing flow in a cylindrical vessel}.
  {\itshape \pof}, 2015, \textbf{27}, 124102.

\bibitem[\protect\citeauthoryear{{Herreman}}{2009}]{herreman2009phd}
{Herreman}, W., Instabilit{\'e} elliptique sous champ magn{\'e}tique \& Dynamo
  d'ondes inertielles. Ph.D. Thesis, Universit{\'e} de Provence Aix-Marseille
  I, Insitut de Recherche sur les Ph{\'e}nom{\`enes Hors {\'E}quilibre}
  Https://tel.archives-ouvertes.fr/tel-00452471/document, 2009.

\bibitem[\protect\citeauthoryear{{Hollerbach} and
  {Kerswell}}{1995}]{hollerbach1995}
{Hollerbach}, R. and {Kerswell}, R.R., {Oscillatory internal shear layers in
  rotating and precessing flows}. {\itshape \jfm}, 1995, \textbf{298},
  327--339.

\bibitem[\protect\citeauthoryear{{Horimoto} and {Goto}}{2017}]{horimoto2017}
{Horimoto}, Y. and {Goto}, S., {Sustaining mechanism of small-scale turbulent
  eddies in a precessing sphere}. {\itshape Physical Review Fluids}, 2017,
  \textbf{2}, 114603.

\bibitem[\protect\citeauthoryear{{Horimoto}
  {\itshape{et~al.}}}{2018}]{horimoto2018}
{Horimoto}, Y., {Simonet-Davin}, G., {Katayama}, A. and {Goto}, S., {Impact of
  a small ellipticity on the sustainability condition of developed turbulence
  in a precessing spheroid}. {\itshape Phys. Rev. Fluids}, 2018, \textbf{3},
  044603.

\bibitem[\protect\citeauthoryear{{Kerswell}}{1993}]{kerswell1993}
{Kerswell}, R.R., {The instability of precessing flow}. {\itshape \gafd}, 1993,
  \textbf{72}, 107--144.

\bibitem[\protect\citeauthoryear{{Kirillov}
  {\itshape{et~al.}}}{2009}]{kirillov2009}
{Kirillov}, O.N., {G{\"u}nther}, U. and {Stefani}, F., {Determining role of
  Krein signature for three-dimensional Arnold tongues of oscillatory dynamos}.
  {\itshape \pre}, 2009, \textbf{79}, 016205.

\bibitem[\protect\citeauthoryear{{Kobine}}{1995}]{kobine1995}
{Kobine}, J.J., {Inertial wave dynamics in a rotating and precessing cylinder}.
  {\itshape \jfm}, 1995, \textbf{303}, 233--252.

\bibitem[\protect\citeauthoryear{{Kobine}}{1996}]{kobine1996}
{Kobine}, J.J., {Azimuthal flow associated with inertial wave resonance in a
  precessing cylinder}. {\itshape \jfm}, 1996, \textbf{319}, 387--406.

\bibitem[\protect\citeauthoryear{Lagrange
  {\itshape{et~al.}}}{2011}]{lagrange2011}
Lagrange, R., Meunier, P., Nadal, F. and Eloy, C., Precessional instability of
  a fluid cylinder. {\itshape \jfm}, 2011, \textbf{666}, 104--145.

\bibitem[\protect\citeauthoryear{{L{\'e}orat}}{2006}]{leorat2006}
{L{\'e}orat}, J., Large scales features of a flow driven by precession.
  {\itshape Magnetohydrodynamics}, 2006, \textbf{42}, 143--151.

\bibitem[\protect\citeauthoryear{L{\'e}orat
  {\itshape{et~al.}}}{2003}]{leorat2003}
L{\'e}orat, J., Rigaud, F., Vitry, R. and Herpe, G., Dissipation in a flow
  driven by precession and application to the design of a MHD wind tunne.
  {\itshape Magnetohydrodynamics}, 2003, \textbf{39}, 321--326.

\bibitem[\protect\citeauthoryear{{Liao} and {Zhang}}{2012}]{liao2012}
{Liao}, X. and {Zhang}, K., {On flow in weakly precessing cylinders: the
  general asymptotic solution}. {\itshape \jfm}, 2012, \textbf{709}, 610--621.

\bibitem[\protect\citeauthoryear{{Lin} {\itshape{et~al.}}}{2016}]{lin2016}
{Lin}, Y., {Marti}, P., {Noir}, J. and {Jackson}, A., {Precession-driven
  dynamos in a full sphere and the role of large scale cyclonic vortices}.
  {\itshape \pof}, 2016, \textbf{28}, 066601.

\bibitem[\protect\citeauthoryear{{Lin} {\itshape{et~al.}}}{2014}]{lin2014}
{Lin}, Y., {Noir}, J. and {Jackson}, A., {Experimental study of fluid flows in
  a precessing cylindrical annulus}. {\itshape \pof}, 2014, \textbf{26},
  046604.

\bibitem[\protect\citeauthoryear{{Lopez} and {Marques}}{2016}]{lopez2016}
{Lopez}, J.M. and {Marques}, F., {Nonlinear and detuning effects of the
  nutation angle in precessionally forced rotating cylinder flow}. {\itshape
  \prf}, 2016, \textbf{1}, 023602.

\bibitem[\protect\citeauthoryear{{Lorenzani} and
  {Tilgner}}{2001}]{lorenzani2001}
{Lorenzani}, S. and {Tilgner}, A., {Fluid instabilities in precessing
  spheroidal cavities}. {\itshape \jfm}, 2001, \textbf{447}, 111--128.

\bibitem[\protect\citeauthoryear{{Lorenzani} and
  {Tilgner}}{2003}]{lorenzani2003}
{Lorenzani}, S. and {Tilgner}, A., {Inertial instabilities of fluid flow in
  precessing spheroidal shells}. {\itshape \jfm}, 2003, \textbf{492}, 363--379.

\bibitem[\protect\citeauthoryear{{Malkus}}{1968}]{malkus1968}
{Malkus}, W.V.R., {Precession of the Earth as the Cause of Geomagnetism}.
  {\itshape Science}, 1968, \textbf{160}, 259--264.

\bibitem[\protect\citeauthoryear{{Manasseh}}{1992}]{manasseh1992}
{Manasseh}, R., {Breakdown regimes of inertia waves in a precessing cylinder}.
  {\itshape \jfm}, 1992, \textbf{243}, 261--296.

\bibitem[\protect\citeauthoryear{{Manasseh}}{1994}]{manasseh1994}
{Manasseh}, R., {Distortions of inertia waves in a rotating fluid cylinder
  forced near its fundamental mode resonance}. {\itshape \jfm}, 1994,
  \textbf{265}, 345--370.

\bibitem[\protect\citeauthoryear{{Marques} and {Lopez}}{2015}]{marques2015}
{Marques}, F. and {Lopez}, J.M., {Precession of a rapidly rotating cylinder
  flow: traverse through resonance}. {\itshape \jfm}, 2015, \textbf{782},
  63--98.

\bibitem[\protect\citeauthoryear{{McEwan}}{1970}]{mcewan1970}
{McEwan}, A.D., {Inertial oscillations in a rotating fluid cylinder}. {\itshape
  \jfm}, 1970, \textbf{40}, 603--640.

\bibitem[\protect\citeauthoryear{{Meunier}
  {\itshape{et~al.}}}{2008}]{meunier2008}
{Meunier}, P., {Eloy}, C., {Lagrange}, R. and {Nadal}, F., {A rotating fluid
  cylinder subject to weak precession}. {\itshape \jfm}, 2008, \textbf{599},
  405--440.

\bibitem[\protect\citeauthoryear{{Mouhali}
  {\itshape{et~al.}}}{2012}]{mouhali2012}
{Mouhali}, W., {Lehner}, T., {L{\'e}orat}, J. and {Vitry}, R., {Evidence for a
  cyclonic regime in a precessing cylindrical container}. {\itshape Exp.
  Fluids}, 2012, \textbf{53}, 1693--1700.

\bibitem[\protect\citeauthoryear{{Noir}
  {\itshape{et~al.}}}{2001{\natexlab{a}}}]{noir2001a}
{Noir}, J., {Brito}, D., {Aldridge}, K. and {Cardin}, P., {Experimental
  evidence of inertial waves in a precessing spheroidal cavity}. {\itshape
  \grl}, 2001{\natexlab{a}}, \textbf{28}, 3785--3788.

\bibitem[\protect\citeauthoryear{{Noir} {\itshape{et~al.}}}{2003}]{noir2003}
{Noir}, J., {Cardin}, P., {Jault}, D. and {Masson}, J.P., {Experimental
  evidence of non-linear resonance effects between retrograde precession and
  the tilt-over mode within a spheroid}. {\itshape \gji}, 2003, \textbf{154},
  407--416.

\bibitem[\protect\citeauthoryear{{Noir} and {C{\'e}bron}}{2013}]{noir2013}
{Noir}, J. and {C{\'e}bron}, D., {Precession-driven flows in non-axisymmetric
  ellipsoids}. {\itshape \jfm}, 2013, \textbf{737}, 412--439.

\bibitem[\protect\citeauthoryear{{Noir}
  {\itshape{et~al.}}}{2001{\natexlab{b}}}]{noir2001b}
{Noir}, J., {Jault}, D. and {Cardin}, P., {Numerical study of the motions
  within a slowly precessing sphere at low Ekman number}. {\itshape \jfm},
  2001{\natexlab{b}}, \textbf{437}, 283--299.

\bibitem[\protect\citeauthoryear{{Nore} {\itshape{et~al.}}}{2011}]{nore2011}
{Nore}, C., {L{\'e}orat}, J., {Guermond}, J.L. and {Luddens}, F., {Nonlinear
  dynamo action in a precessing cylindrical container}. {\itshape \pre}, 2011,
  \textbf{84}, 016317.

\bibitem[\protect\citeauthoryear{{P{\'e}tr{\'e}lis} and
  {Fauve}}{2006}]{petrelis2006}
{P{\'e}tr{\'e}lis}, F. and {Fauve}, S., {Inhibition of the dynamo effect by
  phase fluctuations}. {\itshape Europhys Lett.)}, 2006, \textbf{76}, 602--608.

\bibitem[\protect\citeauthoryear{{P{\'e}tr{\'e}lis}
  {\itshape{et~al.}}}{2007}]{petrelis2007}
{P{\'e}tr{\'e}lis}, F., {Mordant}, N. and {Fauve}, S., {On the magnetic fields
  generated by experimental dynamos}. {\itshape \gafd}, 2007, \textbf{101},
  289--323.

\bibitem[\protect\citeauthoryear{Poincar{\'e}}{1910}]{poincare1910}
Poincar{\'e}, H., {Sur la pr{\'e}cession des corps d{\'e}formables}. {\itshape
  Bulletin Astronomique}, 1910, \textbf{27}, 321--356.

\bibitem[\protect\citeauthoryear{{Stefani}
  {\itshape{et~al.}}}{2012}]{stefani2012}
{Stefani}, F., {Eckert}, S., {Gerbeth}, G., {Giesecke}, A., {Gundrum}, T.,
  {Steglich}, C., {Weier}, T. and {Wustmann}, B., {DRESDYN -- a new facility
  for MHD experiments with liquid sodium}. {\itshape Magnetohydrodynamics},
  2012, \textbf{48}, 103--114.

\bibitem[\protect\citeauthoryear{{Stewartson} and
  {Roberts}}{1963}]{stewartson1963}
{Stewartson}, K. and {Roberts}, P.H., {On the motion of a liquid in a
  spheroidal cavity of a precessing rigid body}. {\itshape \jfm}, 1963,
  \textbf{17}, 1--20.

\bibitem[\protect\citeauthoryear{{Thomson}}{1880}]{kelvin1880}
{Thomson}, W., Vibrations of a columnar vortex. {\itshape Philos. Mag.}, 1880,
  \textbf{10}, 152--165.

\bibitem[\protect\citeauthoryear{{Tilgner}}{1998}]{tilgner1998}
{Tilgner}, A., {On models of precession driven core flow}. {\itshape Studia
  geoph. et geod.}, 1998, \textbf{42}, 232--238.

\bibitem[\protect\citeauthoryear{{Tilgner}}{2005}]{tilgner2005}
{Tilgner}, A., {Precession driven dynamos}. {\itshape \pof}, 2005, \textbf{17},
  034104.

\bibitem[\protect\citeauthoryear{{Tilgner}}{2007}]{tilgner2007b}
{Tilgner}, A., {Zonal Wind Driven by Inertial Modes}. {\itshape Phys. Rev.
  Lett.}, 2007, \textbf{99}, 194501.

\bibitem[\protect\citeauthoryear{{Vanyo}}{1991}]{vanyo1991}
{Vanyo}, J.P., {A geodynamo powered by luni-solar precession}. {\itshape
  \gafd}, 1991, \textbf{59}, 209--234.

\bibitem[\protect\citeauthoryear{{Vanyo} and {Likins}}{1971}]{vanyo1971}
{Vanyo}, J.P. and {Likins}, P.W., {Measurement of Energy Dissipation in a
  Liquid-Filled, Precessing, Spherical Cavity}. {\itshape \jam}, 1971,
  \textbf{38}, 674.

\bibitem[\protect\citeauthoryear{{Wu} and {Roberts}}{2009}]{wu2009}
{Wu}, C.C. and {Roberts}, P., {On a dynamo driven by topographic precession}.
  {\itshape \gafd}, 2009, \textbf{103}, 467--501.

\end{thebibliography}
\end{document}